\documentclass[a4paper,11pt]{article}

\usepackage{a4wide}
\usepackage[latin1]{inputenc}
\usepackage[T1]{fontenc}
\usepackage[english]{babel}
\usepackage{graphicx,subfigure}
\usepackage{geometry}
\usepackage{bm}
\usepackage{color}
\usepackage{latexsym}
\usepackage{amssymb}

\usepackage[numbers]{natbib}

\begin{document}

%
%

\begin{center}
\LARGE{Turbulence without Richardson-Kolmogorov cascade}
\end{center}

\begin{center}
$\mbox{N. Mazellier}^{1}$\footnote{Permanent address: Institut
PRISME, 8, rue L{\'e}onard de Vinci, 45072 Orl{\'e}ans, FRANCE} $\mbox{ \& J.C. Vassilicos}^{1,2}$
\end{center}


\begin{center}
\small{$\mbox{}^1$Turbulence, Mixing and Flow Control Group, Department of Aeronautics\\
$\mbox{}^2$Institute for Mathematical Sciences\\
Imperial College London, London, SW7 2BY, UK\\}
\end{center}


\centerline{\bf Abstract}

\noindent
We investigate experimentally wind tunnel turbulence generated by
multiscale/fractal grids pertaining to the same class of low-blockage
space-filling fractal square grids. These grids are not active and
nevertheless produce very much higher turbulence intensities $u'/U$
and Reynolds numbers $Re_{\lambda}$ than higher blockage regular
grids. Our hot wire anemometry confirms the existence of a protracted
production region where turbulence intensity grows followed by a decay
region where it decreases, as first reported by Hurst \& Vassilicos
\citep{HurstVassilicos2007}. We introduce the wake-interaction
length-scale $x_*$ and show that the peak of turbulence intensity
demarcating these two regions along the centreline is positioned at
about $0.5x_*$. The streamwise evolutions on the centreline of the
streamwise mean flow and of various statistics of the streamwise
fluctuating velocity all scale with $x_*$. Mean flow and turbulence
intensity profiles are inhomogeneous at streamwise distances from the
fractal grid smaller than $0.5 x_*$, but appear quite homogeneous
beyond $0.5 x_*$. The velocity fluctuations are highly non-gaussian in
the production region but approximately gaussian in the decay
region. Our results confirm the finding of Seoud \& Vassilicos
\citep{SeoudVassilicos2007} that the ratio of the integral length-scale
$L_u$ to the Taylor microscale $\lambda$ remains constant even though
the Reynolds number $Re_\lambda$ decreases during turbulence decay in
the region beyond $0.5 x_*$. As a result the scaling $L_{u}/\lambda
\sim Re_{\lambda}$ which follows from the $u'^{3}/L_u$ scaling of the
dissipation rate in boundary-free shear flows and in usual
grid-generated turbulence does not hold here. This extraordinary
decoupling is consistent with a non-cascading and instead
self-preserving single-length scale type of decaying homogeneous
turbulence proposed by George \& Wang \citep{GeorgeWang2009}, but we
also show that $L_{u}/\lambda$ is nevertheless an increasing function
of the inlet Reynolds number $Re_0$. Finally, we offer a detailed
comparison of the main assumption and consequences of the George \&
Wang theory against our fractal-generated turbulence data.

\vskip 1truecm

\section{Introduction}

Which turbulence properties are our current best candidates for
universality or, at least, for the definition of universality classes?
The assumed independence of the turbulence kinetic energy dissipation
rate on Reynolds number Re in the high Re limit is a cornerstone
assumption on which Kolmogorov's phenomenology is built and on which
one-point and two-point closures and LES rely, whether directly or
indirectly \citep{TennekesLumley1972}, \citep{Frisch1995},
\citep{Pope2000}, \citep{SagautCambon2008}, \citep{Lesieur2008}. This
cornerstone assumption is believed to hold universally (at least for
relatively weakly strained/sheared turbulent flows). It is also
related to the universal tendency of turbulent flows to develop sharp
velocity gradients within them and to the apparently universal
geometrical statistics of these gradients \citep{SreeniAntonia1997}, as
to the apparently universal mix of vortex stretching and compression
(described in some detail by Tsinober \citep{Tsinober2009} who
introduced the expression ``qualitative universality" to describe such
ubiquitous qualitative properties). 


Evidence against universality has been reported since the 1970s, if
not earlier, in works led by Roshko, Lykoudis, Wygnanski, Champagne
and George (see for example \citep{George2008} and references therein
as well as the landmark work of Bevilaqua \& Lykoudis
\citep{BevilaquaLykoudis1978} and more recent works such as
\citep{TongWarhaft1994} and \citep{Lavoieetal2005} to cite but a few)
and has often been accounted for by the presence or absence of
long-lived coherent structures. Coherent/persistent flow structures can
actually appear at all scales and can be the carrier of long-range
memory, thus implying long-range effects of boundary/inlet
conditions. 

In summary, kinetic energy dissipation,
vortex stretching and compression, geometrical alignments,
coherent structures
and the universality or non-universality of each one of these
properties are central to turbulent flows with an impact which
includes engineering turbulence modelling and basic Kolmogorov
phenomenology and scalings.
Is it possible to tamper with these properties by systematic
modifications of a flow's boundary and/or inlet/upstream conditions?

To investigate such questions, new classes of turbulent flows have
recently been proposed which allow for systematic and well-controlled
changes in multiscale boundary and/or upstream conditions. These new
classes of flows fall under the general banner of ``fractal-generated
turbulence" or ``multiscale-generated turbulence'' (the term
``fractal'' is to be understood here in the broadest sense of a
geometrical structure which cannot be described by any non-multiscale
way, which is why we refer to fractal and multiscale grids
interchangeably). These flows have such unusual turbulence properties
\citep{HurstVassilicos2007}, \citep{SeoudVassilicos2007} that they may
directly serve as new flow concepts for new industrial flow solutions,
for example conceptually new energy-efficient industrial mixers
\citep{Coffeyetal2009}. 
These same turbulent flow concepts in conjunction with conventional
flows such as turbulent jets and regular grid turbulence have also
been used recently for fundamental research into what determines the
dissipation rate of turbulent flows and even to demonstrate the
possibility of renormalising the dissipation constant so as to make it
universal at finite, not only asymptotically infinite, Reynolds
numbers (see \citep{MazellierVassilicos2008},
\citep{GotoVassilicos2009}). These works have shown that the
dissipation rate constant depends on small-scale intermittency, on
dissipation range broadening and on the large-scale internal
stagnation point structure which itself depends on boundary and/or
upstream conditions. In the case of at least one class of
multiscale-generated homogeneous turbulence, small-scale intermittency
does not increase with Reynolds number \citep{Stresingetal2009} and the
dissipation constant is inversely proportional to turbulence intensity
even though the energy spectrum is broad with a clear power-law shaped
intermediate range \citep{SeoudVassilicos2007},
\citep{HurstVassilicos2007}. In this paper we investigate this
particular class of multiscale-generated turbulent flows: turbulent
flows generated by low-blockage space-filling fractal square grids.

Grid-generated wind tunnel turbulence has been extensively
investigated over more than seventy years \citep{BatchelorTownsend1948}
and is widely used to create turbulence under well controlled
conditions.  This flow has the great advantage of being nearly
homogeneous and isotropic downstream \citep{ComteBellotCorrsin1966}.
However, its Reynolds number is not large enough for conclusive
fundamental studies and industrial mixing applications. Several
attempts have been made to modify the grid so as to increase the
Reynolds number whilst keeping as good homogeneity and isotropy as
possible: for example jet-grids by Mathieu's \citep{MathieuAlcaraz1965}
and Corrsin's \citep{GadElHakCorrsin1974} groups (who may have been
inspired by Betchov's porcupine \citep{Betchov1957}), non-stationary,
so-called active, grids by Makita \citep{Makita1991} followed by
Warhaft's group \citep{MydlarskiWarhaft1996} and others, and most
recently passive grids with tethered spheres attached at each mesh
corner \citep{VonlanthenMonkewitz2008}. Jet-grids and active grids have
been very successful in increasing both the integral length-scale and
the turbulence intensity whilst keeping a good level of homogeneity
and isotropy. The three different families of fractal/multiscale grids
introduced by Hurst \& Vassilicos \citep{HurstVassilicos2007} generate
turbulence which becomes approximately homogeneous and isotropic
considerably further downstream than jet-grids and active grids, but
achieve comparably high Reynolds numbers even though, unlike jet-grids
and active grids, they are passive. However, the most important reason
for studying fractal/multiscale-generated turbulence is that it can
have properties which are clearly qualitatively different from
properties which are believed to be universal to all other
grid-generated turbulent flows and even boundary-free shear flows for
that matter.

In this paper we report the results of an experimental investigation
of turbulent flows generated by four low-blockage space-filling
fractal square grids. The grids used in our study are described in the
next section and the experimental set up (wind tunnels and anemometry)
is presented in section 3. Our results are reported in section
4. Specifically, in subsection 4.1 we introduce the wake-interaction
length-scale $x_*$ and use it to derive and explain the scaling of the
downstream peak in turbulence intensity which was first reported in
\citep{HurstVassilicos2007}. We also show in this subsection that the
streamwise dependence of the streamwise turbulence intensity is
independent of inlet velocity and fractal grid parameters if $x_*$ is
used to scale streamwise distance. In subsection 4.2 we confirm the
far field statistical homogeneity first reported in
\citep{SeoudVassilicos2007} and, for the first time, present near-field
profiles illustrating the evolution from near-field inhomogeneity to
far-field homogeneity. Subsection 4.3 contains a detailed report on
the skewness and flatness of the fluctuating velocities illustrating
how they become gaussian in the far field following a clearly
non-gaussian near-field behaviour which peaks at $0.2 x_*$. Finally,
in subsection 4.4 we report a significant improvement and
generalisation of the single-scale self-preservation theory of George
\& Wang \citep{GeorgeWang2009} which shows that there are many more
single-scale solutions to the spectral energy equation than originally
thought. Subsections 4.5 to 4.8 make use of this multiplicity of
solutions for an analysis of our data that is significantly finer than
in previous studies of fractal-generated turbulence and which confirms
the self-preserving single-scale nature of the far-field decaying
fractal-generated turbulence in terms of the behaviours of the
integral scale, the Taylor microscale, the energy spectrum and the
turbulence intensity.

Finally, in section 5 we conclude and discuss some of the issues
raised by our investigation.


\section{The space-filling fractal square grids}
	
\begin{figure}[htbp]
\centering
\subfigure
{\includegraphics[scale=0.35]{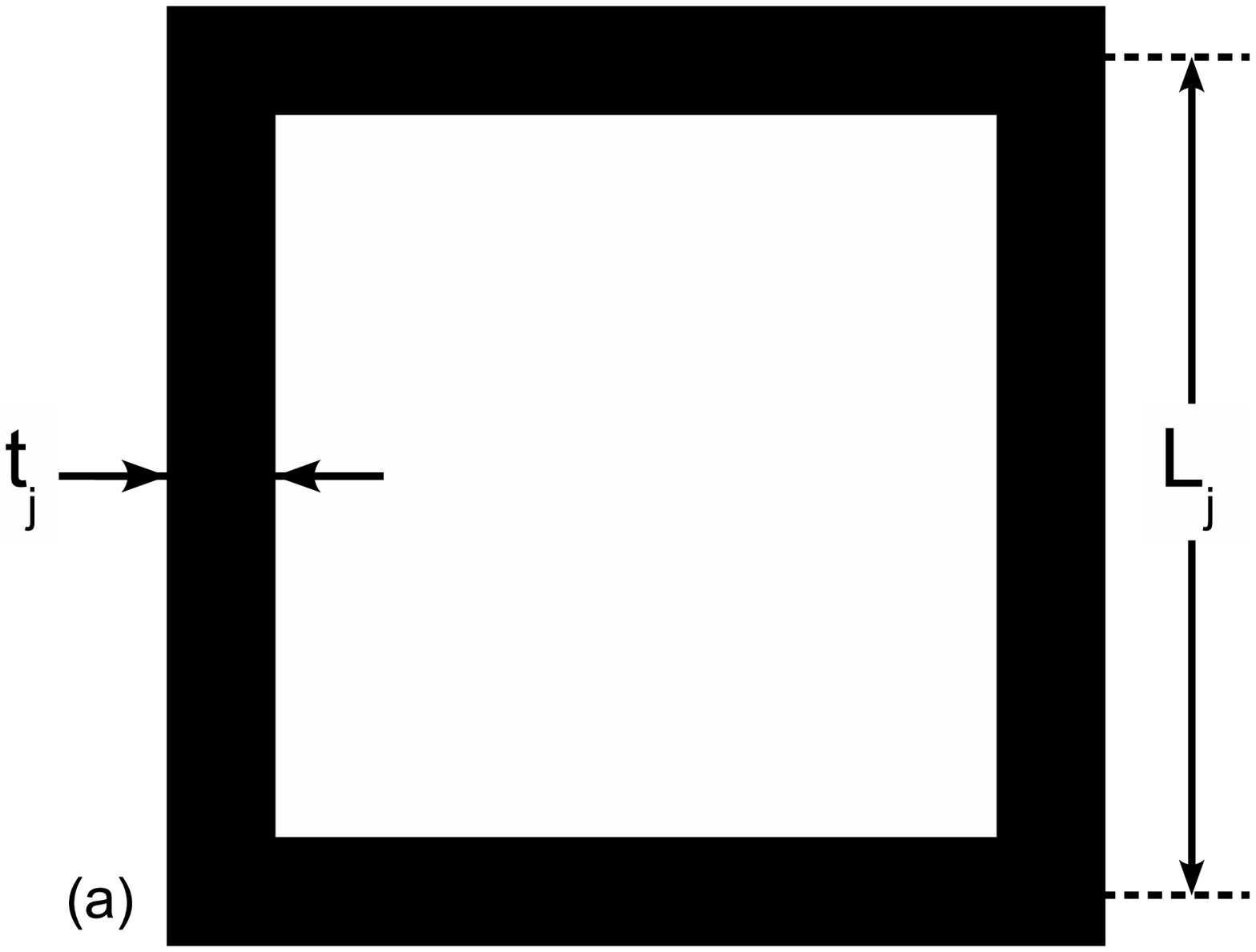}
\label{fig:GridPattern}}
\hspace{0.2cm}
\subfigure
{\includegraphics[scale=0.35]{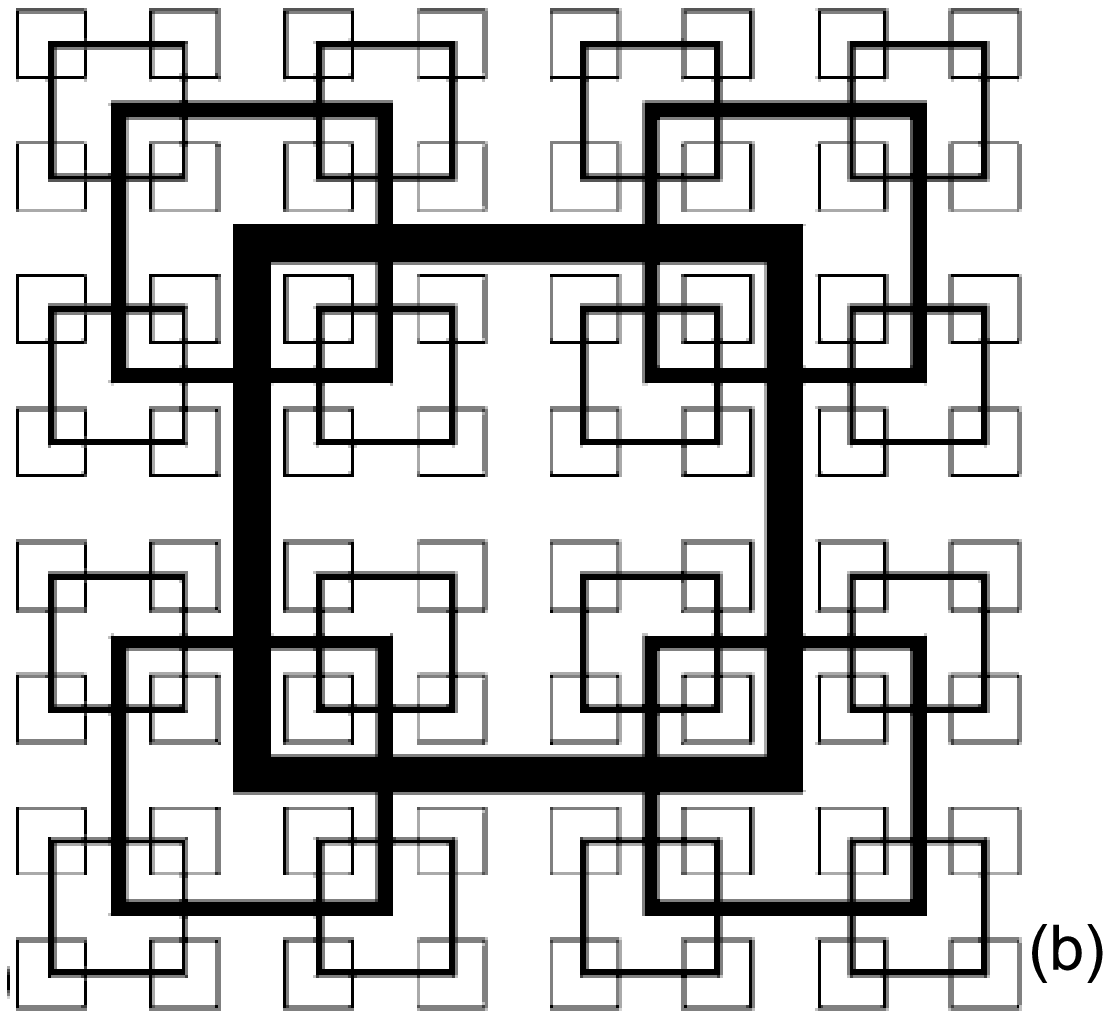}
\label{fig:SFG13}}
\hspace{0.2cm}
\subfigure
{\includegraphics[scale=0.35]{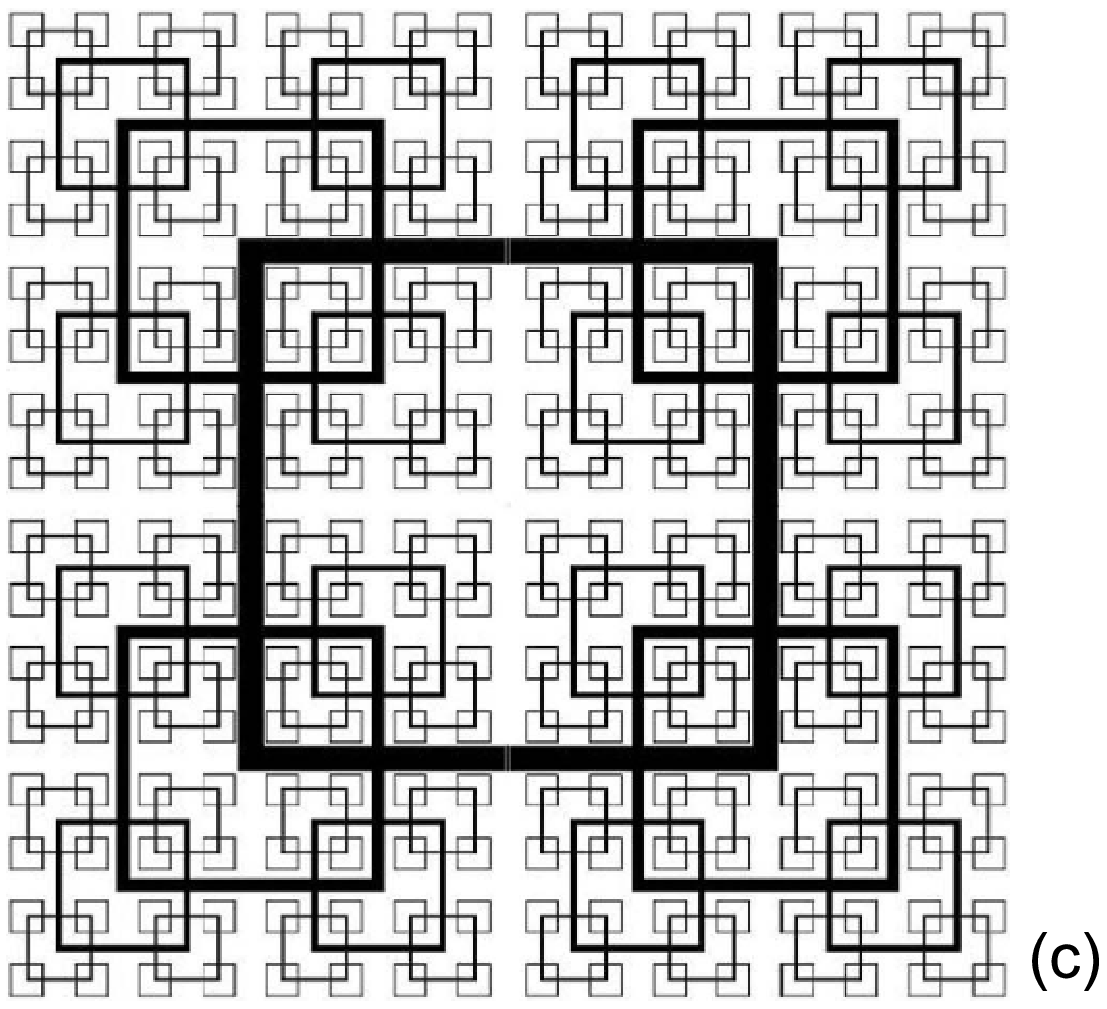}
\label{fig:BFG17}}
\caption{\textit{(a)} Space-filling multiscale/fractal square grid
  generating pattern. Examples of planar multiscale/fractal square
  grids used in the present work with N=4 scales \textit{(b)} and N=5
  scales \textit{(c)}.}
\end{figure}

Turbulent flows are generated in this study by the planar and
space-filling multiscale/fractal square grids first introduced and
described in \citep{HurstVassilicos2007}. The main characteristics of
those grids are summarised as follows. In general, multiscale/fractal
grids consist of a multiscale collection of obstacles/openings which
may be all based on a single specific pattern that is repeated in
increasingly numerous copies at smaller scales. For the present work,
the pattern used is a empty square framed by four rectangular bars as
shown in figure \ref{fig:GridPattern}. Each scale-iteration $j$ is
characterized by a length $L_j$ and a thickness $t_j$ of these
bars. At iteration $j$ there are four times more square patterns that
at iteration $j-1$ ($1\le j\le N$ where $N$ is the total number of
scales) and their dimensions are related by $L_{j} = R_L L_{j-1}$ and
$t_{j} = R_t t_{j-1}$. The scaling factors $R_L$ and $R_t$ are
independent of $j$ and are smaller or equal than $1/2$ and $1$
respectively. As explained in \citep{HurstVassilicos2007}, the fractal
square grid is space filling when its fractal dimension takes the
maximum value 2, which is the case when $R_L = 1/2$.

A total of four different planar space-filling fractal square grids
have been used in the wind tunnel experiments reported here. The
complete planar geometry of these grids is detailed in table
\ref{tab:FractalGridSize}. Scaled-down diagrams of two of these grids
are displayed in figures \ref{fig:SFG13} and
\ref{fig:BFG17}. Multiscale/fractal grids are clearly designed to
generate turbulence by directly exciting a wide range of fluctuation
length-scales in the flow rather than by relying on the non-linear
cascade mechanism for multiscale excitation. The latter approach is
the classical one and is exemplified by the use of regular grids
as homogeneous turbulence generators.



\begin{table}[htbp]
\begin{center}
\begin{tabular}{ccccc}
\hline
Grid	&	\emph{SFG8}	&	\emph{SFG13}	&	\emph{SFG17}	&	\emph{BFG17} \\
\hline
\hline
$L_0 (mm)$	&	237.5	&	237.7	&	237.8	&	471.2	\\
$L_1 (mm)$	&	118.8	&	118.9	&	118.9	&	235.6	\\
$L_2 (mm)$	&	59.4	&	59.4	&	59.5	&	117.8	\\
$L_3 (mm)$	&	29.7	&	29.7	&	29.7	&	58.9	\\
$L_4 (mm)$	&	  -	  &	  -	  &	  -	  &	29.5	\\
$t_0 (mm)$	&	14.2	&	17.2	&	19.2	&	23.8	\\
$t_1 (mm)$	&	6.9	  &	7.3	  &	7.5	  &	11.7	\\
$t_2 (mm)$	&	3.4	  &	3.1	  &	2.9	  &	5.8	\\
$t_3 (mm)$	&	1.7	  &	1.3	  &	1.1	  &	2.8	\\
$t_4 (mm)$	&	  -	  &	  -	  &	  -	  &	1.4	\\
\hline
\end{tabular}
\caption{Geometry of the space-filling fractal square grids.}
\label{tab:FractalGridSize}
\end{center}
\end{table}

As explained in \citep{HurstVassilicos2007}, the
complete design of space-filling grids requires a total of four independent
parameters such as :

			\begin{itemize}			
				\item $N$ : the number of scales
				($N-1$ being the number of scale-iterations),
				\item $L_0$ : the biggest bar length of the grid
				\item $t_0$ : the biggest bar thickness of the grid
				\item $t_{N-1}$ : the smallest bar thickness of the grid
			\end{itemize}

The smallest bar length $L_{N-1}$ on the grid is
determined by $R_L = 1/2$ and $N$. Note, also, that the 
fractal grids are manufactured from an acrylic plate with a constant 
thickness ($5 mm$) in the direction of the mean flow. 
 

Hurst \& Vassilicos \citep{HurstVassilicos2007}
introduced the thickness ratio $t_r = \frac{t_0}{t_{N-1}}$ and the
effective mesh size
\begin{equation}
M_{eff} = \frac{4 T^2}{P} \sqrt{1 - \sigma}
\end{equation}
where $P$ is the fractal perimeter length of the grid, $T^{2}$ is the
tunnel's square cross section and $\sigma$ is the blockage ratio of
the grid defined as the ratio of the area $A$ covered by the grid to
$T^{2}$:
\begin{equation}
\sigma = \frac{A}{T^2} = \frac{L_0 t_0 \sum_{j=0}^{N-1} 4^{j+1} R_L^j R_t^j - t_0^2 \sum_{j=1}^{N-1} 2^{2j+1} R_t^{2j-1}}{T^2}.
\end{equation}
 
These quantities are derivable from the few independent geometrical
parameters chosen to uniquely define the grids. When applied to a
regular grid, this definition of $M_{eff}$ returns its mesh size. When
applied to a multiscale grid where bar sizes and local blockage are
inhomogeneously distributed across the grids, it returns an average
mesh size which was shown in \citep{HurstVassilicos2007} to be fluid
mechanically relevant.




A total of four space-filling fractal square grids have been used in
the present work. They all have the same blockage ratio $\sigma =
0.25$ (low compared to regular grids where, typically, $\sigma$ is
about 0.35 or 0.4 or so \citep{Corrsin1963},
\citep{ComteBellotCorrsin1966}) and turn out to have values of
$M_{eff}$ which are all very close to $26.4mm$. 
Three of these grids, refered to as SFG8, SFG13 and SFG17, differ by
only one parameter, $t_r$, and as a consequence also by the values of
$t_j$ ($0\le j\le N-1$) as $t_r$ was chosen to be one of the four
all-defining parameters along with the fixed parameters $N=4$, $L_3 =
29.7mm$ and $\sigma = 0.25$. The fourth grid, BFG17, has one extra
iteration, i.e. $N=5$ instead of $N=4$ but effectively the same
smallest length, i.e. $L_4 = 29.5mm$, and a value of $t_r$ very close
to that of SFG17. It is effectively very similar to SFG17 but with one
extra fractal iteration. The main characteristics of these grids are
summarized in tables 1 and \ref{tab:FractalGridCharac} which also
includes values for $L_{r}\equiv L_{0}/L_{N-1}$. 
						
\begin{table}[htbp]
\begin{center}
\begin{tabular}{cccccccc}
\hline
Grid	&	$N$	&	$L_0 (mm)$	&	$t_{0} (mm)$	&	$L_r$	&	$t_r$	&	$\sigma$	&	$M_{eff} (mm)$	\\
\hline
\hline
SFG8	&	4	&	237.5	&	14.2	&	8	&	8.5	&	0.25	&	26.4	\\
SFG13	&	4	&	237.7	&	17.2	&	8	&	13.0	&	0.25	&	26.3	\\
SFG17	&	4	&	237.8	&	19.2	&	8	&	17.0	&	0.25	&	26.2	\\
BFG17	&	5	&	471.2	&	23.8	&	16	&	17.0	&	0.25	&	26.6	\\
\hline
\end{tabular}
\caption{Main characteristics of the fractal square grids.}
\label{tab:FractalGridCharac}
\end{center}
\end{table}

In addition to the fractal grids, we have also performed a comparative
study of turbulence generated by a regular grid, refered as
\emph{SRG} hereafter, made of a bi-plane square rod array. Table
\ref{tab:ClassicalGridCharac} presents the main properties of this
grid. Its blockage ratio is higher than that of our fractal grids and
closer to the usual values found in literature for regular grids (see
e.g. \citep{BatchelorTownsend1948}, \citep{Corrsin1963}). The regular
grid \emph{SRG} also has slightly higher mesh size.

\begin{table}[htbp]
\begin{center}
\begin{tabular}{cccccccc}
\hline Grid & $N$ & $L_0 (mm)$ & $t_0 (mm)$ & $L_r$ & $t_r$ & $\sigma$
& $M_{eff} (mm)$ \\ \hline \hline SRG & 4 & 460 & 6 & 8 & 1 & 0.34 &
32 \\ \hline
\end{tabular}
\caption{Main characteristics of the regular grid.}
\label{tab:ClassicalGridCharac}
\end{center}
\end{table}


\section{The experimental set-up}


	\subsection{The wind tunnels}

\begin{figure}[htbp]
\centering
\includegraphics[scale=0.4]{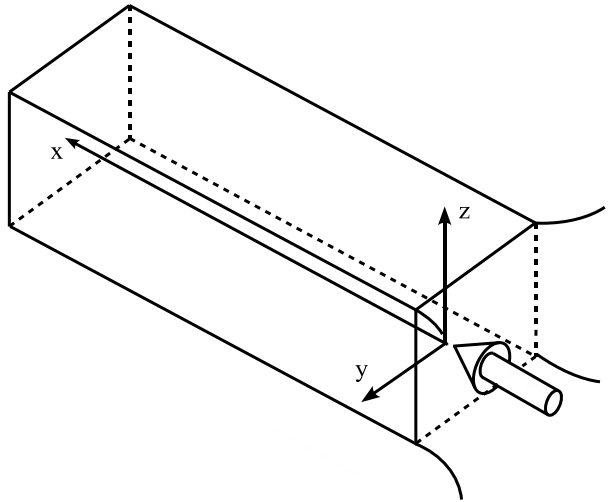}
\caption{Wind-tunnel sketch and coordinate notation}
\label{fig:WindTunnel}
\end{figure}

Measurements are performed in two air wind tunnels, one which is
open-circuit with a $5 m$ long and $T=0.46 m$ wide square test section
and one which is recirculating with a $5.4m$ long and $T=0.91m$ wide
square test section. A generic sketch of a tunnel's square test
section is given in figure \ref{fig:WindTunnel} for the purpose of
defining spatial coordinate notation. The arrow in this figure
indicates the direction of the mean flow and of the inlet velocity
$U_{\infty}$. The turbulence-generating grids are placed at the inlet
of the test section.
 
The fractal grids \emph{SFG8}, \emph{SFG13}, \emph{SFG17} and the
regular grid \emph{SRG} were tested in the open circuit tunnel whereas
the fractal grid BFG17 was tested in the recirculating tunnel.

The maximum flow velocity without a grid or any other obstruction is
$33 m/s$ in the $T=0.46 m$ open-circuit tunnel. Turbulence-generating
grids were tested with three values of the inlet velocity $U_\infty$
in this tunnel: $5.2 m/s$, $10 m/s$ and $15 m/s$. The uniformity of
the inlet velocity at the convergent's outlet, checked with Pitot tube
measurements, is better than 5\%. The residual turbulence intensity in
the absence of a turbulent-generating grid is about 0.4\% along the
axis of the tunnel.

In the $T=0.91m$ recirculating tunnel, the maximum flow velocity
without a grid or any other obstruction is $45 m/s$. The inlet
velocity $U_\infty$ was fixed at $5.2 m/s$ in this facility when
testing the turbulence generated by the BFG17 grid. The entrance flow
uniformity is better than 2\% and a very low residual turbulence
intensity ($\leq 0.05$\%) remains in the test section in the absence
of a turbulence-generating grid or obstacle.
			
In both tunnels, the temperature is monitored during measurement
campaigns thanks to a thermometer sensor located at the end of the
test section. The inlet velocity $U_\infty$ is imposed by measuring
the pressure difference in the tunnel's contractions with a
micromanometer Furness Controls MCD1001. 

	\subsection{Velocity measurements}
		
A single hot-wire, running in constant-temperature mode, was used to
measure the longitudinal velocity component. The DANTEC 55P01 single
probe was driven by a DISA 55M10 anemometer and the probe was mounted
on an aluminum frame allowing 3D displacements in space. A systematic
calibration of the probe was performed at the beginning and at the end
of each measurement campaign and the temperature was monitored for
thermal compensation. The sensing part of the wire (PT-0.1Rd) was $5
\mu m$ in diameter ($d_w$) and about $1 mm$ in length $l_w$ so that
the aspect ratio $\frac{l_w}{d_w}$ was about $200$. Our spatial
resolution $l_w / \eta$ ranges between $2$ and $9$ for all the
measurements. The estimated frequency response of this anemometry
system is $1.5$ to $9$ times higher than the Kolmogorov frequency
$f_\eta = \frac{U}{2 \pi \eta}$. The spatially-varying longitudinal
velocity component $\tilde{u}(x)$ in the direction of the mean flow
was recovered from the time-varying velocity $\tilde{u}(t)$ measured
with the hot-wire probe by means of local Taylor's hypothesis as
defined in \citep{Kahalerrasetal997}.
			
The signal coming from the anemometer was compensated and amplified
with a DISA 55D26 signal conditioner to enhance the signal to noise
ratio which is typically of the order of $45 dB$ for all measurements.
The uncertainties on the estimation of the dissipation
rate $\epsilon$ due to electronic noise occurring at high
frequencies (wavenumbers) is lower than 4\% for all our
measurements. The conditioned signal was low-pass
filtered to avoid aliasing and then sampled by a 16 bits National
Instruments NI9215 USB card. The sampling frequency was adjusted to be
slightly higher than twice the cut-off frequency. The sampled signal
was then stored on the hard-drive of a PC. The signal acquisition was
controlled with the commercial software LabVIEW\copyright, while the
post-processing was carried out with the commercial software
MATLAB\copyright.
			
The range of Reynolds numbers and lengths-scales of the turbulent
flows generated in both tunnels by all our grids are summarized in
table \ref{tab:WindTunnel}. The longitudinal integral length-scale
$L_u$ was obtained by integrating the autocorrelation function
of the fluctuating velocity component $u(x)$ (obtained by subtracting
the average value of $\tilde{u}(x)$ from $\tilde{u}(x)$):

\begin{equation}
L_{u} = \int_0^\infty \frac{\left\langle u(x) u(x+\Delta)\right\rangle}{\left\langle u(x)^2\right\rangle} d\Delta
\end{equation}
where the averages are taken over time, i.e. over $x$ in this
equation's notation, where $x$ is obtained from time $t$ by means of
the local Taylor hypothesis. In this paper we use the notation
$u'\equiv \sqrt{ \left\langle u(x)^2\right\rangle }$.

The Taylor microscale $\lambda$ was computed via the following
expression:

\begin{equation}
\lambda = \sqrt{\frac{\left\langle u^2\right\rangle}{\left\langle \left(\partial u/ \partial x\right)^2\right\rangle}}.
\end{equation}

Finally, using the kinematic viscosity $\nu$ of the fluid (here air at
ambient temperature) we also calculate the length-scale

\begin{equation}
\eta = \left(\frac{\nu^2}{15 \left\langle \left(\partial u/ \partial x\right)^2\right\rangle}\right)^{1/4}
\end{equation}
which is often refered to as Kolmogorov microscale. We have estimated from the turbulent
kinetic energy budget that the uncertainties in the computation of $\lambda$ and $\eta$
are lower than 10\% and 5\% respectively for all our measurements.

\begin{table}[htbp]
\begin{center}
\begin{tabular}{|c|c|c|c|c|c|}
\hline
$T [m]$ & $U_\infty [m/s]$ & $L_{u} [mm]$	&	$\lambda [mm]$	&	$\eta [mm]$ &	$Re_\lambda$ \\
\hline
$0.46$  & $5/10/15$ & $43 - 52$ & $4 - 7.4$ & $0.11 - 0.32$ & $140 - 370$\\
\hline
$0.91$ & $5$ & $50 - 70$ & $7 - 10$ & $0.3 - 0.45$ & $60 - 220$\\
\hline
\end{tabular}
\caption{Main flow characteristics: $T$ is the wind tunnel width,
$L_u$ is the longitudinal integral length-scale, $\lambda$ the
Taylor-microscale, $\eta$ the Kolmogorov scale and $Re_\lambda$ the
Taylor based Reynolds number.}
\label{tab:WindTunnel}
\end{center}
\end{table}


\section{Results}

Hurst \& Vassilicos \citep{HurstVassilicos2007} found that the
streamwise and spanwise turbulence velocity fluctuations generated by
the space-filling fractal square grids used here increase in intensity
along $x$ on the centreline till they reach a point $x=x_{peak}$
beyond which they decay. Thus they defined the production region as
being the region where $x < x_{peak}$ and the decay region as
being the region where $x > x_{peak}$. They also found that various
turbulence statistics collapse when plotted as functions of
$x/x_{peak}$ and they attempted to give an empirical formula for
$x_{peak}$ as a function of the geometric parameters of the fractal
grid. It was also clear in their results that the turbulent
intensities depend very sensitively on parameters of the fractal grids
even at constant blockage ratio, thus generating much higher
turbulence intensities than regular grids. An understanding and
determination of how $x_{peak}$ and turbulence intensities depend on
fractal grid geometry matters critically both for achieving a
fundamental understanding of multiscale-generated turbulence and for
potential applications such as in mixing and combustion. In such
applications, it is advantageous to generate desired high levels of
turbulence intensities at flexibly targeted downstream positions with
as low blockage ratios and, consequently, pressure drop and power
input, as possible. An important question left open, for example, in
\citep{HurstVassilicos2007} is whether $x_{peak}$ does or does not
depend on $U_\infty$.

This section is subdivided in eight subsections. In the first we study
the steamwise profiles of the streamwise mean velocity and turbulence
intensity and, in particular, determine $x_{peak}$. In the second we
offer data which describe how homogeneity of mean flow and turbulence
intensities is achieved when passing from the production to the decay
region. In the third subsection we present results on the turbulent
velocity skewness and flatness. The fourth and eightth subsections are
a careful application of the theory of George \& Wang \citep{GeorgeWang2009}
to our data and the fifth, sixth and seventh
subsections are an investigation of the single-length scale assumption
of this theory and its consequences, in particular the extraordinary
property first reported in \citep{SeoudVassilicos2007} that the ratio
of the integral to the Taylor length-scales is independent of
$Re_{\lambda} \equiv {u'\lambda/\nu}$ in the fractal-generated
homogeneous decaying turbulence beyond $x_{peak}$.




\subsection{The wake-interaction length-scale $x_\star$}

\begin{figure}[htbp]
\centering
\subfigure
{\includegraphics[width=7.5cm]{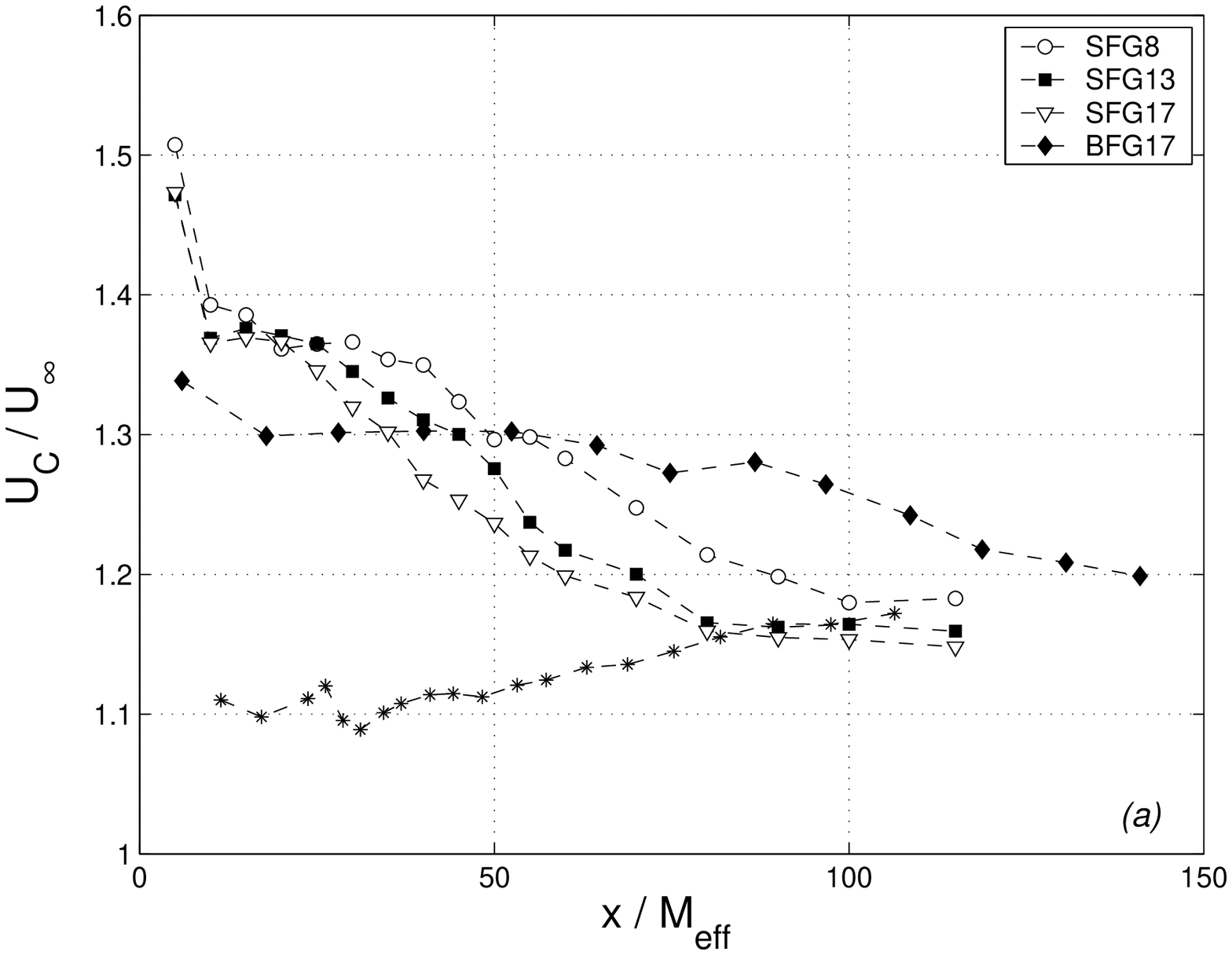}
\label{fig:MeanVel_Tr}}
\hspace{0.1cm}
\subfigure
{\includegraphics[width=7.5cm]{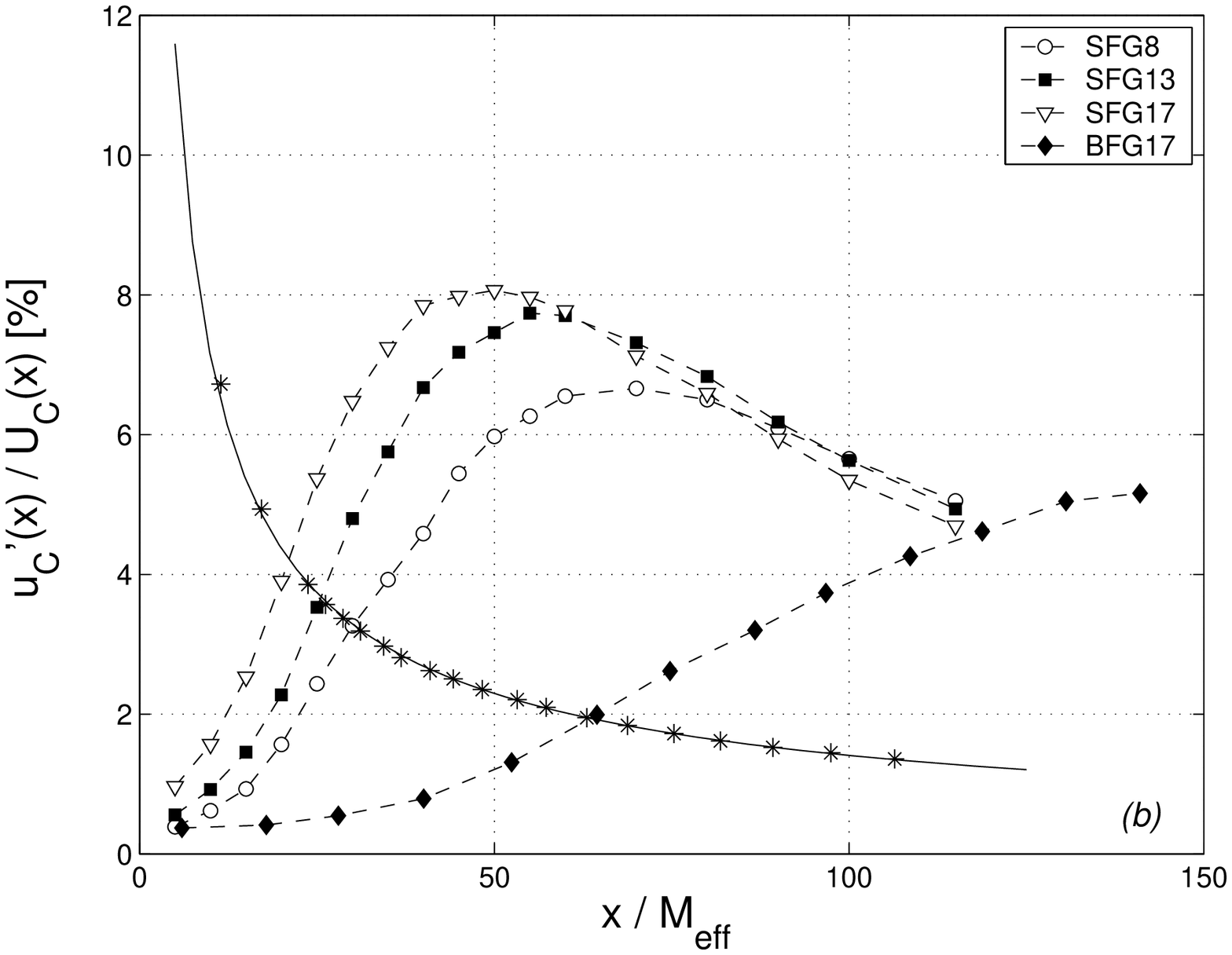}
\label{fig:TurbInt_Tr}}
\vspace{0.1cm}
\subfigure
{\includegraphics[width=7.5cm]{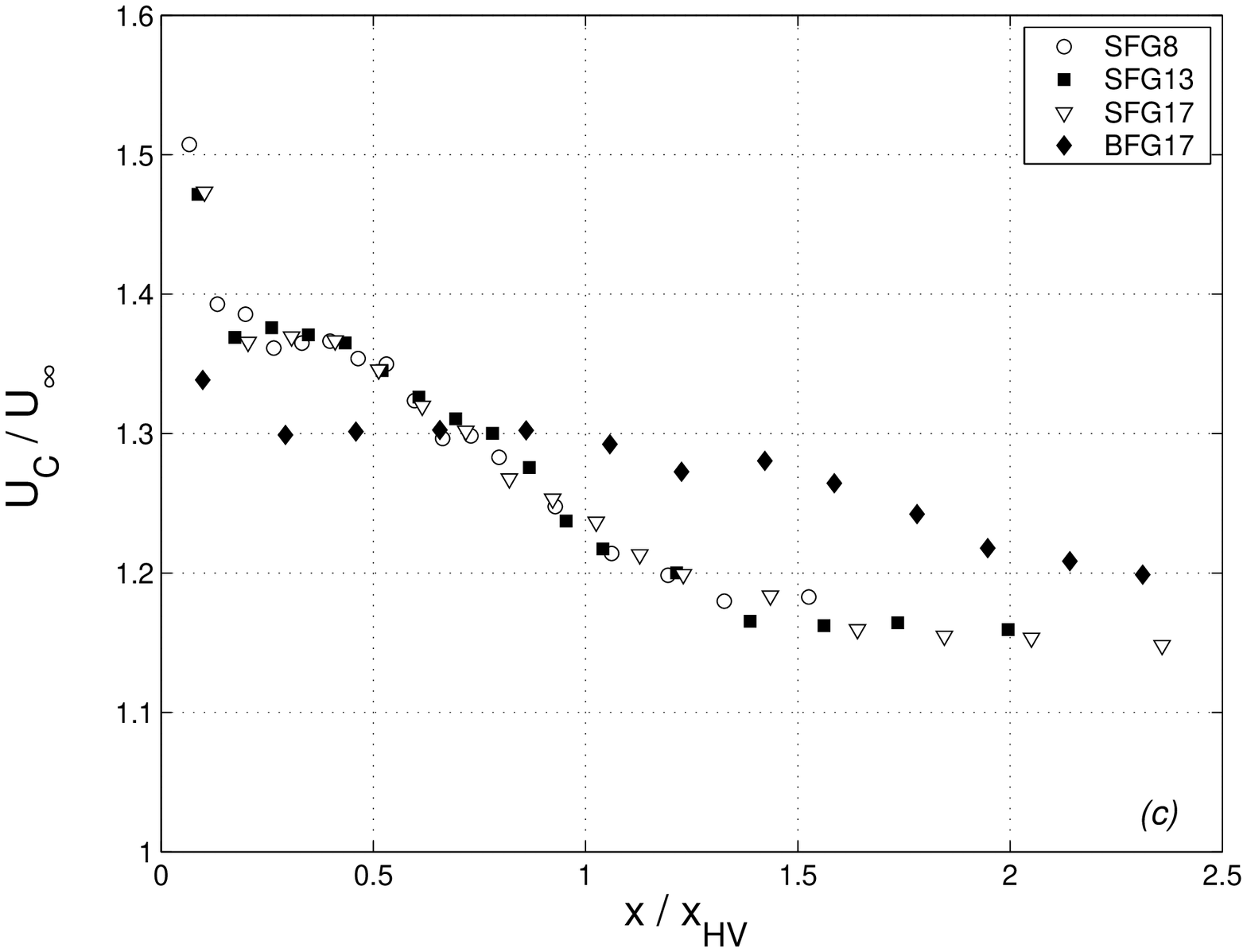}
\label{fig:MeanVel_Tr_HV}}
\hspace{0.1cm}
\subfigure
{\includegraphics[width=7.5cm]{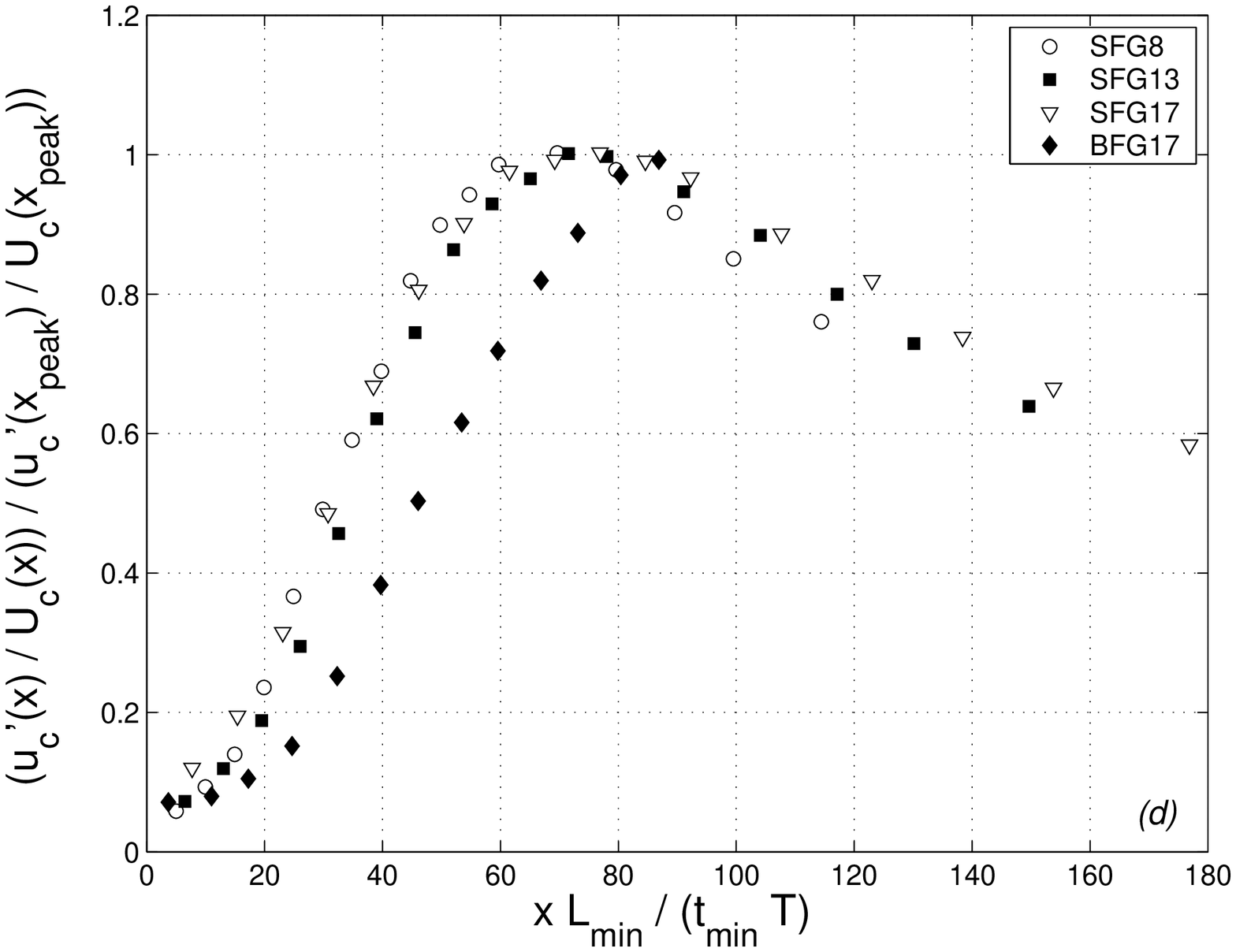}
\label{fig:TurbInt_Tr_NormHV}}
\caption{Streamwise evolution of \textit{(a)} the centreline mean
velocity $U_C / U_\infty$ and of \textit{(b)} the turbulence intensity
$u'_c / U_C$ for all fractal grids with respect to the streamwise
distance $x$ scaled by $M_{eff}$. Streamwise evolution of \textit{(c)}
the centreline mean velocity $U_C / U_\infty$ and of \textit{(d)} the
turbulence intensity $u'_c / U_C$ normalised by its value at
$x=x_{peak}$ for all fractal grids with respect to the streamwise
distance $x$ scaled by $x_{HV} = 75 t_{min}T/L_{min}$
\citep{HurstVassilicos2007} ($U_\infty = 5.2 m/s$). For reference, data
obtained with the regular grid \emph{SRG} and $U_\infty = 10 m/s$, are
also reported ($\ast$ symbols in \textit{(b)}). The power law $ (u'_c
/ U_C )^{2} = A \left(\frac{x -x_0}{M_{eff}}\right)^{-1.41}$ which is
also shown (solid line in \textit{(b)}, $A=0.129$, $x_{0}=0$) fits
this \emph{SRG} data very well.}
\end{figure}

The dimensionless centreline mean velocity $U_C / U_\infty$ and the
centreline turbulence intensity $u'_c/ U_C$ are plotted in
figures \ref{fig:MeanVel_Tr} and \ref{fig:TurbInt_Tr} for all
space-filling fractal square grids as well as for the regular grid
\textit{SRG}. For the latter, we have fitted the turbulence intensity
$u'_c/ U_C$ with the well-known power-law $A \left(\frac{x -
x_0}{M_{eff}}\right)^{-n}$ where the dimensional parameter $A$, the
exponent $n$ and the virtual origin $x_0$ have been empirically
determined following the procedure introduced by Mohamed \& LaRue
\citep{MohamedLaRue1990}. Our results are in excellent agreement with
similar results reported in the literature for regular grids, e.g. $n
= 1.41$ is very close to the usually reported empirical exponent (see
\citep{MohamedLaRue1990} and references therein).

Figure \ref{fig:TurbInt_Tr} confirms that a protracted production
region exists in the lee of space-filling fractal square grids, that
it extends over a distance which depends on the thickness ratio $t_r$
and that it is followed by a region (the decay region first
identified in \citep{HurstVassilicos2007}) where the turbulence
decays. The existence of a distance $x_{peak}$ where the turbulence
intensity peaks is clear in this figure. Figure \ref{fig:MeanVel_Tr}
shows that the production region where the turbulence increases is
accompanied by significant longitudinal mean velocity gradients which
progressively decrease in amplitude till about after $x=x_{peak}$
where they more or less vanish and the turbulence intensity decays.
	
Our data show that the centreline mean velocity is quite high compared
to $U_{\infty}$ on the close downstream side of our fractal square
grids and remains so over a distance which depends on fractal grid
geometry before decreasing towards $U_{\infty}$ further
downstream. This centreline jet-like behavior seems to result from the
relatively high opening at the grid's centre where blockage ratio,
which is inhomogeneously distributed on the grid, seems to be locally
small compared to the rest of the grid. The initial plateau is
therefore characterized by a significant velocity excess ($U_C /
U_\infty > 1.35$). One can also see in figure \ref{fig:MeanVel_Tr}
that the mean velocity remains larger than $U_\infty$ even far away
from the grids. We have checked that this effect is consistent with
the small downstream growth of the boundary layer on the tunnel's
walls.

The space-filling fractal grids \emph{SFG8}, \emph{SFG13} and
\emph{SFG17} are identical in all but one parameter: the thickness
ratio $t_r$. It is therefore clear from figures \ref{fig:MeanVel_Tr} and
\ref{fig:TurbInt_Tr} that $t_r$ plays an important role because, even
though the streamwise profiles of $U_C / U_\infty$ and $u'_c / U_C$
are of identical shape for \emph{SFG8}, \emph{SFG13} and \emph{SFG17},
$U_C/U_\infty$ decreases, $u'_c / U_C$ increases and $x_{peak}$
decreases when increasing $t_r$ whilst keeping all other independent
parameters of the grid constant.

However, the parameter $t_r$ cannot account alone for the differences
between the \emph{SFG17} and \emph{BFG17} grids. These two grids have
the same blockage ratio $\sigma$ and very close values of $t_r$ and
effective mesh size $M_{eff}$.  What they do mainly differ by, are
their values of $L_0$ (by a factor of 2), the number of fractal
iterations ($N=4$ for \emph{SFG17}, $N=5$ for \emph{BFG17}) and the
largest thickness $t_0$. Figures \ref{fig:MeanVel_Tr} and
\ref{fig:TurbInt_Tr} show clearly that when $t_r$ is kept roughly
constant and other grid parameters are varied (such as $L_0$), then
$x_{peak}$ and the overall streamwise profiles of $U_C / U_\infty$ and
of $u'_c / U_C$ change in ways not accounted for by the changes
between \emph{SFG8},\emph{SFG13} and \emph{SFG17}.

Comparing data obtained downstream from different space-filling
fractal square grids, Hurst \& Vassilicos \citep{HurstVassilicos2007}
suggested that the streamwise evolution of turbulence intensity,
i.e. $x_{peak}$, can be scaled by the length-scale $x_{HV} = 75
\frac{t_{min} T}{L_{min}}$. Their empirical formula might appear to
account for the difference between the \emph{SFG17} and \emph{BFG17}
grids in figure 3(b)
because $T$ is double and $t_{min}$ is larger by a factor 1.3 for
\emph{BFG17} compared to \emph{SFG17}. However, Hurst \& Vassilicos
\citep{HurstVassilicos2007} did not attempt to collapse data from
different wind tunnels, and we now show how such a careful collapse
exercise involving both the mean flow and the turbulence intensity
leads to a different formula for $x_{peak}$.


In figures \ref{fig:MeanVel_Tr_HV} and \ref{fig:TurbInt_Tr_NormHV} we
plot the streamwise evolution of $U_C / U_\infty$ and of $u'_c / U_C$
(scaled by its value at $x_{peak}$) using the scaling $x / x_{HV}$
introduced by Hurst \& Vassilicos \citep{HurstVassilicos2007}. One can
clearly see that whilst use of $x_{HV}$ collapses the data obtained in
the \textit{T=0.46m} tunnel, a large discrepancy remains with the
\textit{T=0.91m} tunnel data. In particular, $x_{peak}$ differs for
\emph{BFG17} and \emph{SFG17}.

\begin{figure}[htbp]
\centering
\subfigure
{\includegraphics[scale=0.3]{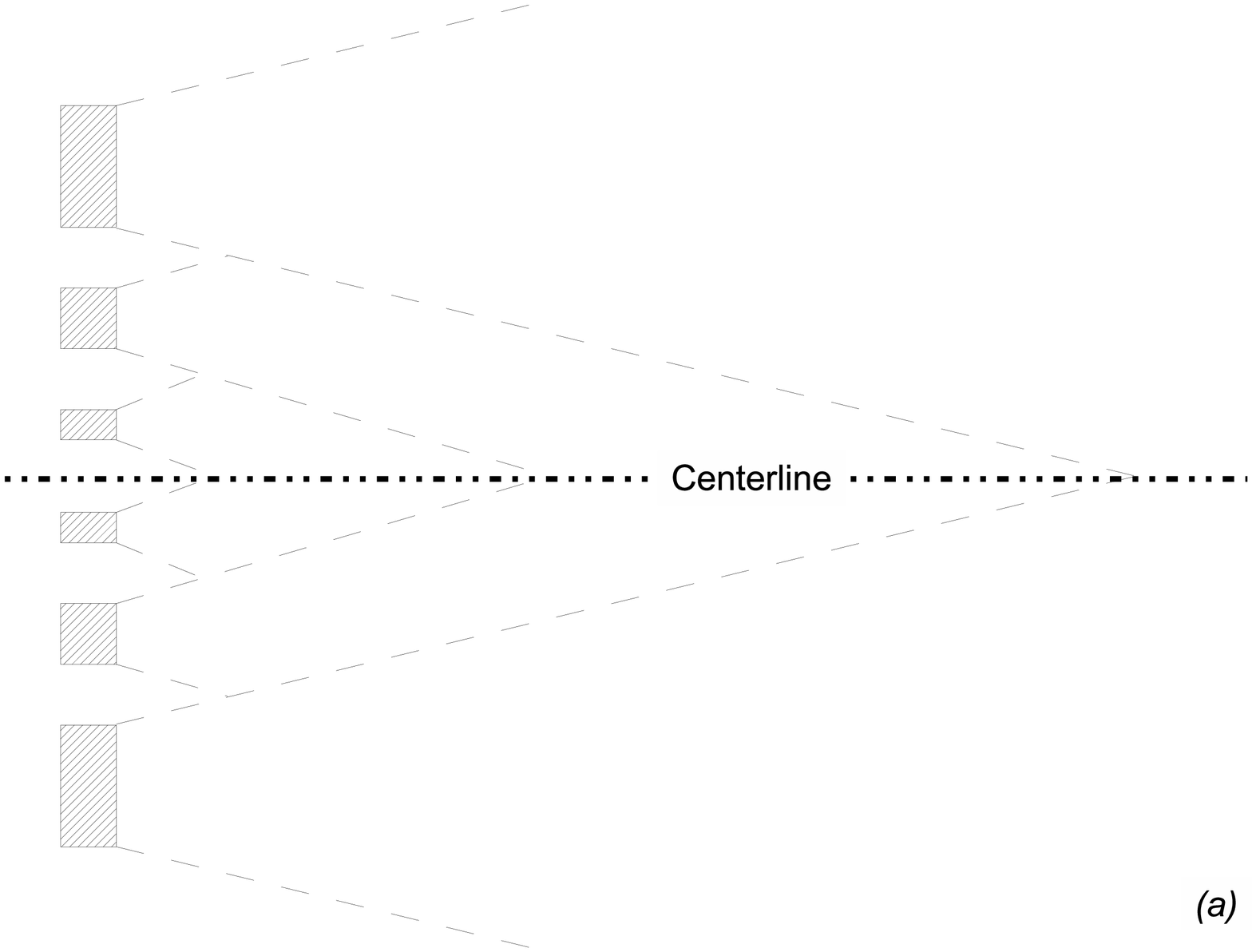}
\label{fig:Xhom}}
\hspace{0.5cm}
\subfigure
{\includegraphics[scale=0.4]{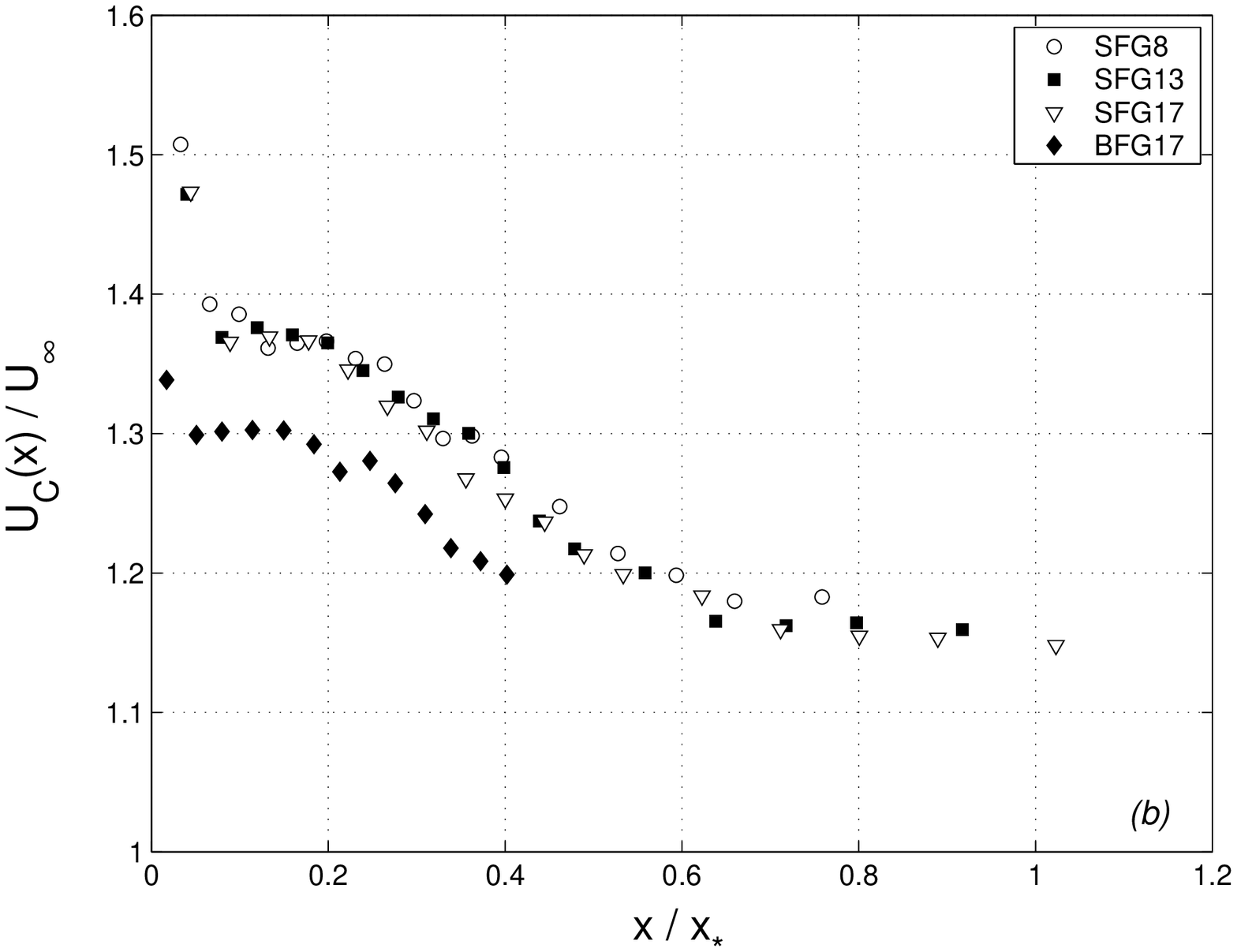}
\label{fig:MeanVel_Tr_Norm1}}
\caption{\textit{(a)} Schematic of wake interactions resulting from
the fractal grid's bars. \textit{(b)} Centreline mean streamwise
velocity vs distance normalised by the wake-interaction length scale
$x_{*} = L_{0}^{2}/t_{0}$ ($U_\infty = 5.2 m/s$).}
\end{figure}

The turbulence generated by either regular or multiscale/fractal grids
with relatively low blockage ratio results from the interactions
between the wakes of the different bars. In the case of fractal grids
these bars have different sizes and, as a result, their wakes interact
at different distances from the grid according to size and position on
the grid (see direct numerical simulations of turbulence generated by
fractal grids in \citep{Nagataetal2008}, \citep{LaizetVassilicos2009}
and \citep{Laizetetal2009}). Assuming that the typical wake-width
$\ell$ at a streamwise distance $x$ from a wake-generating bar of
width/thickness $t_{j}$ ($j=0,...,N-1$) to be $\ell \sim \sqrt{t_{j}
x}$ \citep{Townsend1956}, the largest such width corresponds to the
largest bars on the grid, i.e.  $\ell \sim \sqrt{t_{0} x}$. Assuming
also that this formula can be used even though the bars are surrounded
by other bars of different sizes, then the furthermost interaction
between wakes will be that of the wakes generated by the largest bars
placed furthermost on the grid (see figure \ref{fig:Xhom}). This will
happen at a streamwise distance $x = x_\star$ such that $L_{0} \sim
\ell \sim \sqrt{t_{0} x_\star}$. We therefore introduce

\begin{equation}
x_\star = \frac{L_0^2}{t_0}
\end{equation}
as a characteristic length-scale of interactions between the wakes of
the grid's bars which might bound $x_{peak}$. We stress that the
assumptions used to define $x_\star$ are quite strong and care should
be taken in extrapolating this presumed bound on $x_{peak}$ to any
space-filling fractal square grid beyond those studied here, let alone
any fractal/multiscale grid.

\begin{figure}[htbp]
\centering
\subfigure
{\includegraphics[width=7.5cm]{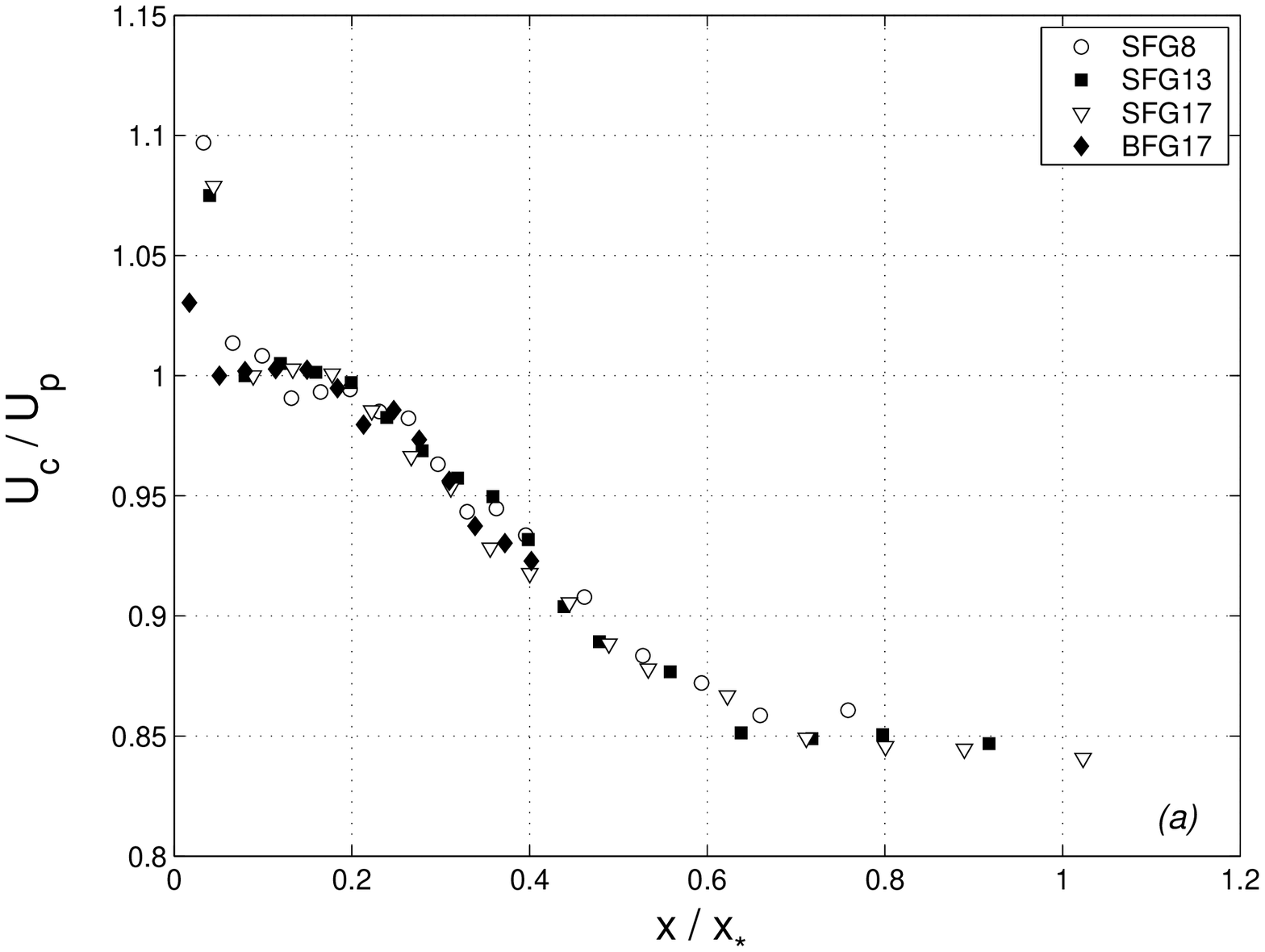}
\label{fig:MeanVel_Tr_Norm2}}
\hspace{0.1cm}
\subfigure
{\includegraphics[width=7.5cm]{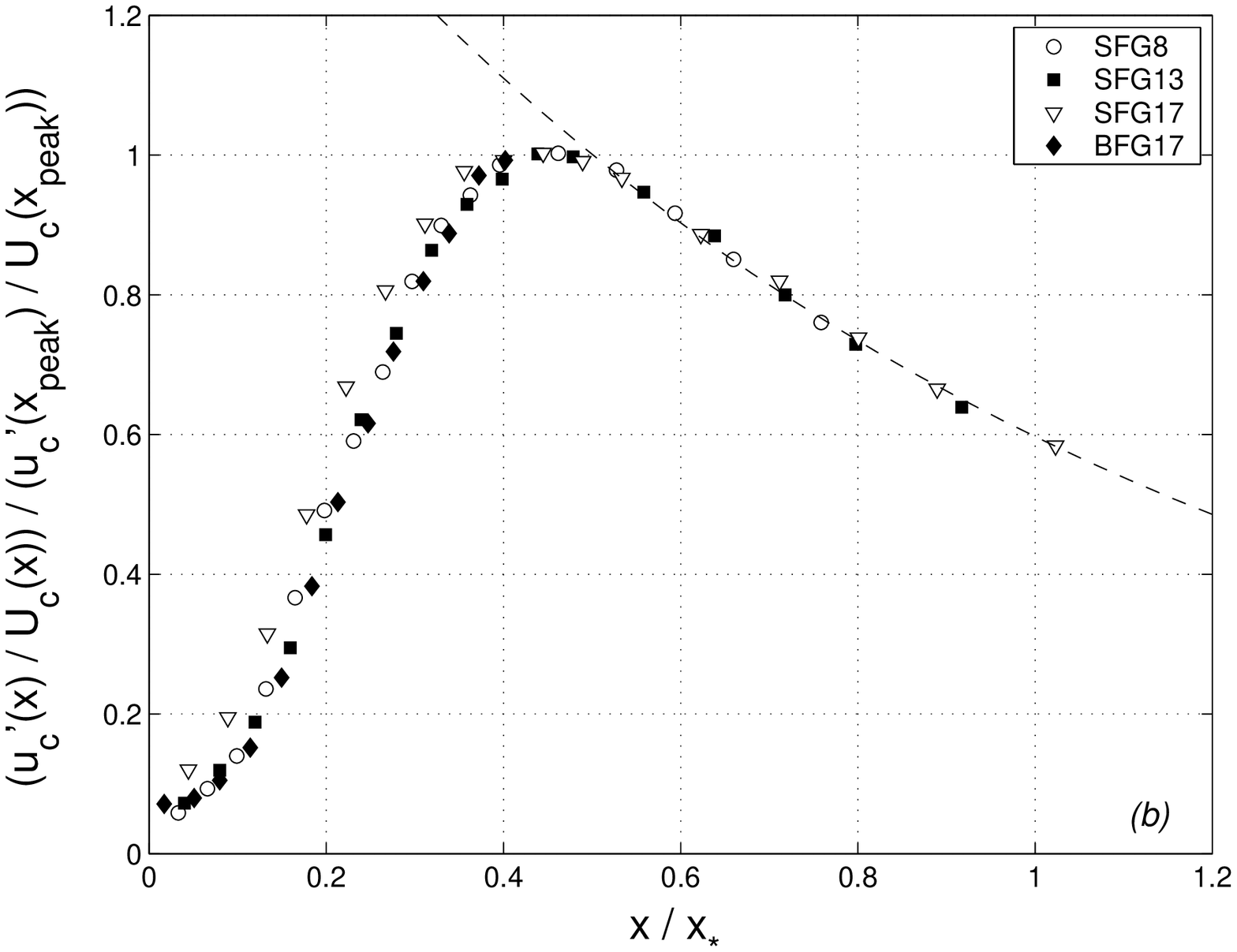}
\label{fig:TurbInt_Tr_Norm}}
\caption{Streamwise evolution of \textit{(a)} the centreline mean
velocity $U_C$ normalized by the initial mean velocity plateau $U_p$
and \textit{(b)} centreline turbulence intensity normalised by its
value at $x=x_{peak}$. Both plots are given as functions of $x$ scaled
by the wake-interation length-scale $x_\star = L_{0}^{2}/t_{0}$
($U_\infty = 5.2 m/s$). In \textit{(b)} the dashed line represents
equation (9) with $B=2.06$ and $A=2.82$.}
\end{figure}

Figure \ref{fig:MeanVel_Tr_Norm1} is a plot of the normalised
centreline mean velocity $U_C / U_\infty$ as a function of
dimensionless distance $x / x_\star$ for all our four space-filling
fractal square grids. All the data from the \emph{T-0.46m} tunnel
collapse in this representation. However the \emph{BFG17} data from
the larger wind tunnel do not. They fall on a similar curve but at
lower values of $U_C / U_\infty$. This systematic difference can be
explained by the fact that the air flow causes the \emph{BFG17} grid
in the large wind tunnel to bulge out a bit and adopt a slightly
curved but steady shape. The flow rate distribution through this grid
must be slightly modified as a result. To compensate for this effect
we introduce the mean velocity $U_p$ characterizing the constant mean
velocity plateau in the vicinity of the fractal grids.  In figure
\ref{fig:MeanVel_Tr_Norm2} we plot the normalised centreline mean
velocity $U_C / U_p$ as a function of $x / x_\star$ for all fractal
grids and both tunnels and find an excellent collapse onto a single
curve.

In figure \ref{fig:TurbInt_Tr_Norm} we plot $u'_c / U_C$ normalised by
its value at $x_{peak}$ as a function of $x/x_\star$. We also find an
excellent collapse onto a single curve irrespective of fractal grid
and tunnel. It is also clear from figures \ref{fig:MeanVel_Tr_Norm2} and
\ref{fig:TurbInt_Tr_Norm} that the longitudinal mean velocity gradient
$\frac{\partial U}{\partial x}$ becomes insignificant where $x \ge 0.6
x_\star$ and that the streamwise turbulence intensity peaks at
 
\begin{equation}
x_{peak} \approx 0.45 x_\star = 0.45 \frac{L_0^2}{t_0}.
\end{equation}
The wake-interaction length-scale $x_\star$ appears to be the
appropriate length-scale characterising the first and second order
statistics of turbulent flows generated by space-filling fractal
square grids, at least on the centreline and for the range of grids
tested in the present work.

\begin{figure}[htbp]
\centering
\subfigure
{\includegraphics[width=7.5cm]{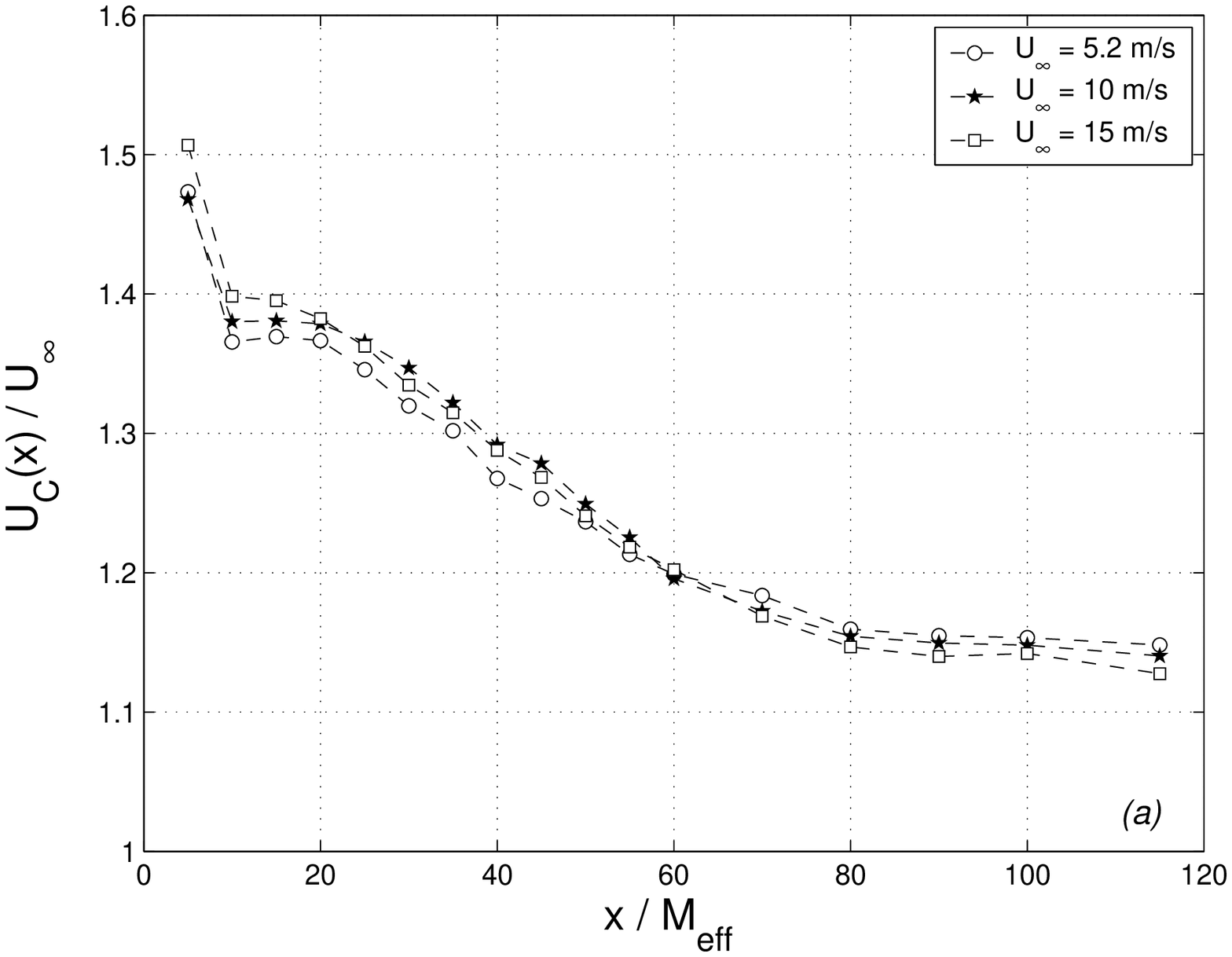}
\label{fig:MeanVel_Uinlet}}
\hspace{0.1cm}
\subfigure
{\includegraphics[width=7.5cm]{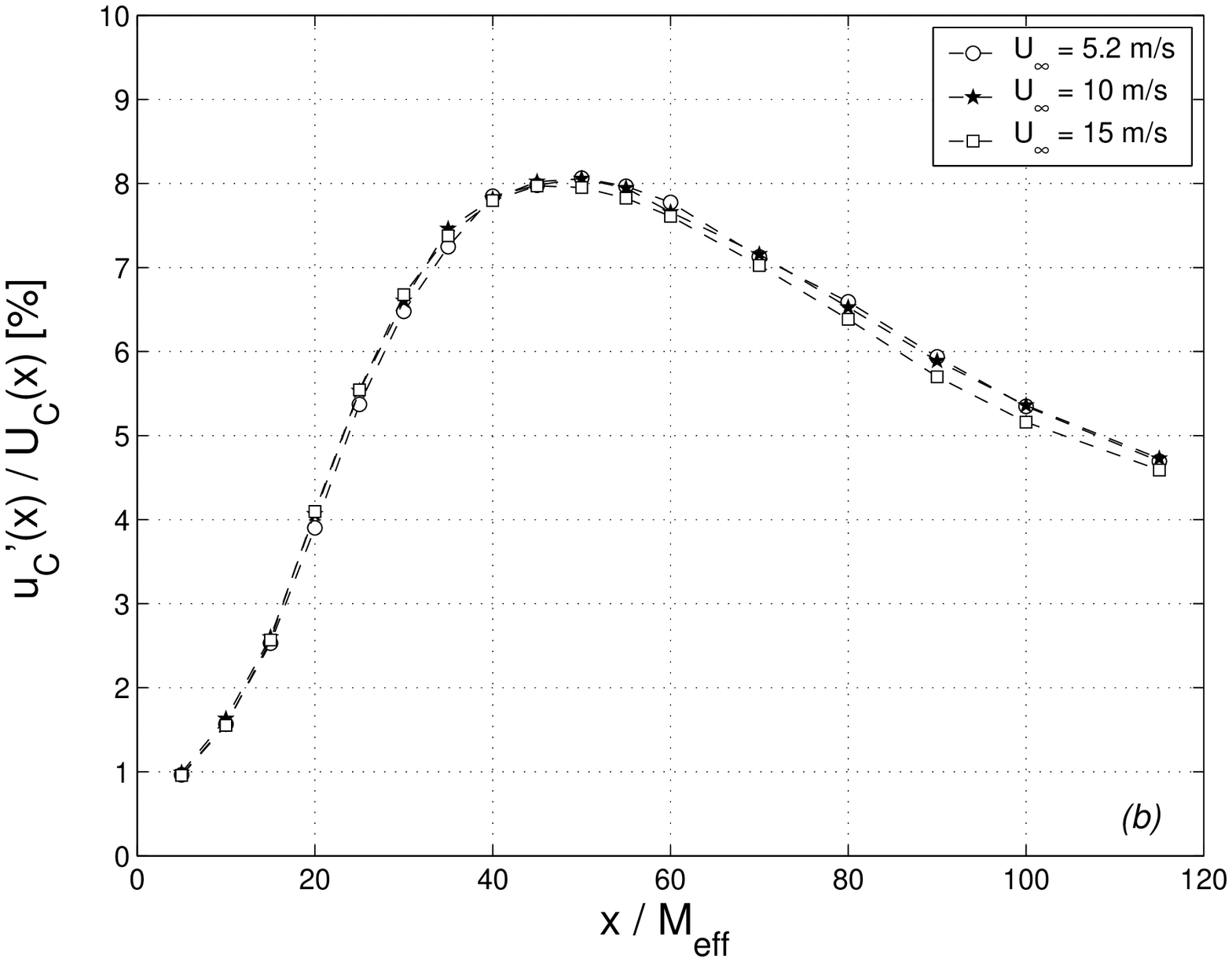}
\label{fig:TurbInt_Uinlet}}
\caption{\textit{(a)} Streamwise evolution of the centreline ($y=z=0$)
mean velocity in the \emph{T-0.46} tunnel for the fractal grid
\emph{SFG17}. \textit{(b)} Streamwise evolution of the turbulence
intensity for the same grid and in the same tunnel.}
\end{figure}

In figures \ref{fig:MeanVel_Uinlet} and \ref{fig:TurbInt_Uinlet} we
plot the streamwise evolutions of the dimensionless centreline
velocity $U_C / U_\infty$ and the centreline turbulence intensity
$u'_c / U_C$ for various inlet velocities $U_\infty$. These particular
results have been obtained for the fractal grid \emph{SFG17} but they
are representative of all our space-filling fractal square grids. One
can clearly see that $x_{peak}$ is independent of
$U_\infty$. Moreover, our data show that the entire streamwise
profiles of both $U_C / U_\infty$ and $u'_c / U_C$ are also
independent of the inlet velocity $U_\infty$ in the range studied.

Hurst \& Vassilicos \citep{HurstVassilicos2007} have shown that the
centreline turbulence intensity decays exponentially in the
decay region $x>x_{peak}$ and that the length-scale $x_{peak}$ can be 
used to collapse this decay for different space-filling fractal square grids
as follows: 
\begin{equation}
{\frac{u'^{2}_{c}}{U_{C}^{2}}} = {u'^{2}_{c}(x_{0})\over U_{C}^{2}(x_{0})}
exp\left(-B' \left(\frac{x - x_0}{x_{peak}}\right)\right)
\end{equation}
where $x_0$ is a virtual origin and $B'$ an empirical dimensionless
parameter. We confirm this scaling decay form, specifically by fitting 
\begin{equation}
{u'^{2}_{c}\over U_{C}^{2}} = A
exp\left(-B \left(\frac{x}{x_{*}}\right)\right)
\end{equation}
to our experimental data with a slightly modified dimensional
parameter $B$ and an extra dimensionless parameter $A$ which does not
have much influence on the quality of the fit except for shifting it
all up or down. We have arbitrarily set $x_{0}=0$, which we are
allowed to do because the virtual origin does not affect the value of
$B$. It only affects the value of $A$. 

As shown in figure \ref{fig:TurbInt_Tr_Norm} the exponential decay law
(9) is in excellent agreement with our data for all the space-filling
fractal square grids used in the present work. In particular, the
parameter $B$ seems to be the same for all the fractal square grids we
used. Using a least-mean square method we find $B = 2.06$ and
$A=2.82$.

Evidence in support of the idea that the decay region around the
centreline downstream of $x_{peak}$ is approximately homogeneous and
isotropic was given in \citep{SeoudVassilicos2007}. The exponential
turbulence decay observed in this region (\citep{HurstVassilicos2007},
\citep{SeoudVassilicos2007}) is therefore remarkable because it differs
from the usual power-law decay of homogeneous isotropic turbulence. We
have already reported in this subsection that the mean flow becomes
approximately homogeneous in the streamwise direction where $x > 0.6
x_\star$, i.e. in the decay region. In the following subsection we
investigate the spanwise mean flow and turbulence fluctuation profiles
downstream from space-filling fractal square grids and show how a
highly inhomogeneous flow near the fractal grid morphs into a
homogeneous one beyond $0.6 x_\star$.


%
%
%
%

\subsection{Homogeneity}

\begin{figure}[htbp]
\centering
\subfigure
{\includegraphics[width=7.5cm]{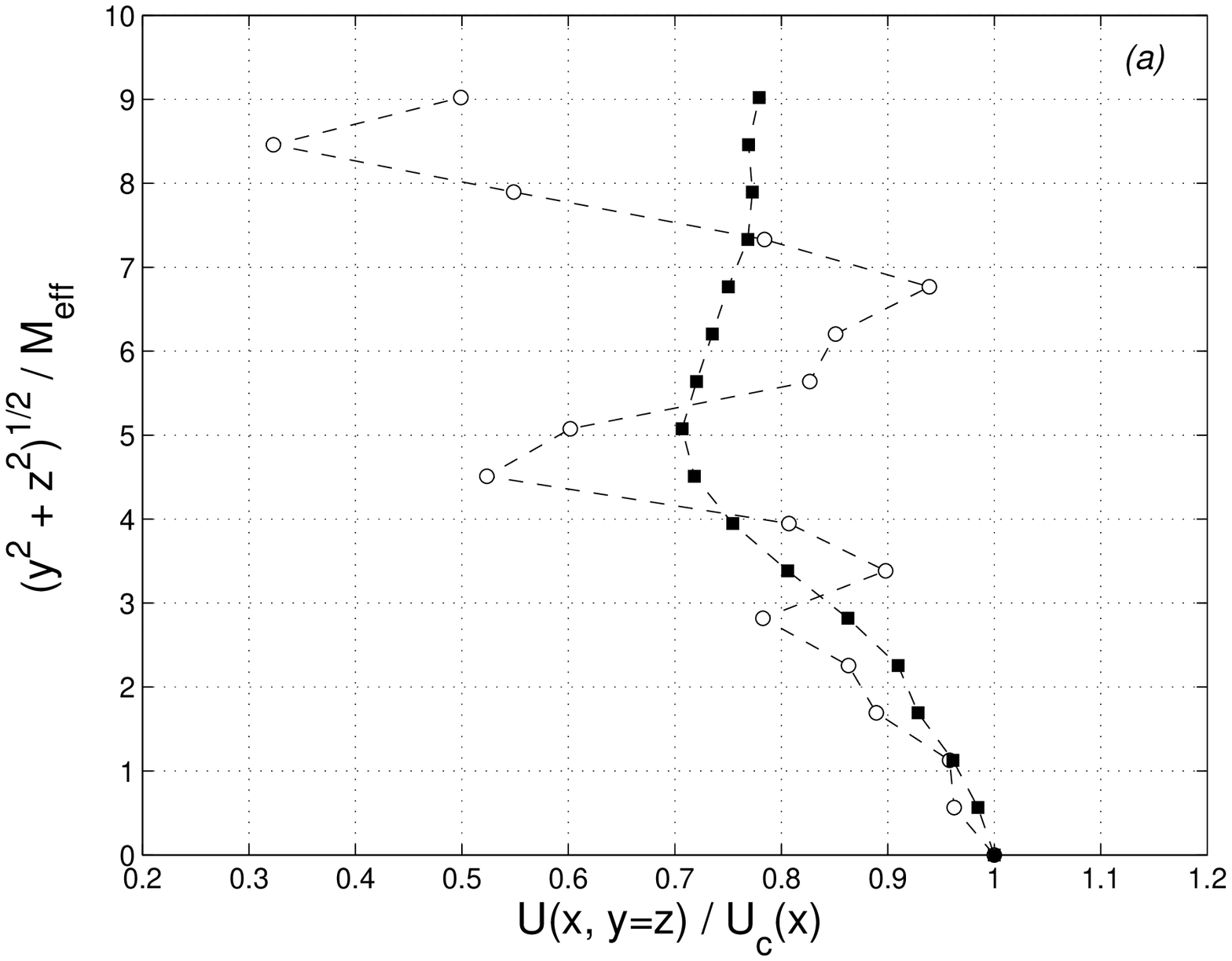}
\label{fig:ProfileMeanVel_BFG17}}
\hspace{0.1cm}
\subfigure
{\includegraphics[width=7.5cm]{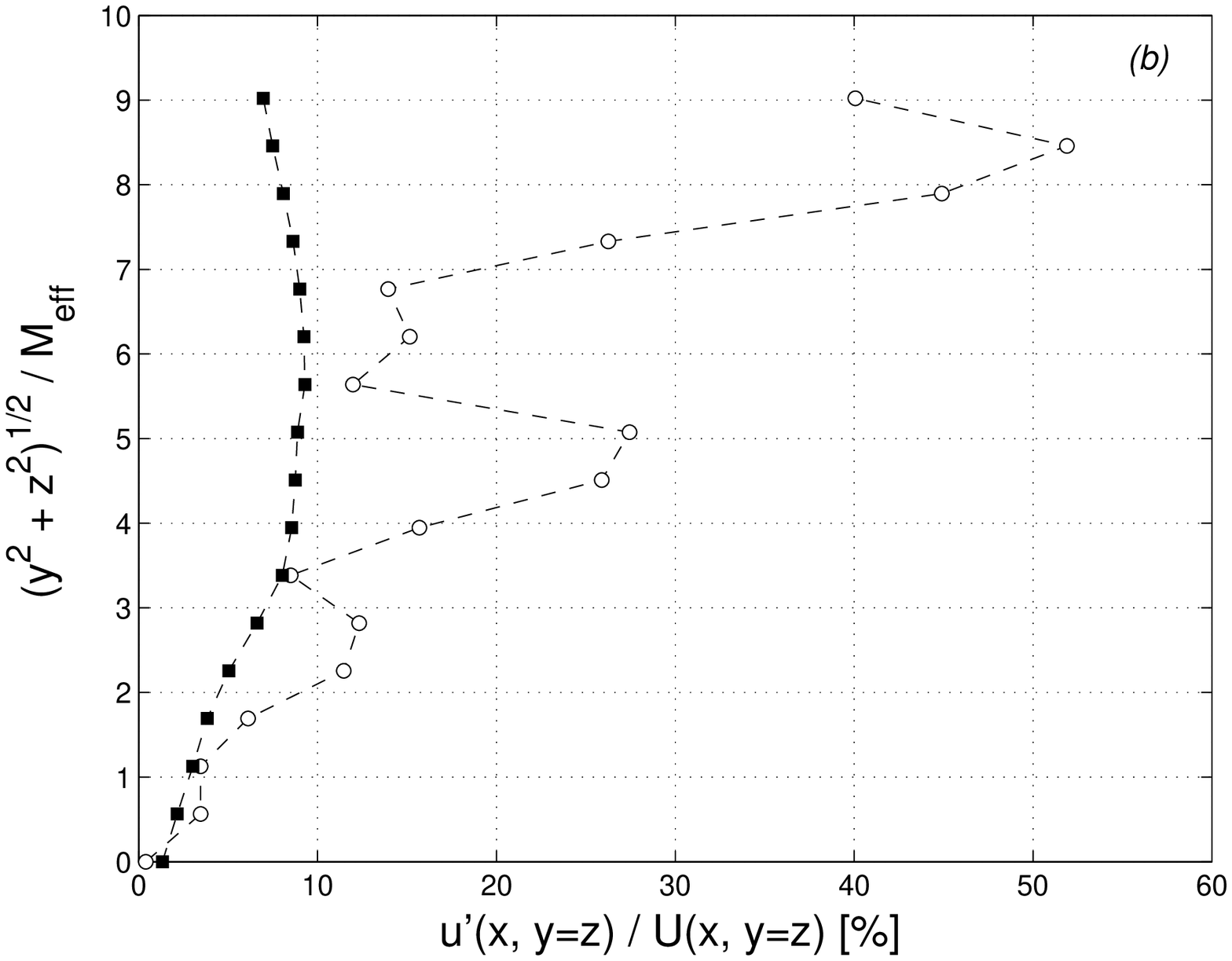}
\label{fig:ProfileTurbInt_BFG17}}
\caption{ \textit{(a)} Normalised mean velocity profiles and
\textit{(b)} turbulence intensity profiles along the diagonal line in
the \emph{T-0.91m} tunnel. The fractal grid is \emph{BFG17}. Symbols:
$\bigcirc$, $x/M_{eff} \approx 7$ (i.e. $x/x_{*}\approx 0.02$);
$\blacksquare$, $x/M_{eff} \approx 53$ (i.e. $x/x_{*}\approx 0.15$).}
\end{figure}

Like regular grids, the flow generated in the lee of space-filling
fractal square grids is marked by strong inhomogeneities near the grid
which smooth out further downstream under the action of turbulent
diffusion. This is evidenced in figures \ref{fig:ProfileMeanVel_BFG17}
and \ref{fig:ProfileTurbInt_BFG17} which show scaled mean velocity $U
/ U_c$ and turbulence intensity $u' / U$ profiles along the diagonal
in the $y - z$ plane, i.e. along the line parameterised by $(y^{2}
+z^{2})^{1/2}$ in that plane. Close to the grid, the mean velocity
profile is very irregular, especially downstream from the grid's bars
where large mean velocity deficits are clearly present. These deficits
are surrounded by high mean flow gradients where the intense
turbulence levels reach local maxima as shown in figure
\ref{fig:ProfileTurbInt_BFG17}. Further downstream, both mean velocity
and turbulence intensity profiles become much smoother supporting the
view that the turbulence tends towards statistical homogeneity. Note
that figures \ref{fig:ProfileMeanVel_BFG17} and
\ref{fig:ProfileTurbInt_BFG17} show diagonal profiles at $x/M_{eff}
\approx$ $7$ and $53$ in the lee of the BFG17 grid which is a long way
before $x_{peak}$ (see figure 3b). The profiles are quite uniform in
the decay region as shown by Seoud \& Vassilicos
\citep{SeoudVassilicos2007}. Our evidence for homogeneity complements
theirs in two ways: they concentrated only on the decay region whereas
we report profiles in the production region and how they smooth out
along the downstream direction; and we report diagonal profiles
whereas the profiles in \citep{SeoudVassilicos2007} are all along the
$y$ coordinate.

\begin{figure}[htbp]
\centering
\subfigure
{\includegraphics[width=5.5cm]{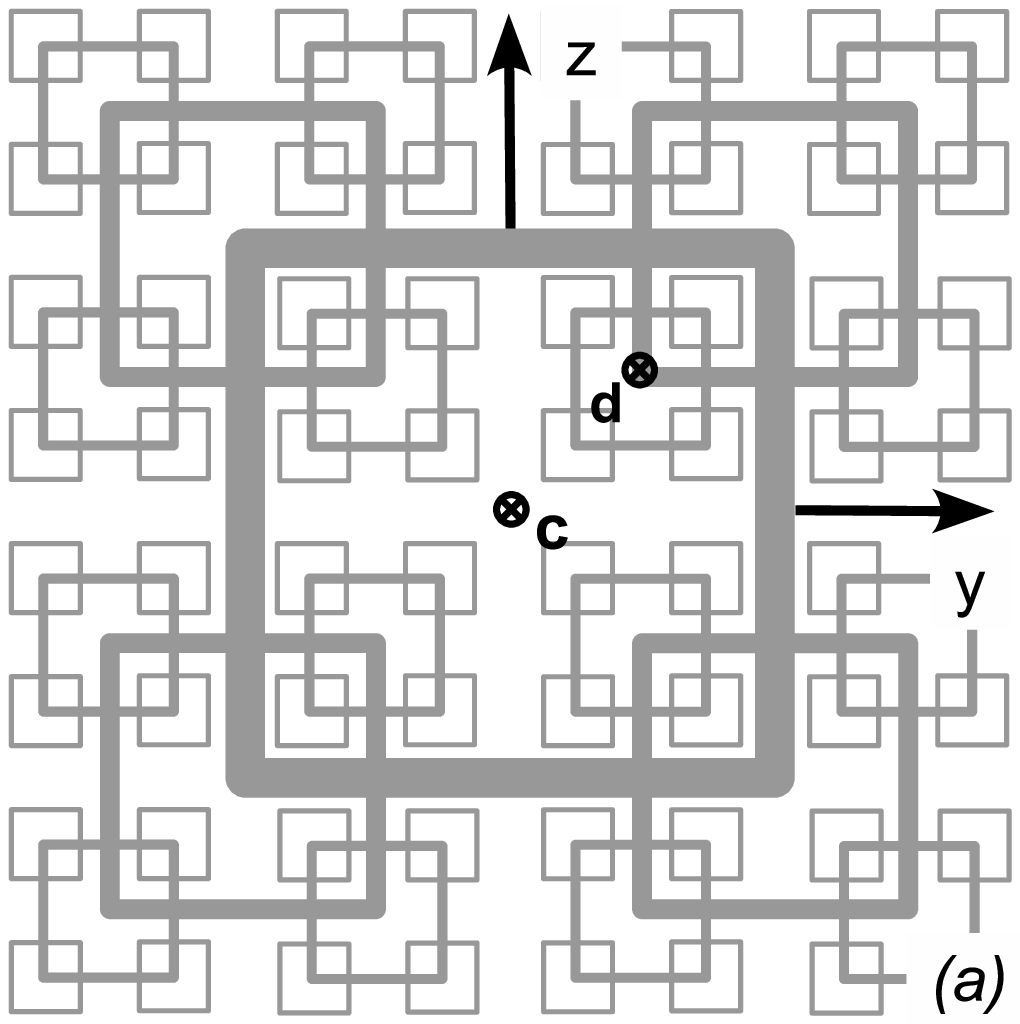}
\label{fig:MeasLoc_SmallTunnel}}
\hspace{0.2cm}
\subfigure
{\includegraphics[width=7cm]{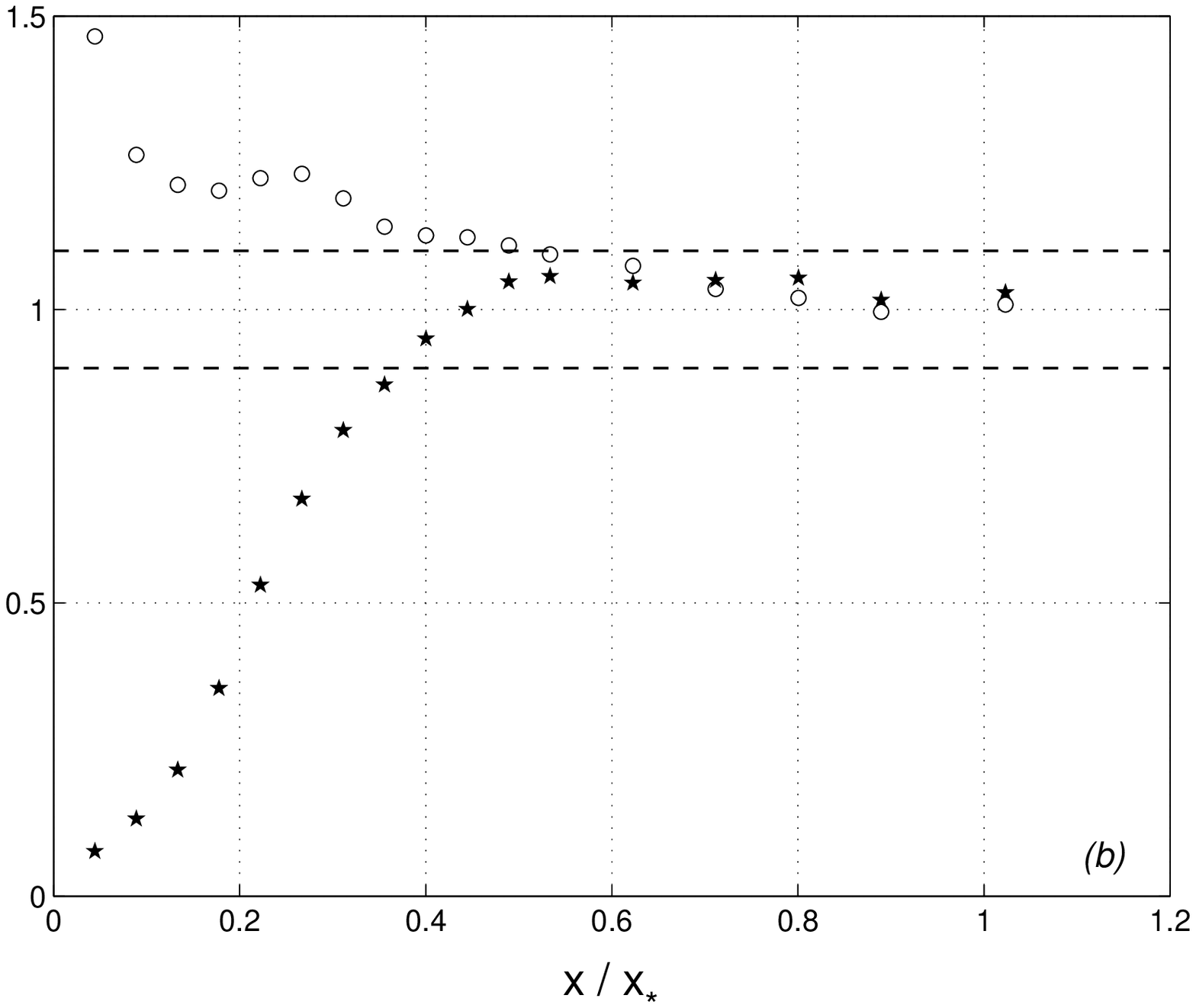}
\label{fig:RatioUandu}}
\caption{\textit{(a)} The measurements in (b) are taken in the
\emph{T-0.46m} tunnel along the straight lines which run in the
$x$-direction and cut the plane of the pictured \emph{SFG17} grid at
points $c$ and $d$. \textit{(b)} Streamwise evolution of the ratios
$U_c / U_d$ ($\circ$) and $u'_c / u'_d$ ($\star$) measured for the 
\emph{SFG17} grid in the \emph{T-0.46m} tunnel ($U_\infty = 5.2
m/s$). The horizontal dashed lines represent the range $\pm 10$\%.}
\end{figure}

To evaluate the distance from the grid where the inlet inhomogeneities
become negligible, we introduce the ratios $U_c/U_d$ and $u'_c / u'_d$
where subscripts $c$ and $d$ denote respectively the centreline ($y =
z = 0$) and the streamwise line cutting through the second iteration
corner ($y = z = 6cm$ in the SFG17 case) as shown in
figure \ref{fig:MeasLoc_SmallTunnel}. These two straight lines meet the
inlet conditions at two different points, their difference being
representative of the actual inhomogeneity on the fractal grid itself:
one point is at the central empty space whilst the other is at the
corner of one of the second iteration squares.
The streamwise evolutions of $U_c/U_d$ and $u'_c / u'_d$ are reported
in figure \ref{fig:RatioUandu}. As expected, large differences are
observable in the vicinity of the space-filling fractal square grid
for both the mean and the fluctuating velocities. One can see that the
mean velocity ratio $U_c/U_d$ is bigger than unity. This reflects the
mean velocity excess on the centreline where the behaviour is jet-like
because of the local opening compared to the mean velocity deficit
downstream from the second iteration corner where the behaviour is
wake-like because of the local blockage. This difference between
jet-like and wake-like local behaviours also explains why the
fluctuating velocity ratio $u'_c / u'_d$ is almost null close to the
grid where the centreline is almost turbulence free. Further
downstream, both ratios $U_c/U_d$ and $u'_c / u'_d$ tend to unity as
the flow homogenises. One can see that inhomogeneities become
negligible by these criteria beyond $x / x_\star = 0.6$, which is
quite close to $x_{peak}$.



A main consequence of statistical homogeneity is that the small-scale
turbulence is not sensitive to mean flow gradients. For this to be the
case, the time scales defined by the mean flow gradients must be much
larger than the largest time-scale of the small-scale turbulence. From
our measurements and those in \citep{SeoudVassilicos2007} (see figure 3
in \citep{SeoudVassilicos2007}), $(\partial U / \partial x)^{-1}$ and
$(\partial U / \partial y)^{-1}$ are always larger than about 1 and
0.15 second, respectively, at and beyond $x_{peak}$ where the
time-scale of the energy-containing eddies is well below 0.07
seconds. The ratios between the smallest possible estimate of a mean
shear time scale and any turbulence fluctuation time scale are
therefore well above 2 (in the worst of cases) at $x_{peak}$ and far
larger (by one or two orders of magnitude typically) beyond it for all
three fractal grids SFG8, SFG13 and SFG17 and all inlet velocities
$U_{\infty}$ in the range tested. By this time-scale criterion, from
$x_{peak}$ onwards, the small-scale turbulence generated by our
fractal grids, including the energy-containing eddies, is not affected
by the typically small mean flow gradients which are therefore
negligible in that sense.

We close this subsection with figure 9 which illustrates in yet another
way the homogeneity of the turbulence intensity at $x \ge 0.5 x_*$ and
suggests that equation (9) can be extended beyond the centreline in
the $y-z$ plane in that homogeneous region $x \ge 0.5 x_*$, i.e.
\begin{equation}
{u'^{2}\over U^{2}} = A
exp\left(-B \left(\frac{x}{x_{*}}\right)\right)
\end{equation}
with the same values $A \approx 2.82$ and $B \approx 2.06$
independently of $U_{\infty}$ and space-filling fractal square grid.
It is worth pointing out, however, that figure 9 also shows quite
clearly that $x_{peak}$ can vary widely across the $y-z$ plane and
that a very protracted production region does not exist
everywhere. Formula (7) gives $x_{peak}$ on the centreline but
$x_{peak}$ can be very much smaller than $0.45 x_*$ at other $y-z$
locations. This is a natural consequence of the inhomogeneous blockage
of fractal grids and the resulting combination of wake-like and
jet-like regions in the flows they generate.

\begin{figure}[htbp]
\centering
{\includegraphics[scale=0.5]{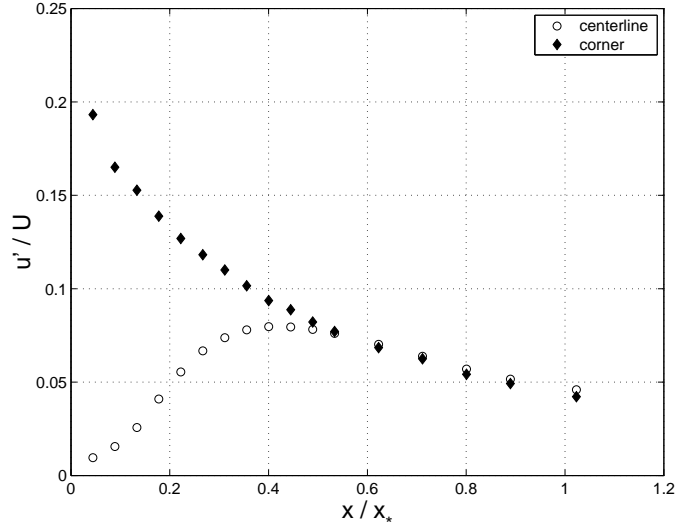}
\label{fig:RmsVelCenterDiag}}
\caption{Streamwise turbulence intensities as functions of $x/x_*$
along the centreline and along the straight streamwise line which
crosses point $d$ in Fig. 8(a). SFG17 grid in the \emph{T-0.46m}
tunnel with $U_{\infty} = 15m/s$.  This plot remains essentially the
same for the other values of $U_{\infty}$ that we tried.}
\end{figure}

\subsection{Skewness and flatness of the velocity fluctuations}


Previous wind tunnel investigations of turbulence generated by
space-filling fractal square grids \citep{HurstVassilicos2007},
\citep{SeoudVassilicos2007} have not reported results on the
gaussianity/non-gaussianity of turbulent velocity fluctuations. We
therefore study here this important aspect of the flow, mostly in
terms of the skewness $S_u = \frac{\left\langle
u^3\right\rangle}{\left\langle u^2\right\rangle^{3/2}}$ and flatness
$F_u = \frac{\left\langle u^4\right\rangle}{\left\langle
u^2\right\rangle^{2}}$ of the longitudinal fluctuating velocity
component $u$ along the centreline. This skewness is also a measure of
one limited aspect of large-scale isotropy, namely mirror symmetry, as
it vanishes when statistics are invariant to the transformation $u$ to
$-u$, but not otherwise. Isotropy was studied in much more detail in
\citep{HurstVassilicos2007} and \citep{SeoudVassilicos2007} where
x-wires were used. These previous works reported good small-scale
isotropy in the decay region \citep{SeoudVassilicos2007} and levels of
large-scale anisotropy before and after $x_{peak}$ on the centreline
\citep{HurstVassilicos2007} which, for the turbulence generated by the
grids SFG13, SFG17 and BFG17 in particular, are very similar to the
levels of large-scale anisotropy in turbulence generated by active
grids \citep{Makita1991}, \citep{MydlarskiWarhaft1996}.

We first check that both $S_u$ and $F_u$ do not depend on the inlet
velocity $U_{\infty}$, and this is indeed the case as shown in figures
10(a) and 10(b). These plots are particularly interesting in that they
reveal the existence of large values of both $S_u$ and $F_u$ at about
the same distance from the grid on the centreline. This distance
scales with the wake-interaction length-scale $x_*$ as shown in
figures 11 and 12. 
Indeed, the profiles of both $S_u$ and $F_u$ along the centreline
collapse for all four fractal square grids (SFG8, SFG13 and SFG17 in
the \emph{T-0.46m} tunnel and BFG17 in the \emph{T-0.91m} tunnel) when
plotted against $x/x_*$. The alternative plots against $x/M_{eff}$
clearly do not collapse (see figures 11(a) and 12(a)).

\begin{figure}[htbp]
\centering \subfigure
{\includegraphics[width=7.5cm]{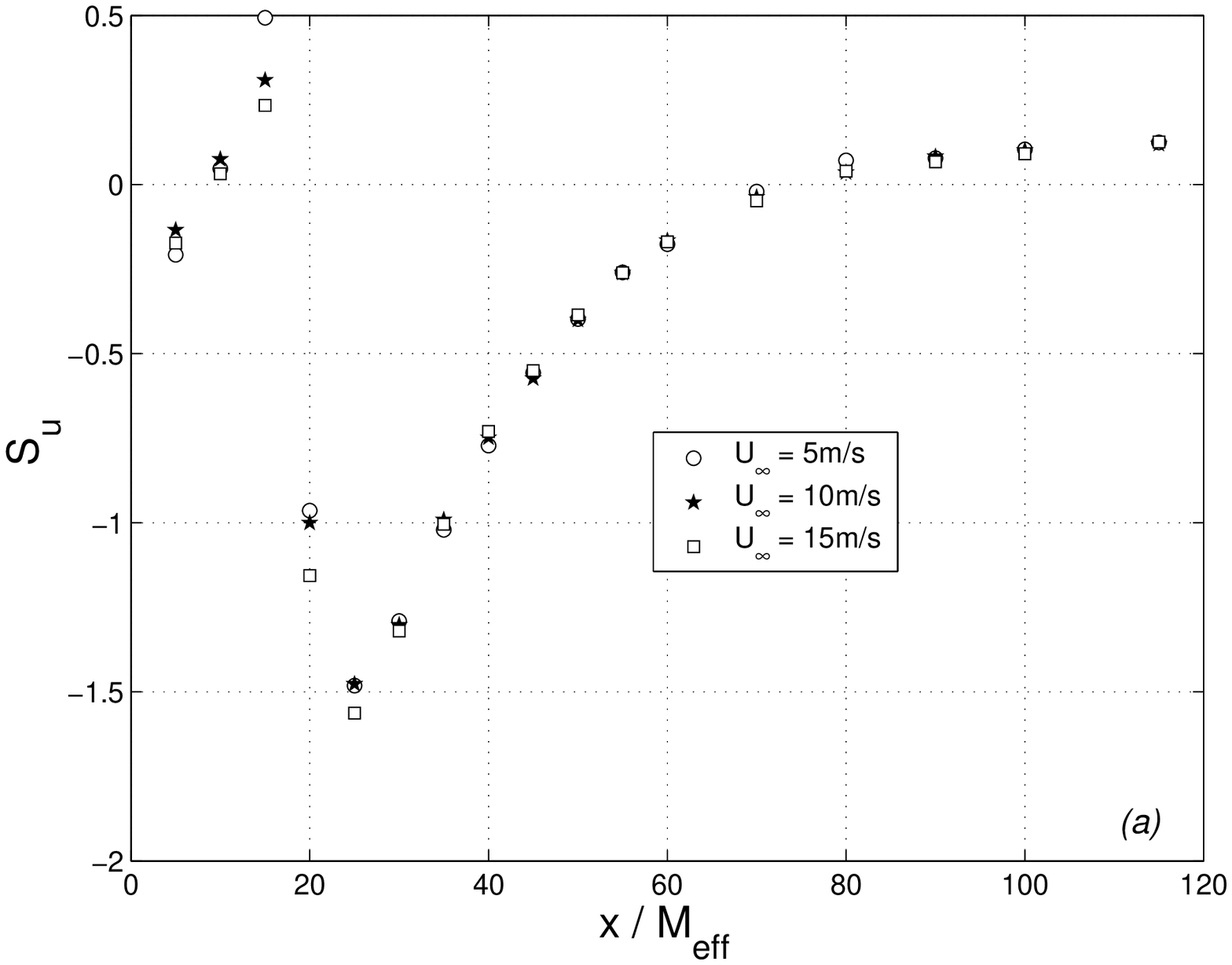}
\label{fig:SkewnessSFG17}}
\hspace{0.1cm}
\subfigure
{\includegraphics[width=7.5cm]{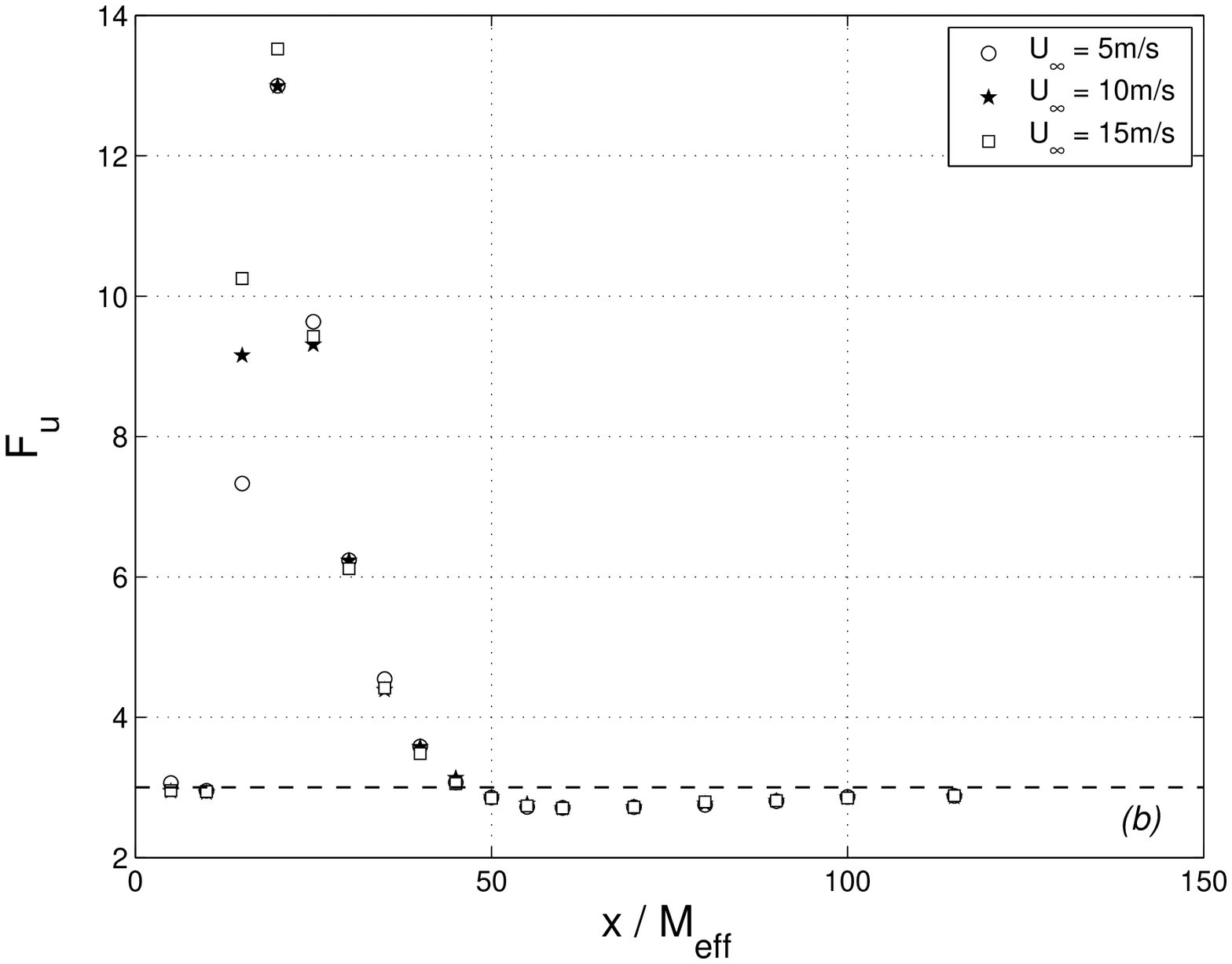}
\label{fig:FlatUinletBis}}
\caption{\textit{(a)} Skewness and \textit{(b)} flatness of the
longitudinal fluctuating velocity component $u$ as functions of
$x/M_{eff}$ along the centreline. SFG17 grid in the \emph{T-0.46m}
tunnel. The dashed lined in the right plot is $F_{u}=3$.}
\end{figure}

For comparison, figure 11(a) contains data of $S_u$ obtained with the
regular grid \emph{SRG} which are in fact in good agreement with usual
values reported in the literature (see e.g. \citep{BennettCorrsin1978},
\citep{MohamedLaRue1990}). It is well known that regular grids generate
small, yet non-zero, positive values of the velocity skewness $S_u$
and Maxey \citep{Maxey1987} explained how their non-vanishing values
are in fact a consequence of the free decay of homogeneous isotropic
turbulence. Whilst the velocity skewness $S_u$ generated by fractal
grids takes values which are also close to zero yet clearly positive
in the decay region, $S_u$ behaves very differently in the production
region on the centreline.


\begin{figure}[htbp]
\centering
\subfigure
{\includegraphics[width=7.5cm]{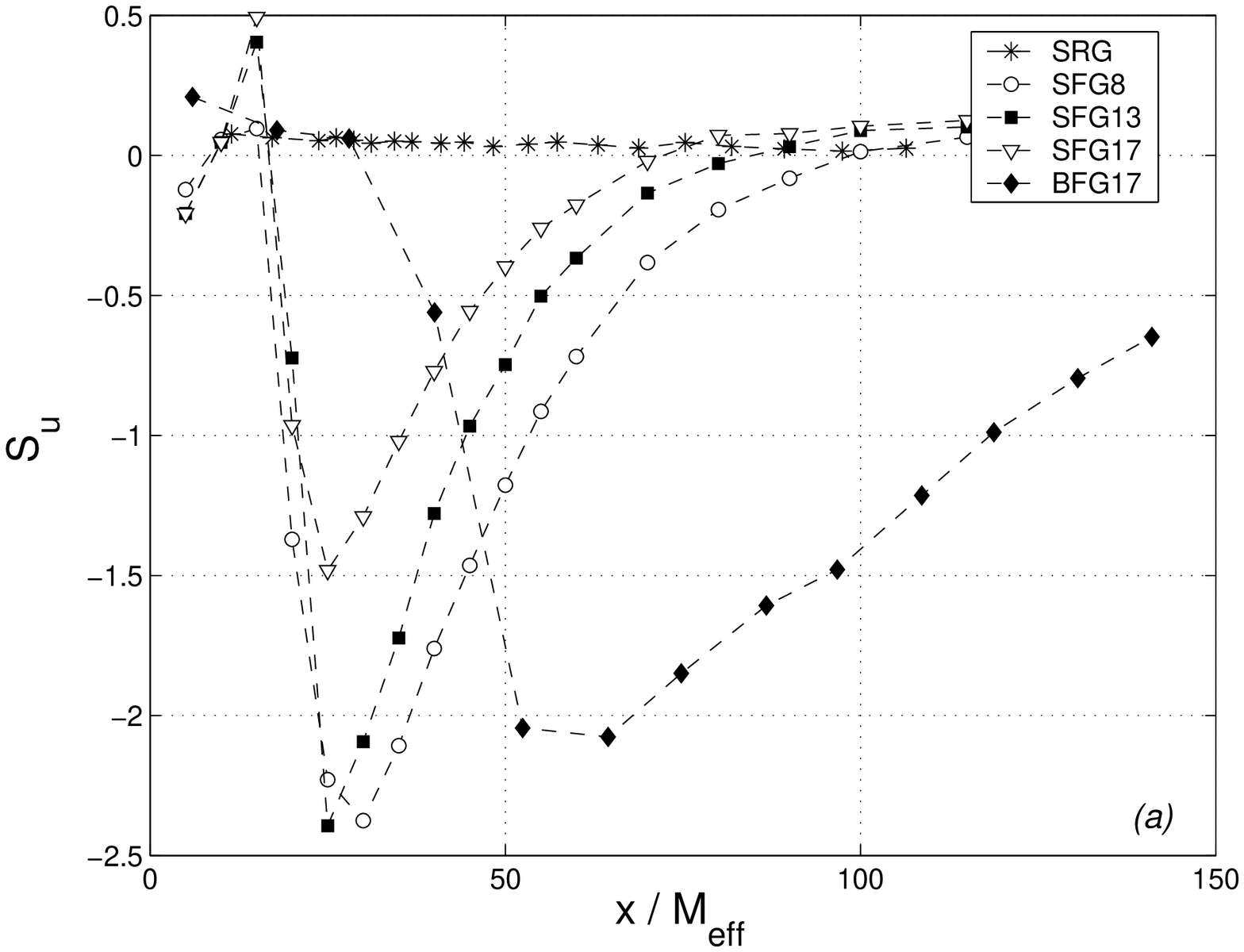}
\label{fig:Skew_Tr}}
\hspace{0.1cm}
\subfigure
{\includegraphics[width=7.5cm]{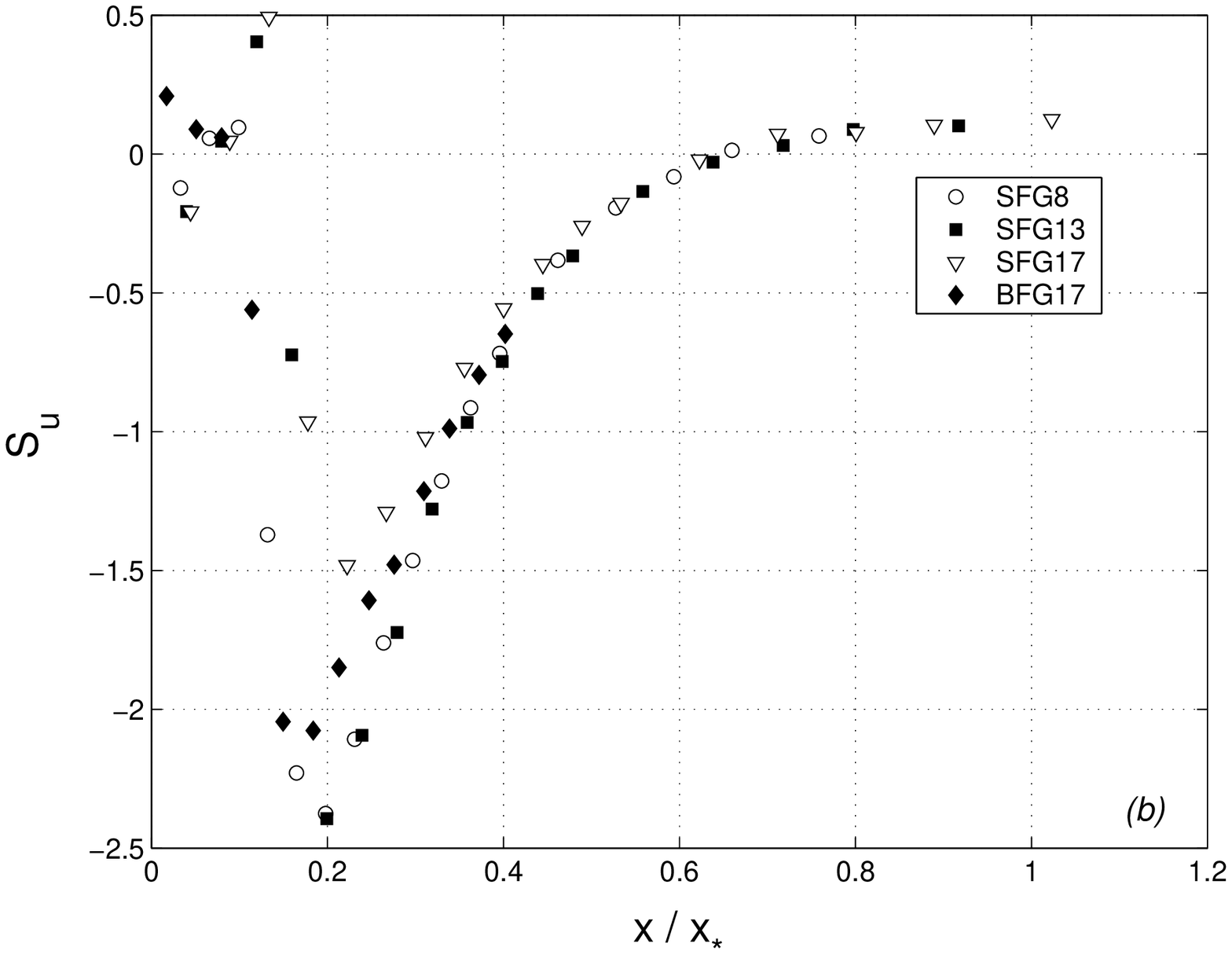}
\label{fig:Skew_Tr_Norm}}
\caption{Velocity skewness $S_u$ as function of \textit{(a)}
$x/M_{eff}$ and \textit{(b)} $x/x_*$. $U_\infty = 5.2 m/s$. Data
corresponding to all fractal grids and to the regular grid \emph{SRG}
as per insert.}
\end{figure}

\begin{figure}[htbp]
\centering
\subfigure
{\includegraphics[width=7.5cm]{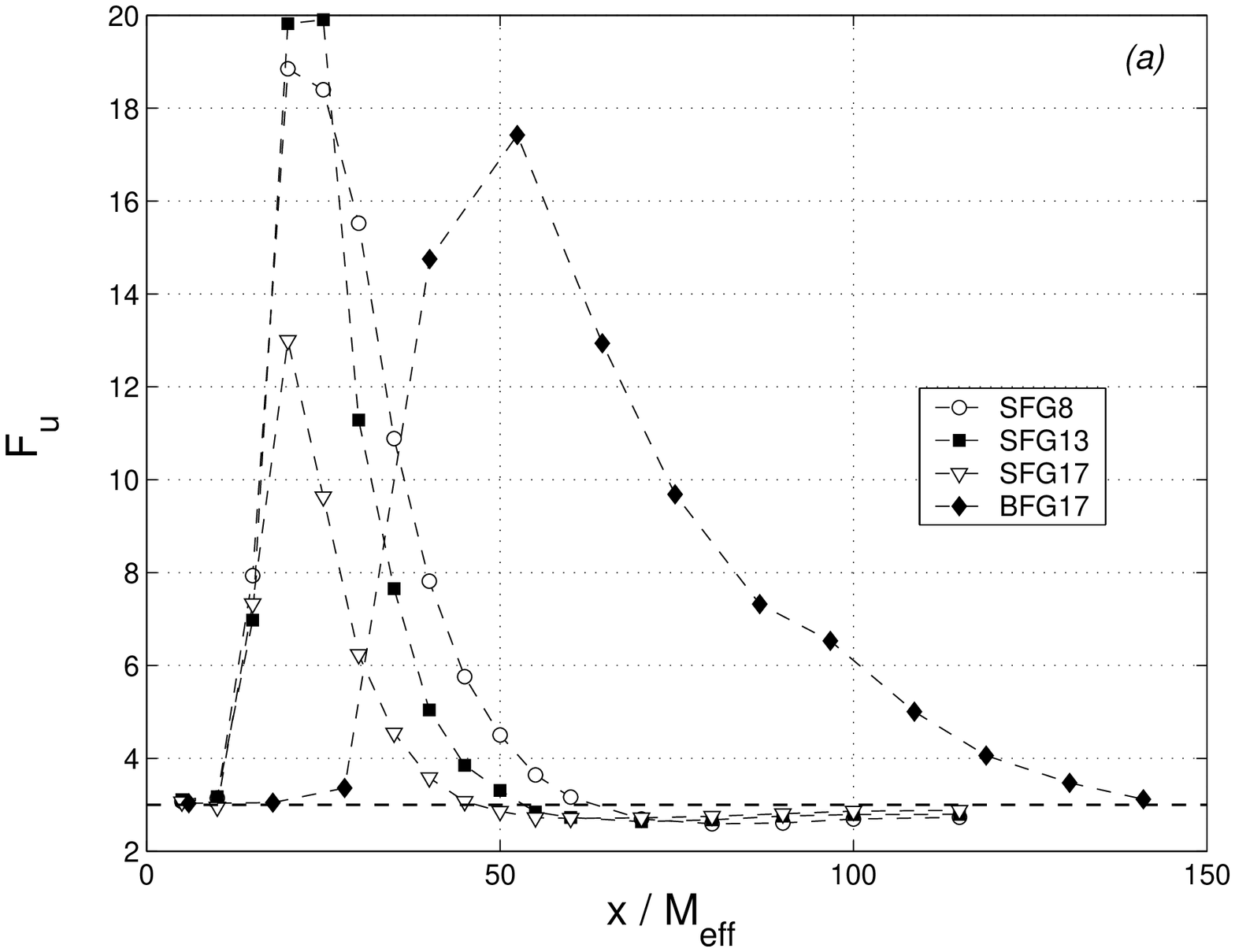}
\label{FlatTrBis}}
\hspace{0.1cm}
\subfigure
{\includegraphics[width=7.5cm]{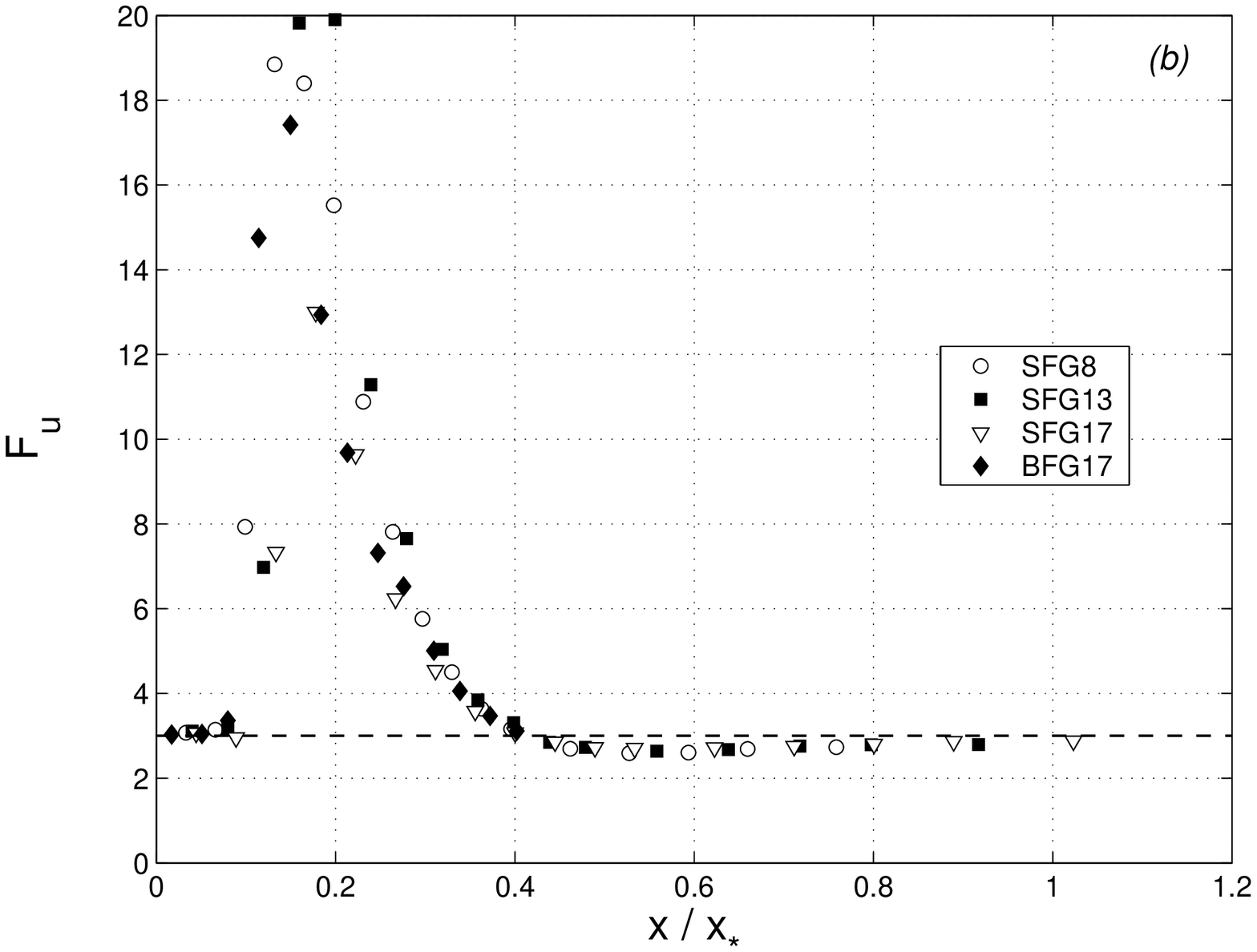}
\label{fig:FlatTr.eps}}
\caption{Velocity flatness $F_u$ as function of \textit{(a)}
$x/M_{eff}$ and \textit{(b)} $x/x_*$. $U_\infty = 5.2 m/s$. Data
corresponding to all fractal grids as per insert.  The dashed line in
both plots is $F_{u}=3$.}
\end{figure}

The behaviours of $S_u$ and $F_u$ in the production region on the
centreline are both highly unusual as can be seen in figures 10, 11
and 12. Clearly some very extreme/intense events occur at $x\approx
0.2 x_*$ and it is noteworthy that the location of these extreme
events scales with $x_*$ even though it is clearly different from
$x_{peak} \approx 0.45 x_*$. The scatter observed at and around this
location is due to lack of convergence because these intense events
are quite rare as clearly seen in time traces such as the one given in
figure 13(a). These intense events are so high in magnitude
($\left|S_u\right| \sim 1$) that they cannot be attributed to
experimental uncertainties. The negative signs of $S_u$ and of these
intense events on the time traces demonstrate clearly that these
extreme events correspond to locally decelerating flow. We leave their
detailed analysis for future study.

\begin{figure}[htbp]
\centering
\subfigure
{\includegraphics[width=7.5cm]{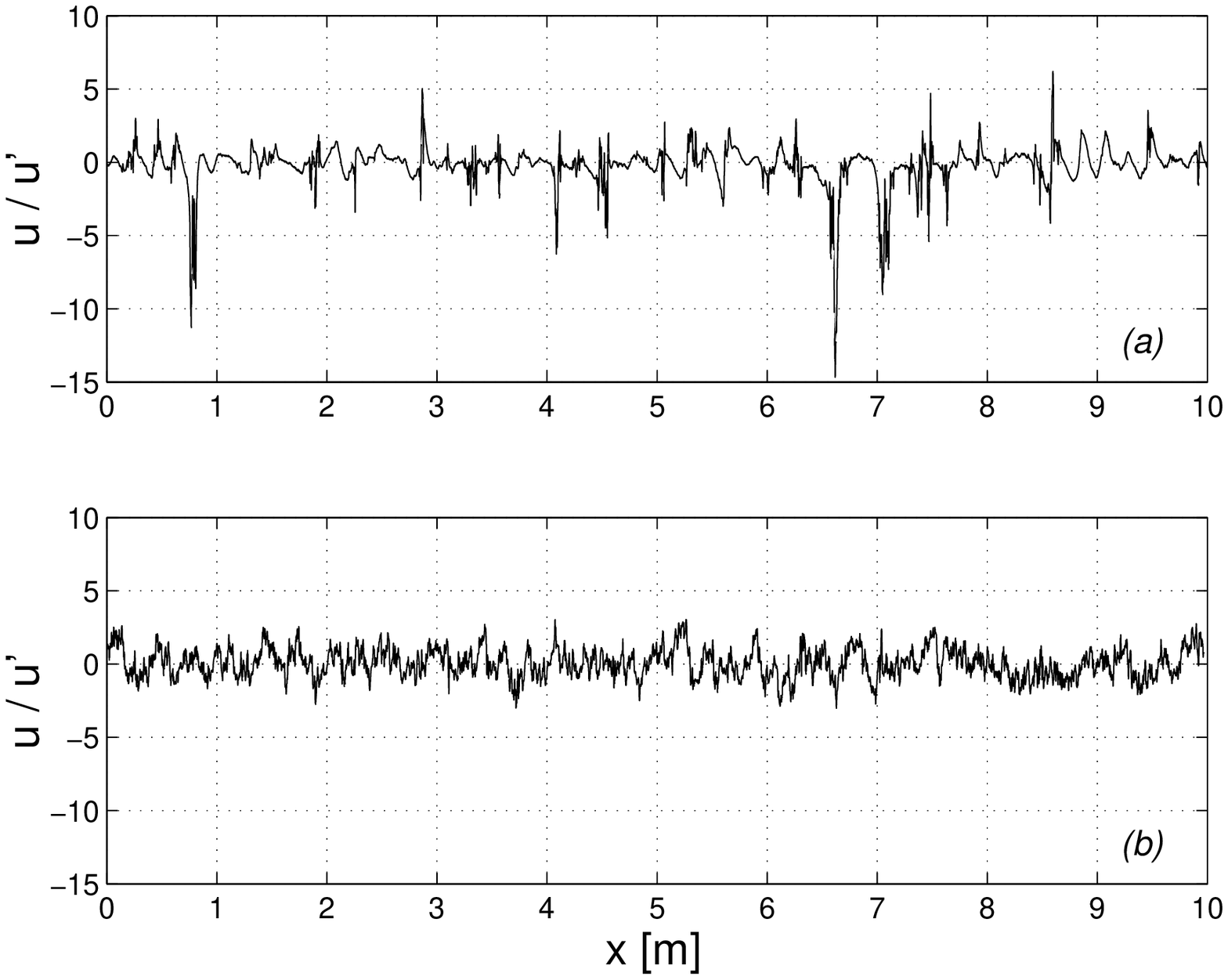}
\label{fig:VelExample_SFG17}}
\hspace{0.1cm}
\subfigure
{\includegraphics[width=7.5cm]{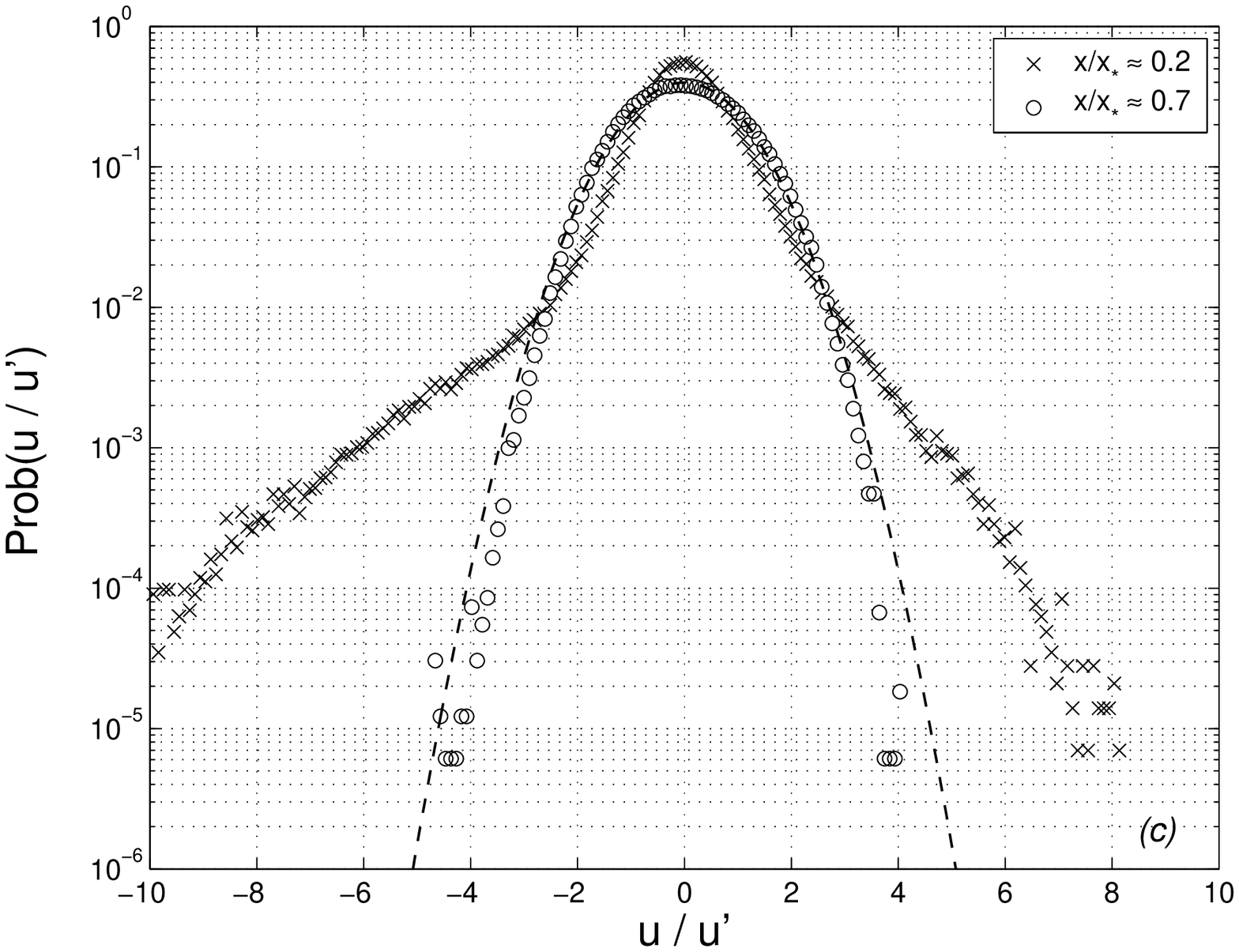}
\label{fig:Pdf_SFG17.eps}}
\caption{\textit{(a)} and \textit{(b)} Normalised velocity samples
recorded downstream of the SFG17 fractal grid at \textit{(a)} $x
\approx 0.2 x_{*}$ and \textit{(b)}$x \approx 0.7 x_{*}$. \textit{(c)}
Probability Density Function (PDF) computed from signals corresponding
to the SFG17 grid such as those shown on the left. The dashed line is
the Gaussian distribution. $U_\infty = 5.2 m/s$ for all the plots on 
this figure.}
\end{figure}

These extreme decelerating flow events are also reflected in the Probability Density Function (PDF)
of u which is clearly non-gaussian and skewed to the left
(i.e. towards negative $u$-values) at $x=0.2x_*$ whereas it is very
closely gaussian in the decay region (see figure 13(b)). The
flatness $F_u$ takes values close to the usual gaussian value of 3 in
the decay region and remains close to 3 for all $x \ge x_{peak}$
(see figures 10(b) and 12(b)).

Note, finally, that (10) and $F_{u} \approx 3$ in the homogeneous
region $x \ge 0.5 x_*$ imply that
\begin{equation}
{<u^{4}>\over U^{4}} \approx 3A^{2} 
exp\left(-2B \left(\frac{x}{x_{*}}\right)\right)
\end{equation}
in that region, with the same values $A \approx 2.82$ and $B \approx
2.06$ independently of $U_{\infty}$ and space-filling fractal square
grid.

\subsection{Spectral energy budget in the decay region}

In the region beyond $x_{peak}$ where the turbulence is approximately
homogeneous and isotropic, the energy spectrum has been reported in
previous studies \citep{SeoudVassilicos2007},
\citep{HurstVassilicos2007} to be broad with a clear power-law shaped
intermediate range where the power-law exponent is not too far from
-5/3. For this region we can follow George \& Wang
\citep{GeorgeWang2009} who found a solution of the spectral energy
equation
\begin{equation}
{\partial \over \partial t}E(k,t) = T(k,t) - 2\nu k^{2} E(k,t)
\end{equation}
that implies an exponential rather than power-law turbulence decay. In
this spectral energy equation, $E(k,t)$ is the energy spectrum and
$T(k,t)$ is the spectral energy transfer at time $t$. The energy
spectrum, if integrated, gives ${3\over 2} u'^{2}$, i.e.  
\begin{equation}
{3\over 2} u'^{2} = \int_{0}^{\infty} E(k,t) dk. 
\end{equation}
The correspondence between the time dependencies in these equations and
the dependence on $x$ in our experiments is made via Taylor's
hypothesis.

George \& Wang \citep{GeorgeWang2009} showed that (12) admits solutions of
the form
\begin{equation}
E(k,t) = E_{s} (t) f(kl(t), *)
\end{equation}
\begin{equation}
T(k,t) = T_{s} (t) g(kl(t), *)
\end{equation}
where the functions $f$ and $g$ are dimensionless, and that such
solutions can yield an exponential turbulence decay. These solutions
are single-length scale solutions (the length-scale $l(t)$) and
therefore differ fundamentally from the usual Kolmogorov picture
\citep{Frisch1995} which involves two different length scales, one
outer and one inner, their ratio being an increasing function of
Reynolds number. The argument $*$ in (14) and (15) represents any
dependencies that there might be on boundary/inlet/initial conditions.

The exponential decay reported by George \& Wang \citep{GeorgeWang2009}
exists provided that the length-scale $l(t)$ is independent of time,
i.e. ${d\over dt} l =0$, and takes the form $u'^{2}\sim \exp [-10\nu
t/l^{2}]$. Unless $l^{2} \propto \nu$, this form does not obviously
compare well with the exponential decay (10) because (10) is
independent of Reynolds number. We now carefully apply the approach of
George \& Wang \citep{GeorgeWang2009} to our data by making explicit
use of all potential degrees of freedom and confirm that an
exponential decay which perfectly fits (10) can indeed follow from
their approach.

Consider 
\begin{equation}
E(k,t) = E_{s} (t, U_{\infty}, Re_{0}, *) f(kl(t), Re_{0}, *)
\end{equation}
\begin{equation}
T(k,t) = T_{s} (t,  U_{\infty}, Re_{0}, *) g(kl(t), Re_{0}, *)
\end{equation}
where $Re_{0} \equiv {U_{\infty} t_{0}\over \nu}$ is the Reynolds
number which characterises the thickest bars on the fractal grid,
$l(t) = l(t, Re_{0}, *)$ and the argument $*$ includes various
dimensionless ratios of bar lengths and bar thicknesses on the
fractal grid, as appropriate. The functions $f$ and $g$ are again
dimensionless. The conditions for (16) and (17) to solve (12) are (see
\citep{GeorgeWang2009} for the solution method)
 \begin{equation}
{d\over dt} E_{s} = -a {2\nu \over l^{2}} E_{s}
\end{equation}
\begin{equation}
T_{s} = b {d\over dt} E_{s}
\end{equation}
\begin{equation}
{dl\over dt}/l = c {d E_{s}\over dt} / E_{s} 
\end{equation}
where $a$, $b$ and $c$ are dimensionless functions of $Re_{0}$ and $*$
(note that $a>0$). Under these solvability conditions, the spectral
energy equation (12) collapses onto the dimensionless form 
\begin{equation}
f (\kappa, Re_{0}, *) + c(Re_{0}, *)\kappa {d\over d\kappa}f (\kappa,
Re_{0}, *) = b(Re_{0}, *)g (\kappa, Re_{0}, *) + {\kappa^{2}f (\kappa,
Re_{0}, *)\over a (Re_{0}, *)}
\end{equation}
where $\kappa \equiv kl(t, Re_{0}, *)$. 

Two different families of solutions exist according to whether $c$
vanishes or not. If $c\not = 0$, then $c$ must be negative and
\begin{equation}
l(t)=l(t_{0}) [1+{4\nu a \vert c \vert \over
l^{2}(t_{0})}(t-\tau_{0})]^{1/2}
\end{equation}
\begin{equation}
E_{s}(t)=E_{s}(t_{0}) [1+{4\nu a \vert c \vert \over
l^{2}(t_{0})}(t-\tau_{0})]^{1/2c}
\end{equation}
in terms of a virtual origin $\tau_0$. However, if $c=0$, then ${dl\over dt} = 0$
and 
\begin{equation}
E_{s}(t) \sim exp\left(-2a {\nu t \over l^{2}}\right). 
\end{equation}
It is this second set of solutions, the one corresponding to $c=0$,
with which George \& Wang \citep{GeorgeWang2009} chose to explain the
form (10).

From (24), (16) and (13) and making use of $t=x/U_{\infty}$ it follows
that

\begin{equation}
u'^{2} = u'^{2}_{0} exp \left(-2a{\nu x \over l^{2} U_{\infty}}\right)
\end{equation}

where $u'_{0} = u'_{0} (U_{\infty}, Re_{0}, *)$, $a= a (Re_{0}, *)$
and $l(Re_{0}, *)$ is independent of $x$. Our wind tunnel measurements
in subsection 4.1 and those leading to (10) and its range of validity
suggest that (24) and (10) are the same provided that
$u'_{0}/U_{\infty}$ is a dimensionless function of the geometric inlet
parameters $*$ and nothing else, and that
\begin{equation}
a= 1.03 Re_{0} l^{2}(Re_{0},*)/L_{0}^{2}
\end{equation}
if use is made of (6) and $B \approx 2.06$. In other words, the
single-scale solution of George \& Wang \citep{GeorgeWang2009} fits our
data provided that the single length-scale $l (Re_{0}, *)$ is
independent of $x$ (i.e. $c=0$), that the dimensionless coefficient
$a$ in (18) depends on $Re_0$ and $*$ as per (26) and that $u'_{0}$
scales with $U_{\infty}$. Under these conditions, it follows from
equation (21) that the dimensionless spectral functions $f$ and $g$
satisfy
\begin{equation}
f (kl, Re_{0}, *) = b(Re_{0}, *)g (kl, Re_{0}, *) + 2 Re_{0}^{-1}
\left({L_{0}\over \sqrt{B} l(Re_{0},*)}\right)^{2} (kl)^{2}f (kl,
Re_{0}, *)
\end{equation}
where $B\approx 2.06$. Integrating this dimensionless balance over
$\kappa = kl$ yields
\begin{equation}
\int_{0}^{\infty} f (\kappa, Re_{0}, *) d\kappa = 2 \left({L_{0}\over
\sqrt{B} l(Re_{0},*)}\right)^{2} Re_{0}^{-1} \int_{0}^{\infty}
\kappa^{2}f (\kappa, Re_{0}, *) d\kappa
\end{equation}
because the spectral energy transfer integrates to zero. This equality
can be used in conjunction with (16), (13) and (6) to obtain a formula
for the kinetic energy dissipation rate per unit mass, $\epsilon =
2\nu \int_{0}^{\infty} k^{2} E(k,t) dk$:
\begin{equation}
\epsilon = {3B\over 2} {u'^{2} U_{\infty} \over x_{*}} \approx 3.1 {u'^{2}
U_{\infty} \over x_{*}} .
\end{equation}

This is an important reference formula which we have been able to
reach by applying the George \& Wang \citep{GeorgeWang2009} theory and
by confronting it with new measurements which we obtained for three
different yet comparable fractal grids and three different inlet
velocities $U_{\infty}$. These are the new measurements reported in
subsections 4.1. and 4.2 and summarised by (10) and (6) along with the
observation that $A$ and $B$ in (10) do not depend on $U_{\infty}$ and
on the different parameters of the space-filling fractal square grids
used.

\subsection{Multiscale-generated single-length scale turbulence} 

No sufficiently well-documented boundary-free shear flow
\citep{TennekesLumley1972}, \citep{Pope2000} nor wind tunnel turbulence
generated by either regular or active grids \citep{MohamedLaRue1990},
\citep{MydlarskiWarhaft1996} has turbulence properties comparable to
those discussed here, namely an exponential turbulence decay (10), a
dissipation rate $\epsilon$ proportional to $u'^2$ rather than the
usual $u'^{3}$, and spectra which can be entirely collapsed with a
single length-scale. It is therefore important to subject our data to
further and more searching tests.


The downstream variation of the Reynolds number $Re \equiv
{u'L_{u}\over \nu}$ is different for different boundary-free shear
flows. However, it is always a power-law of the normalised streamwise
distance ${x-x_{0}\over L_{B}}$ where $L_B$ is a length characterising
the inlet and $x_0$ is an effective/virtual origin. For example, $Re
\sim ({x-x_{0}\over L_{B}})^{-1/3}$ for axisymmetric wakes, $Re \sim
({x-x_{0}\over L_{B}})^{0}$ for plane wakes and axisymmetric jets and
$Re \sim ({x-x_{0}\over L_{B}})^{1/2}$ for plane jets
\citep{TennekesLumley1972}, \citep{Pope2000}. The turbulence intensity's
downstream dependence on ${x-x_{0}\over L_{B}}$ is $\sim
({x-x_{0}\over L_{B}})^{-2/3}$ for axisymmetric wakes, $\sim
({x-x_{0}\over L_{B}})^{-1/2}$ for plane wakes and jets and $\sim
({x-x_{0}\over L_{B}})^{-1}$ for axisymmetric jets
\citep{TennekesLumley1972}, \citep{Pope2000}. In wind tunnel turbulence
generated by either regular or active grids the downstream turbulence
also decays as a power law of ${x-x_{0}\over L_{B}}$ and so does $Re$
\citep{MohamedLaRue1990}, \citep{MydlarskiWarhaft1996}. In all these
flows, as in fact in all well-documented boundary-free shear flows,
the integral length scale $L_u$ and the Taylor microscale $\lambda$
grow with increasing $x$, and in fact do so as power laws of
${x-x_{0}\over L_{B}}$. Their ratio $L_{u}/\lambda$ is a function of
${x-x_{0}\over L_{B}}$ and of an inlet Reynolds number $Re_{0} \equiv
{U_{\infty} L_{B}\over \nu}$ where $U_{\infty}$ is the appropriate
inlet velocity scale. Estimating $\lambda$ from $\nu
u'^{2}/\lambda^{2} \sim \epsilon \sim u'^{3}/L_{u}$
\citep{Batchelor1953}, \citep{TennekesLumley1972}, \citep{Pope2000}, it
follows that $L_{u}/\lambda \sim Re_{0}^{1/2} ({x-x_{0}\over
L_{B}})^{-1/6}$ for axisymmetric wakes, $L_{u}/\lambda \sim
Re_{0}^{1/2} ({x-x_{0}\over L_{B}})^{0}$ for plane wakes and
axisymmetric jets, $L_{u}/\lambda \sim Re_{0}^{1/2} ({x-x_{0}\over
L_{B}})^{1/4}$ for plane jets and $L_{u}/\lambda \sim Re_{0}^{1/2}
({x-x_{0}\over L_{B}})^{-p}$ with $p>1/2$ for wind tunnel
turbulence. These downstream dependencies on $Re_{0}$ and
${x-x_{0}\over L_{B}}$ can be collapsed together as follows:
\begin{equation} 
{L_{u}\over \lambda} \sim Re_{\lambda} 
\end{equation}
where $Re_{\lambda} \equiv {u'\lambda\over \nu}$. This means that
values of $L_u/\lambda$ obtained from measurements at different
downstream locations $x$ but with the same inlet velocity $U_{\infty}$
and values of $L_u/\lambda$ obtained for different values of
$U_{\infty}$ but the same downstream location fall on a single
straight line in a plot of $L_u/\lambda$ versus $Re_{\lambda}$. This
conclusion can in fact be reached for all sufficiently well-documented
boundary-free shear flows \citep{TennekesLumley1972} as for decaying
homogeneous isotropic turbulence \citep{Pope2000}, \citep{Frisch1995} if
the cornerstone assumption $\epsilon \sim u'^{3}/L_{u}$ is used.

The relation $L_{u}/\lambda \sim Re_{\lambda}$ is a direct expression
of the Richardson-Kolmogorov cascade and, in particular, of the
existence of an inner and an outer length-scale which are decoupled,
thus permiting the range of all excited turbulence scales to grow with
increasing Reynolds number. This relation is therefore in direct
conflict with the one-length scale solution of George \& Wang
\citep{GeorgeWang2009}. Seoud \& Vassilicos \citep{SeoudVassilicos2007}
found that $L_u/\lambda$ is independent of $Re_\lambda$ in the decay
region of the turbulent flows generated by our space-filling fractal
square grids (in which case $Re_0$ is defined in terms of the wind
tunnel inlet velocity $U_{\infty}$ and $L_{B} = t_0$). Here we
investigate further and refine this claim and also show that it is
compatible with the theory of George \& Wang \citep{GeorgeWang2009}.

The essential ingredient in the previous subsection's considerations
is the single-length scale form of the spectrum (16). Our hot wire
anemometry can only access the 1D longitudinal energy spectrum $E_{u}
(k_{x})$ of the longitudinal fluctuating velocity component $u$. The
single-length scale form of $E_{u} (k_{x})$ is $E_{u} (k_{x})= E_{su}
f_{u} (k_{x} l, Re_{0}, *)$ which can be rearranged as follows if use
is made of ${1\over 2} u'^{2} = \int_{0}^{\infty} E_{u}(k_{x})
dk_{x}$:
\begin{equation}
E_{u} (k_{x}) = u'^{2} l F_{u} (k_{x} l, Re_{0}, *)
\end{equation}
where $F_{u} (k_{x} l, Re_{0}, *) = {1\over 2} f_{u} (k_{x} l, Re_{0},
*)/\int dk_{x} l f_{u} (k_{x} l, Re_{0}, *)$. In the case where
$u'^{2} = u'^{2} (x, U_{\infty}, *)$ decays exponentially (equations
(24) to (29)), the length-scale $l$ is independent of the streamwise
distance $x$ from the grid.

An important immediate consequence of the single length-scale form of
the energy spectrum is that both the integral length-scale $L_u$ and
the Taylor microscale $\lambda$ are proportional to $l$
\citep{GeorgeWang2009}, \citep{SeoudVassilicos2007}. Specifically,
\begin{equation}
L_u = l \int d\kappa_{x} \kappa_{x}^{-1} F_{u} (\kappa_{x} , Re_{0},
*) = \alpha l
\end{equation}
and 
\begin{equation}
\lambda= l /\sqrt{\int d\kappa_{x} \kappa_{x}^{2} F_{u} 
(\kappa_{x} , Re_{0}, *)} = \beta l 
\end{equation}
where $\alpha$ and $\beta$ are dimensionless functions of $Re_{0}$ and
$*$. This implies, in particular, that both $L_u$ and $\lambda$ should
be independent of $x$ (as was reported in \citep{HurstVassilicos2007}
and \citep{SeoudVassilicos2007}) if $l$ is independent of $x$ in the
decay region. Using (29) and $\nu u'^{2}/\lambda^{2} \sim
u'^{2}U_{\infty}/x_{*}$ we obtain
\begin{equation}
\lambda \sim L_{0} Re_{0}^{-1/2}
\end{equation}
which is fundamentally incompatible with the usual $\epsilon \sim
u'^{3}/L_{u}$. As noted in \citep{SeoudVassilicos2007}, $\epsilon \sim
u'^{3}/L_{u}$ is in fact straightforwardly incompatible with an
exponential turbulence decay such as (10) and an integral length-scale
$L_u$ independent of $x$.

We now report measurements of $L_u$, $\lambda$ and $E_{u} (k_{x})$
which we use to test the single-length scale hypothesis and its
consequences. These measurements also provide some information on the
dependencies of $L_u$ and $\lambda$ on $Re_0$ and $*$.

\begin{figure}[htbp]
\centering
\subfigure
{\includegraphics[width=7.5cm]{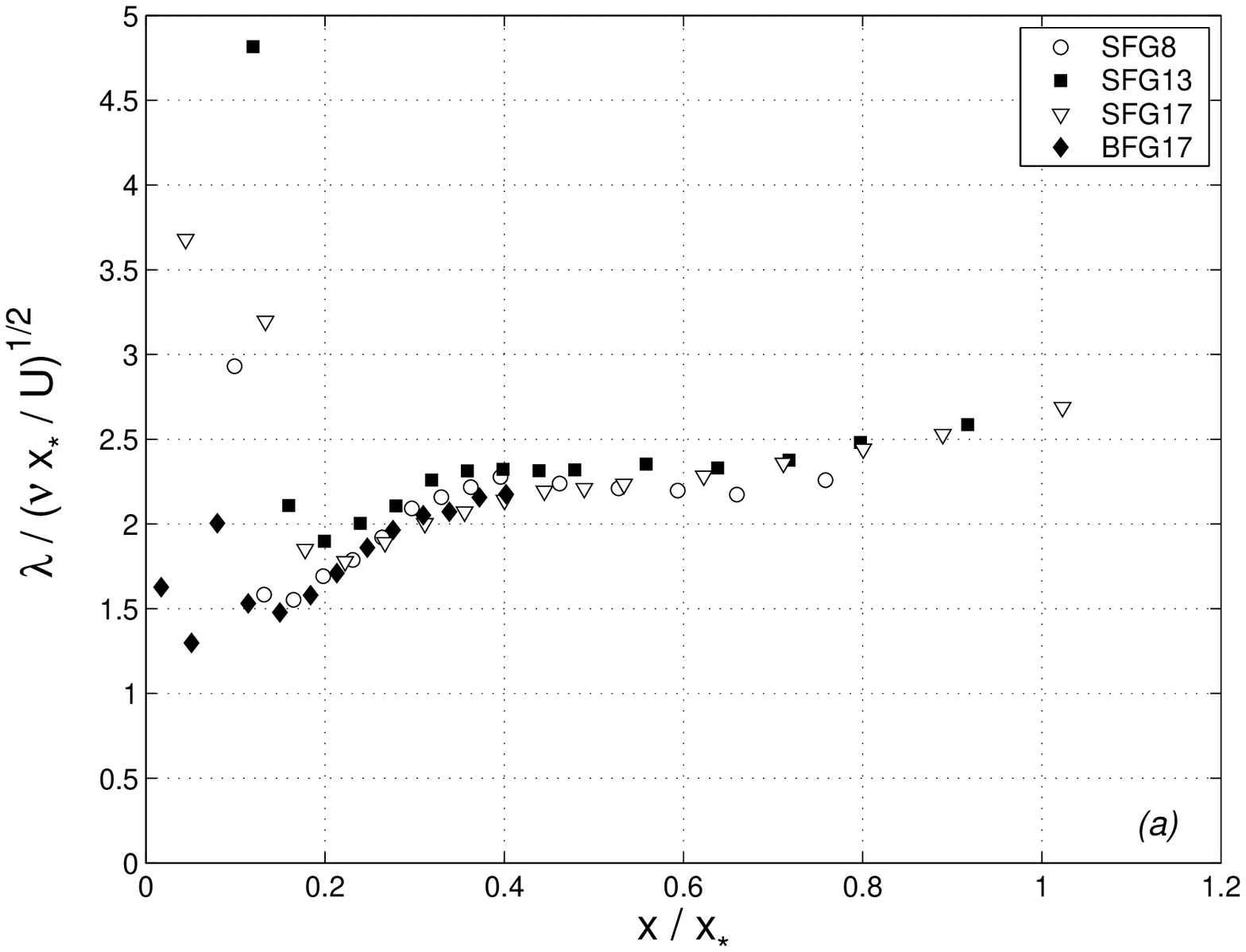}
\label{fig:LambdaScaling}}
\hspace{0.1cm}
\subfigure
{\includegraphics[width=7.5cm]{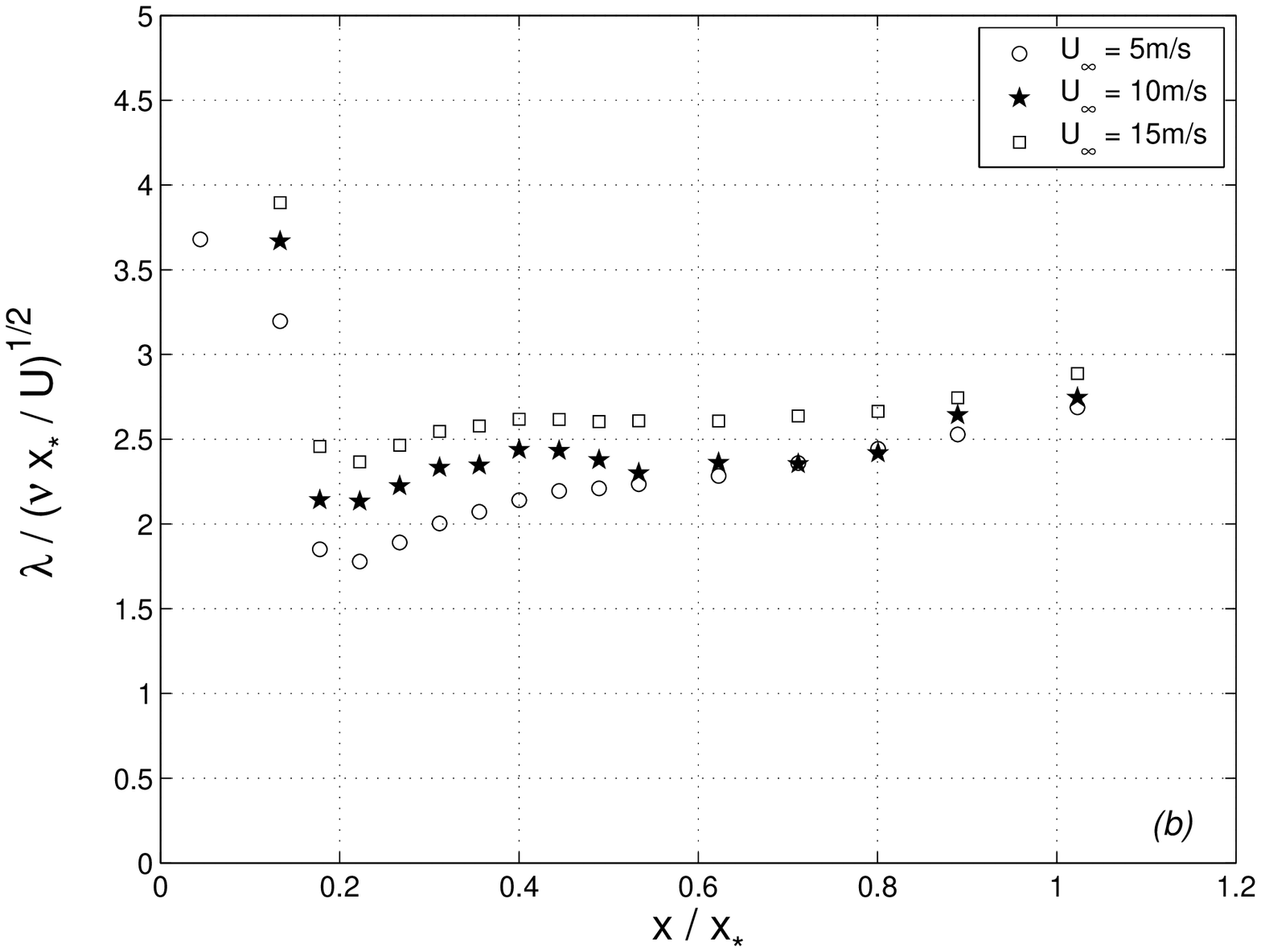}
\label{fig:LambdaUinf}}
\caption{$\lambda/\sqrt{\nu x_{*}/U}$ versus $x/x_*$ along the
centreline.  \textit{(a)} Data for the four different fractal grids
and $U_\infty = 5.2 m/s$. \textit{(b)} Data for the fractal grid SFG17
and three different inlet velocities $U_{\infty}$.}
\end{figure}

Firstly, we test the validity of (34). In figures
\ref{fig:LambdaScaling} and \ref{fig:LambdaUinf} we plot
$\lambda/\sqrt{\nu x_{*}/U}$ versus $x/x_{*}$ along the
centreline. These figures do not change significantly if we plot
$\lambda/\sqrt{\nu x_{*}/U_{\infty}} = (\lambda/L_{0}) Re_{0}^{1/2}$
versus $x/x_{*}$. It is clear that (34) and scaling $x$ with $x_*$
offers a good collapse between the different fractal grids where $x >
0.2x_{*}$, and that $\lambda$ does indeed seem to be approximately
independent of $x/x_{*}$ in the decay region as reported in
\citep{HurstVassilicos2007} and \citep{SeoudVassilicos2007} and as
predicted by (33) and (34). However the collapse for different values
of $U_{\infty}$ is not perfect and there seems to be a residual
dependence on $Re_{0}$ which is not taken into account by (34). It is
worth noting here that our centreline measurements for the regular
grid SRG with $U_{\infty}=5.2 m/s$ produced data which are very well
fitted by $(\lambda/M_{eff})^{2} = 3 \cdot 10^{-4} x/M_{eff}$ in
agreement with previous results \citep{BatchelorTownsend1948},
\citep{MohamedLaRue1990} and usual expectations \citep{Batchelor1953}.

\begin{figure}[htbp]
\centering
\subfigure
{\includegraphics[width=7.5cm]{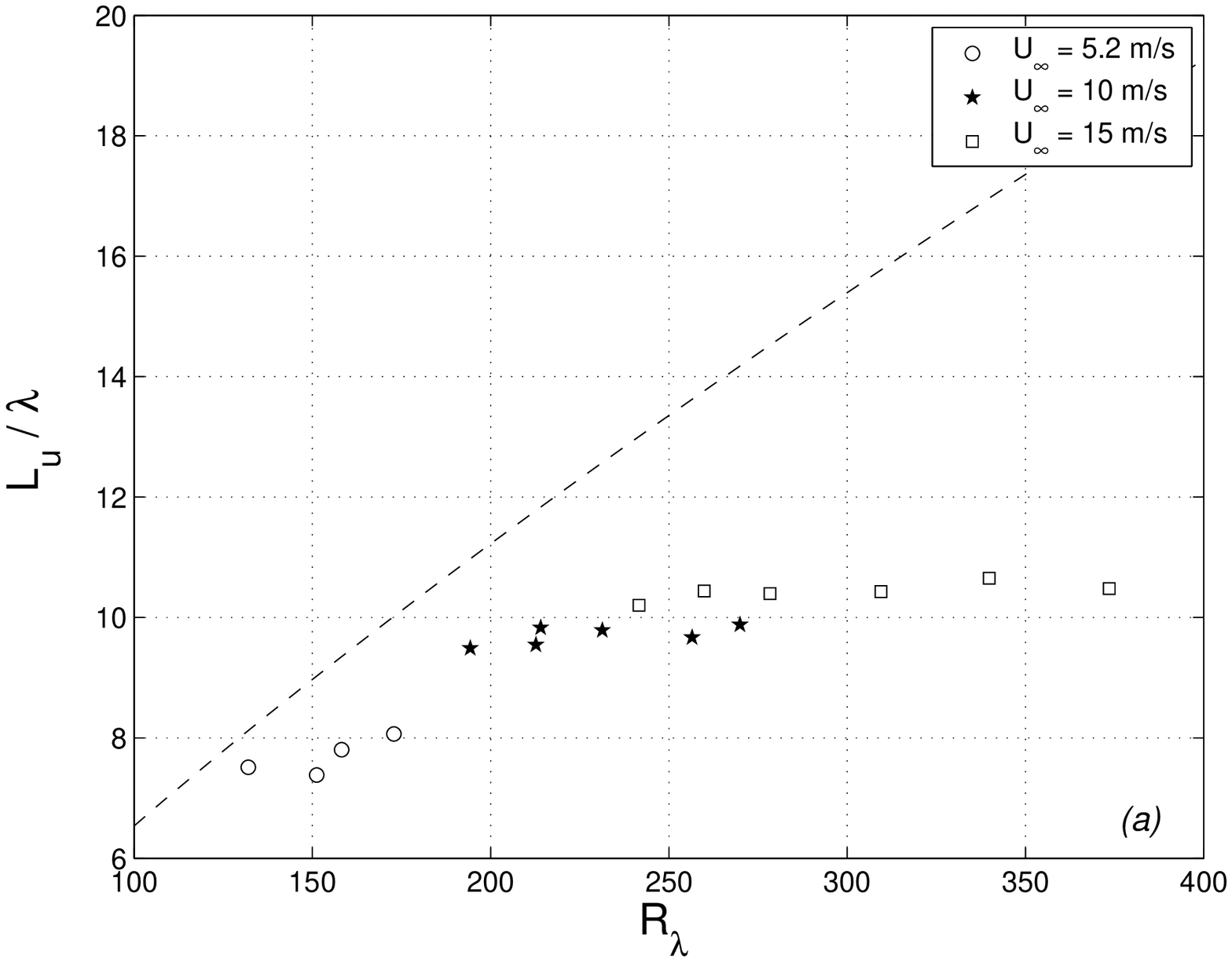}
\label{fig:LoverLambda}}
\hspace{0.1cm}
\subfigure
{\includegraphics[width=7.5cm]{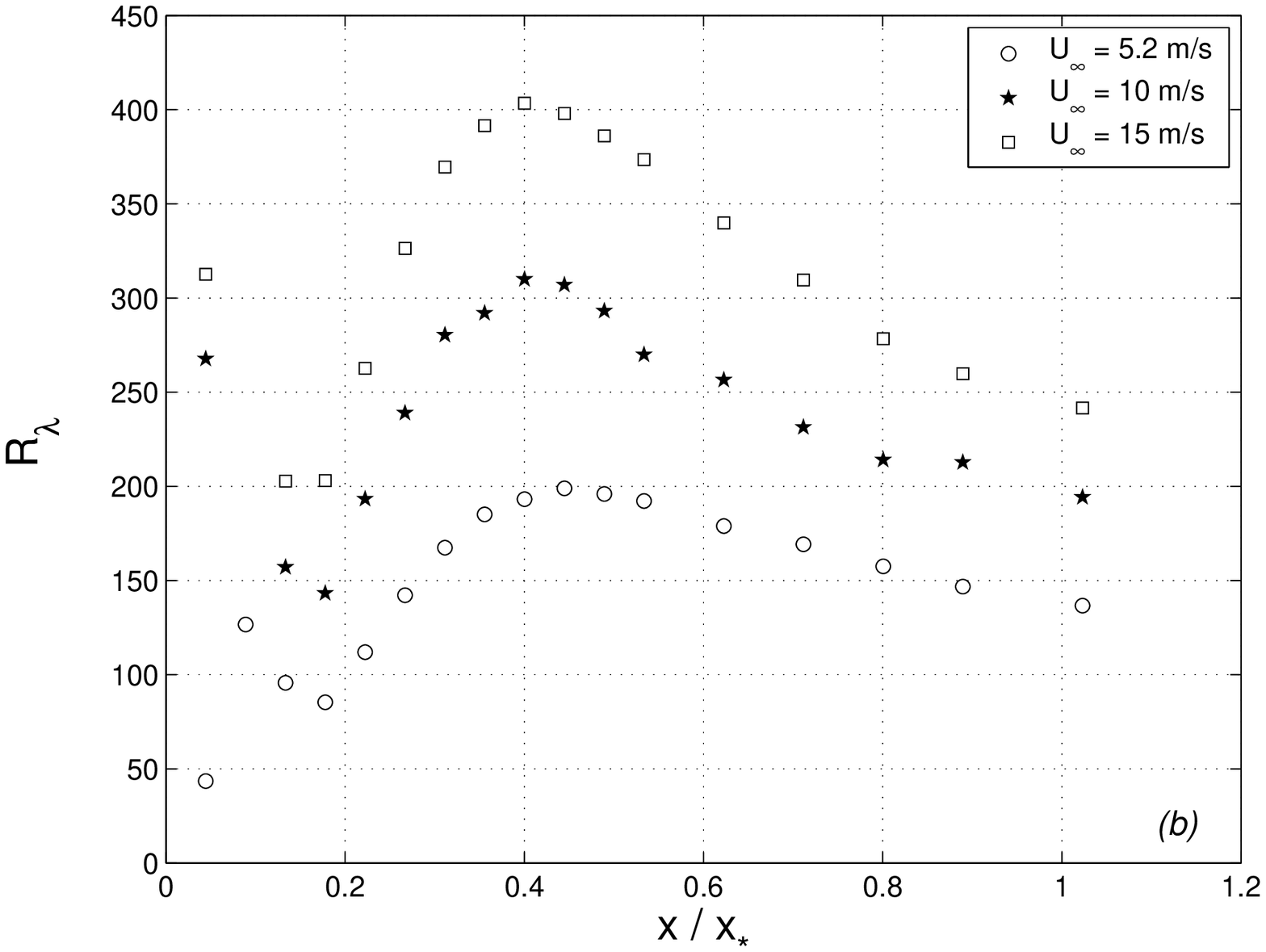}
\label{fig:Rlambda}}
\caption{\textit{(a)} $L_{u} / \lambda$ as a function of $R_\lambda$
for different inlet velocities $U_{\infty}$ and different positions on
the centreline in the decay region in the lee of the \emph{SFG17}
fractal grid. The dashed line represents the empirical law obtained by
fitting experimental data from various turbulent flows none of which
is generated by fractal square grids. This is the data compiled in
\citep{MazellierVassilicos2008} (which includes jet, regular grid, wake
and chunk turbulence), and the dashed line is an excellent
representation of $L_{u} / \lambda \sim R_{\lambda}$ particularly at
$Re_{\lambda}>150$. \textit{(b)} $R_\lambda$ as a function of
$x/x_{*}$ for three different inlet velocities $U_{\infty}$ in the lee
of the \emph{SFG17} fractal grid.}
\end{figure}

Before investigating Reynolds number corrections to (34), and
therefore (29) which (34) is a direct expression of, we check that the
turbulence generated by low-blockage space-filling fractal square
grids is indeed fundamentally different from other turbulent
flows. For this, we plot $L_{u}/\lambda$ versus $Re_{\lambda}$ in
figure \ref{fig:LoverLambda}. Whilst (30), which follows from
$\epsilon \sim u'^{3}/L_{u}$, is very well satisfied in turbulent
flows not generated by fractal square grids, it is clearly violated by
an impressively wide margin in the decay region of turbulent
flows generated by low-blockage space-filling fractal square
grids. This is not just a matter of a correction to usual laws; it is
a matter of dramatically different laws. 

\begin{figure}[htbp]
\centering
\subfigure
{\includegraphics[width=7.5cm]{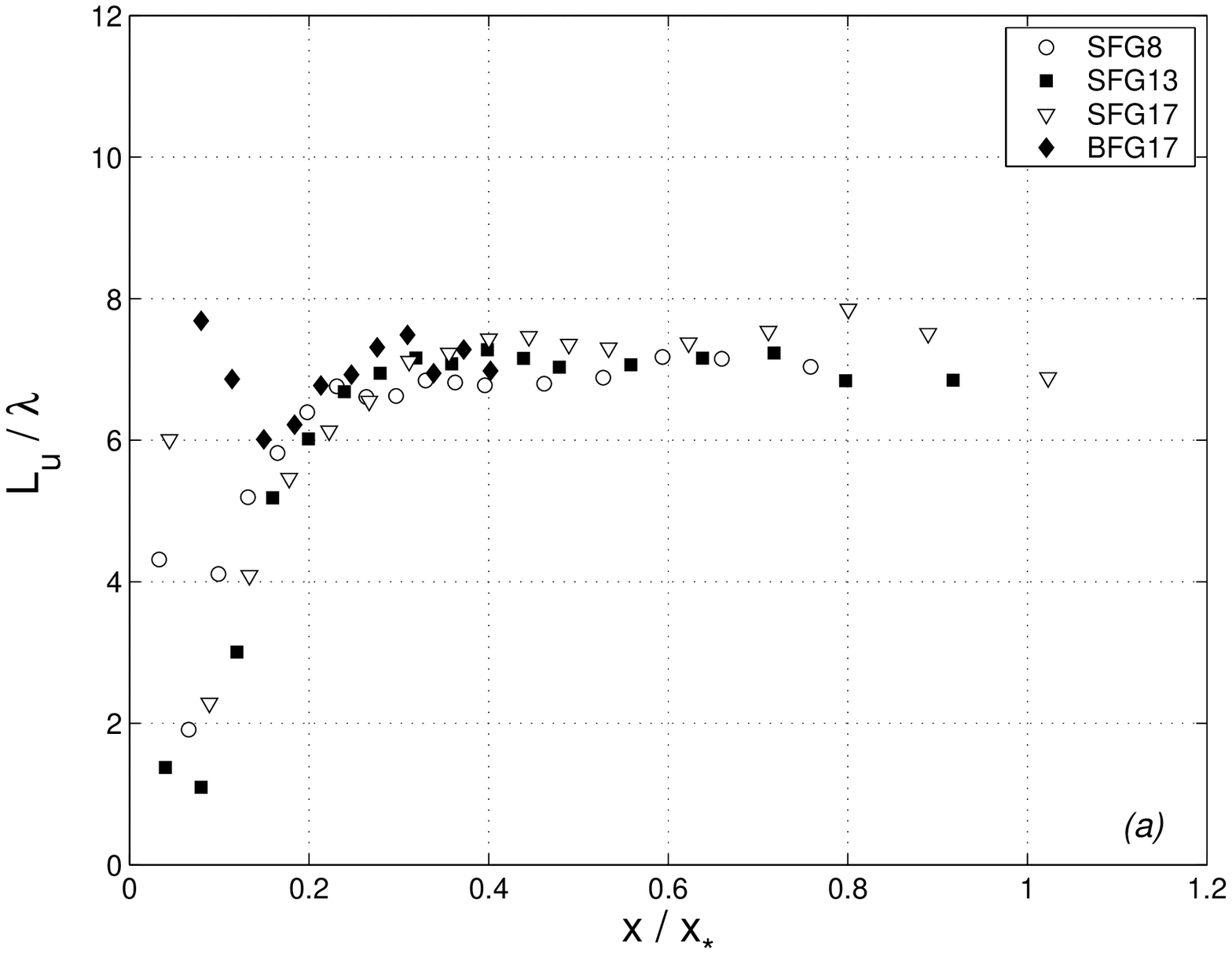}
\label{fig:Lambda_Tr}}
\hspace{0.1cm}
\subfigure
{\includegraphics[width=7.5cm]{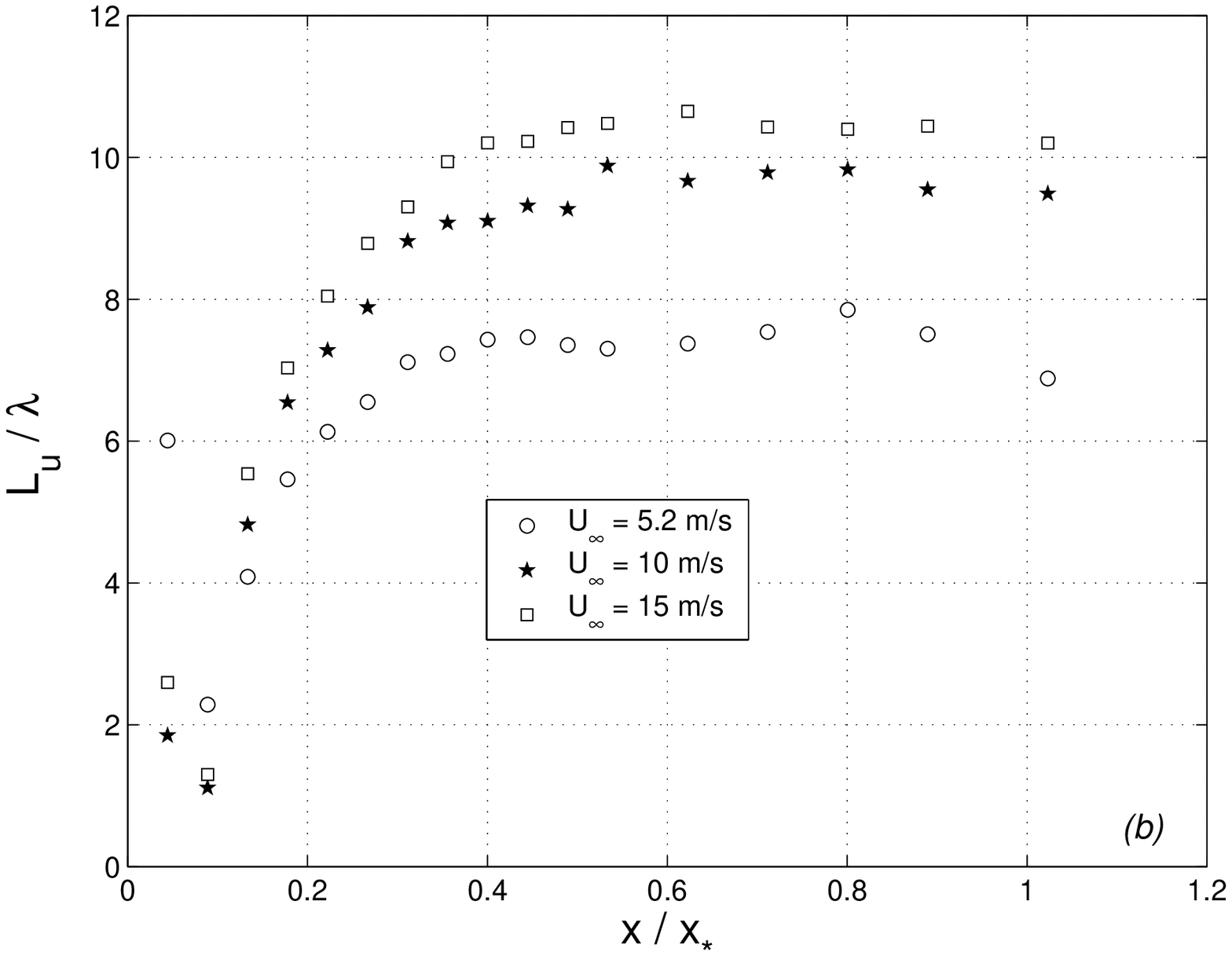}
\label{fig:Lambda_Uinlet}}
\caption{$L_{u} / \lambda$ as a function of $x/x_{*}$ along the
centreline \textit{(a)} for the four different fractal grids and
$U_\infty = 5.2 m/s$ and \textit{(b)} for the fractal grid SFG17 and
three different inlet velocities $U_{\infty}$.}
\end{figure}

One important aspect of (30) is that it collapses onto a single curve
the $x$ and $Re_0$ dependencies of $L_{u}/\lambda$ for many turbulent
flows. Figures \ref{fig:Lambda_Tr} and \ref{fig:Lambda_Uinlet} show
clearly that, in the decay region, $L_{u} / \lambda$ is independent of
$x$ and also not significantly dependent on fractal square grid, but
is clearly dependent on $U_{\infty}$. There are other turbulent flows
where $L_{u}/\lambda$ is independent of $x$, notably plane wakes and
axisymmetric jets. However, the important difference is that
$Re_{\lambda}$ is also independent of $x$ in plane wake and
axisymmetric jet turbulence whereas it is strongly varying with $x$ in
turbulent flows generated by fractal square grids (see figure
\ref{fig:Rlambda}). As a result, (30) holds for plane wakes and
axisymmetric jets but not for turbulent flows generated by fractal
square grids where, instead, $L/\lambda$ is independent of
$Re_{\lambda}$ in the decay region (see figure \ref{fig:LoverLambda}),
as previously reported in \citep{SeoudVassilicos2007}. It is not fully
clear from figure \ref{fig:LoverLambda} if $L/\lambda$ is or is not a
constant independent of $Re_0$ for large enough values of $Re_0$
(specifically for values of $U_{\infty}$ larger or equal to $10 m/s$
in the case of \ref{fig:LoverLambda}). The results in
\citep{SeoudVassilicos2007} might suggest that $L/\lambda$ is
independent of $Re_0$ for large enough values of $Re_0$, but figure
\ref{fig:Lambda_Uinlet} does not comfortably support such a
conclusion. More data are required for a conclusive assessment of this
issue which is therefore left for future study.

\begin{figure}[htbp]
\centering
\subfigure
{\includegraphics[width=7.5cm]{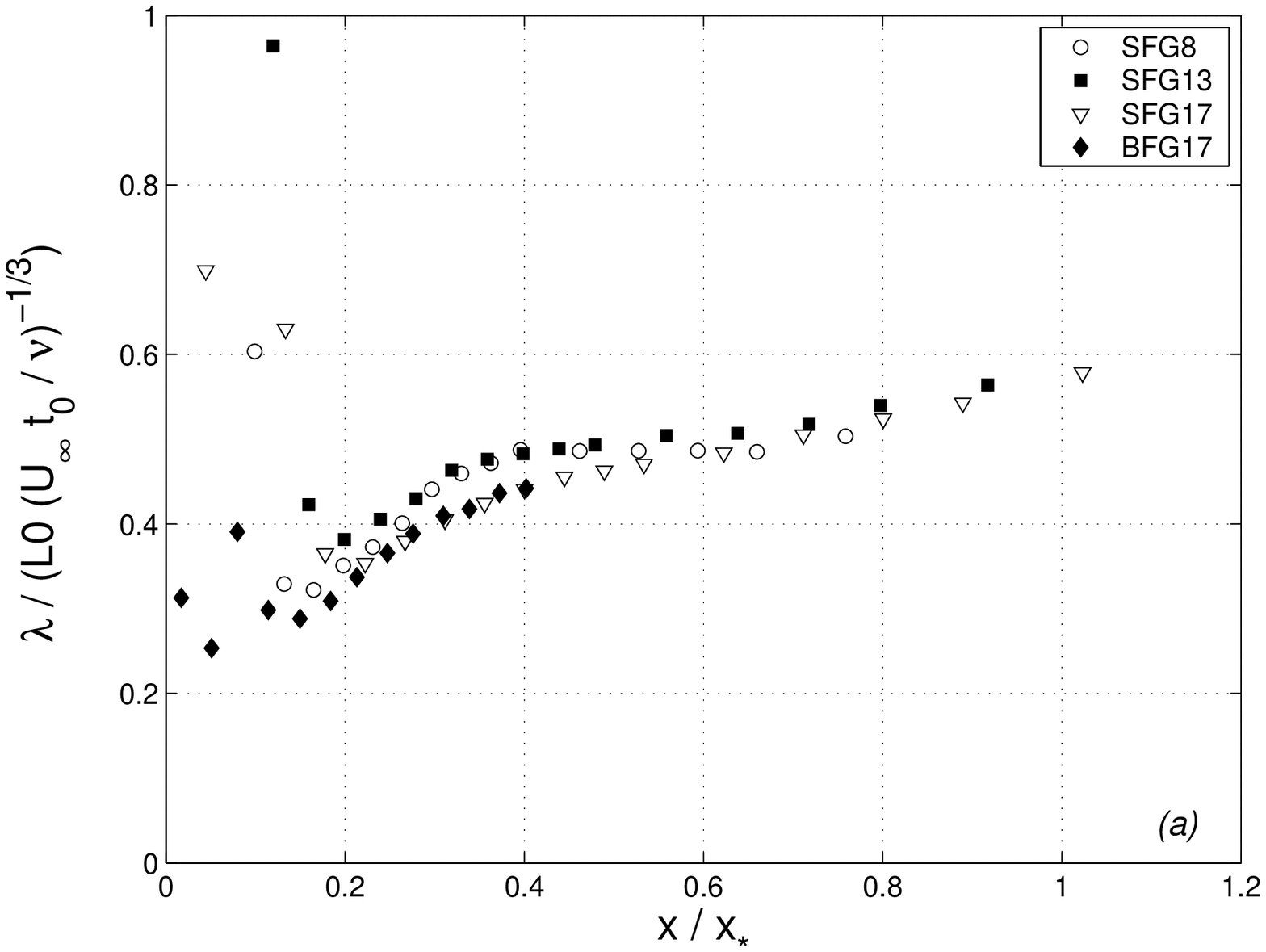}
\label{fig:LambdaScaling_new2}}
\hspace{0.1cm}
\subfigure
{\includegraphics[width=7.5cm]{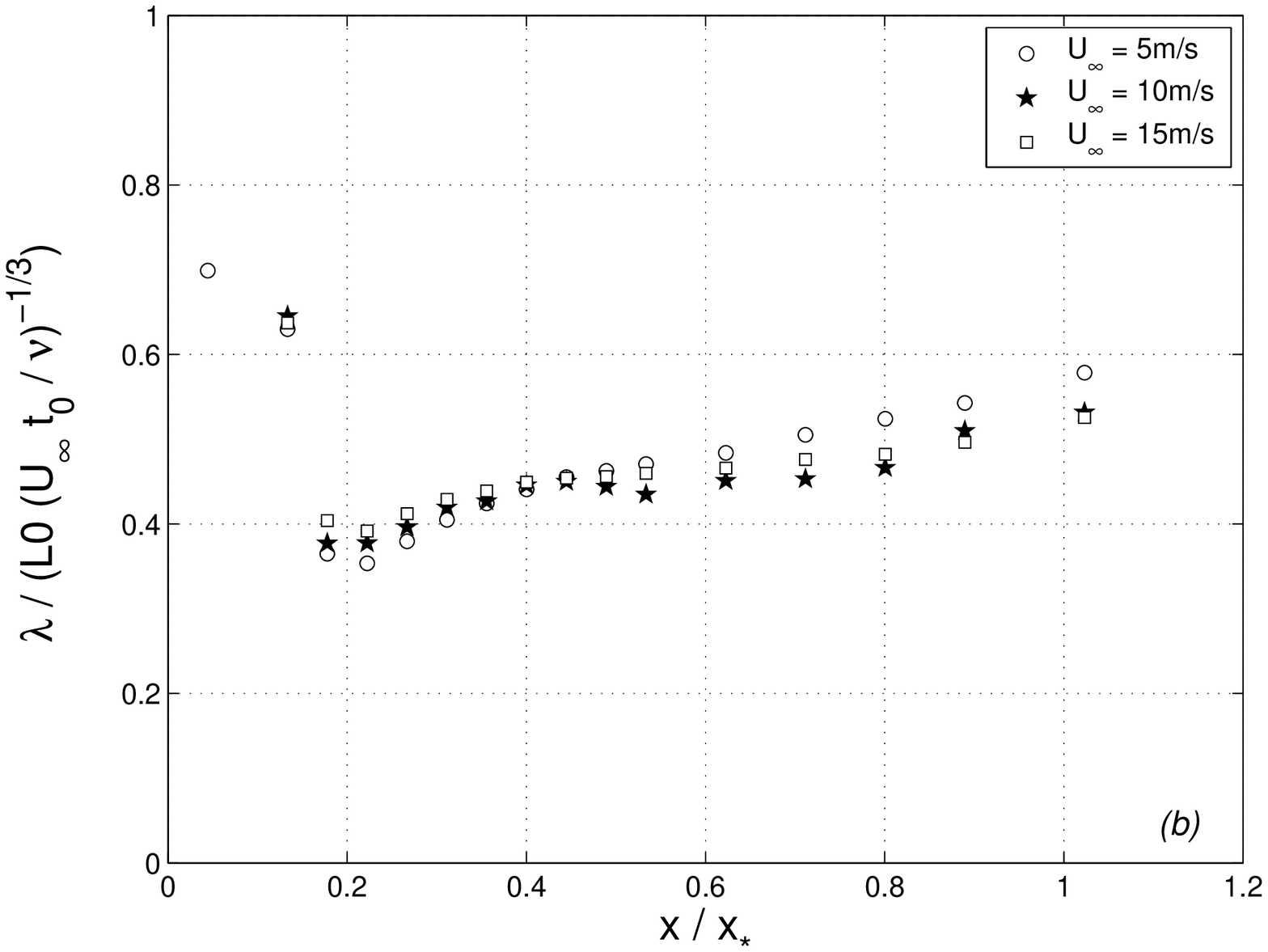}
\label{fig:LambdaUinf_new2}}
\caption{$\lambda/ \left( L_{0} ({U_{\infty} t_{0}\over \nu})^{-n}
\right)$ versus $x/x_*$ along the centreline for $n=1/3$ (various
values of $n$ were tried and $n=1/3$ offers a clear best fit).
\textit{(a)} Data for the four different fractal grids and $U_\infty =
5.2 m/s$. \textit{(b)} Data for the fractal grid SFG17 and three
different inlet velocities $U_{\infty}$. The results look very similar
when plotting $\lambda/ \left( L_{0} ({U t_{0}\over
\nu})^{-n} \right)$.}
\end{figure}

We now turn to the Reynolds number corrections which our measurements
suggest may be needed in (34). Figures \ref{fig:LambdaScaling_new2} and
\ref{fig:LambdaUinf_new2} show that 
\begin{equation}
\lambda \sim L_{0} Re_{0}^{-1/3}
\end{equation}
is a better approximation than (34) as it collapses the different
$U_{\infty}$ data better without altering the quality of the collapse
between different fractal grids. From $\epsilon \sim \nu
u'^{2}/\lambda^{2}$, (35) implies
\begin{equation}
\epsilon \sim {u'^{2}U_{\infty}\over x_{*}}Re_{0}^{-1/3}
\end{equation}
which is an important $Re_0$-deviation from (29) and carries with it
the extraordinary implication that $\epsilon$ tends to $0$ as $Re_{0}
\to \infty$. Of course, this implication is an extrapolation of our
results and must be dealt with care. In fact, we show in subsection
4.8, that this extrapolation is actually not supported by our data and
the single-length scale theoretical framework of our work.

There is only two ways in which George \& Wang's \citep{GeorgeWang2009}
single-scale theory can account for these deviations from (29) and
(34). Either (i) these deviations are an artifact of the different
large-scale anisotropy conditions for different values of $Re_0$, or
(ii) the single-length scale solution of (12) which in fact describes
our fractal-generated turbulent flows belongs to the family for which
$c \not = 0$, not the family for which $c=0$ (see subsection 4.4).

(i) Large-scale anisotropy affects the dimensionless coefficient
required to replace the scaling $\nu u'^{2}/\lambda^{2} \sim \epsilon
(= {3\over 2}B {u'^{2} U_{\infty}\over x_{*}}$ according to (29)) by an exact
equality. This issue requires cross-wire measurements at many values
of $U_{\infty}$ to be settled and must be left for future study.

(ii) If $c\not = 0$, i.e. $c<0$, then 
\begin{equation}
u'^{2} = {2E_{s}(x_{0})\over 3l(x_{0})} [1+{4\nu a \vert c \vert \over
l^{2}(x_{0})U_{\infty}}(x-x_{0})]^{(1-c)/2c}
\end{equation}
(where we have used $x=U_{\infty}t$ and $x_{0}=U_{\infty}t_{0}$) and
(32)-(33) remain valid but with 
\begin{equation}
l(x, Re_{0},*)=l(x_{0}, Re_{0},*)[1+{4\nu a \vert c \vert \over
l^{2}(x_{0})U_{\infty}}(x-x_{0})]^{1/2}
\end{equation} 
not with a length-scale $l$ independent of $x$. Additionally, the
following estimates for the Taylor microscale and the dissipation rate
$\epsilon$ can be obtained, respectively, from ${d\over dt} u'^{2}
\sim \nu u'^{2}/\lambda^{2}$ and from an integration over $\kappa$ of
(21):
\begin{equation}
\lambda \sim {l(x_{0}, Re_{0},*)\over\sqrt{2a(1-c)} }[1+{4\nu a \vert c
\vert \over l^{2}(x_{0})U_{\infty}}(x-x_{0})]^{1/2},
\end{equation}
\begin{equation}
\epsilon = 3\nu a(Re_{0},*) \left(1+\vert c(Re_{0},*) \vert \right)
{u'^{2}\over l^{2}}.
\end{equation}
These two equations replace (34) and (29) which follow from $c=0$. It
is easily seen that the power-law form (37) tends to the exponential
form (25) and that the Taylor microscale $\lambda$ becomes
asymptotically independent of $x-x_{0}$ in the limit $c\to 0$.

The new forms (37)-(40) depend on two length-scales, $l(x_{0},
Re_{0},*)$ and $x_0$, one kinetic energy scale $E_{s}(x_{0})/l(x_{0})$
and two dimensionless numbers, $\nu a (Re_{0}, *)/ \left(l(x_{0},
Re_{0},*)U_{\infty} \right)$ and $c(Re_{0},*)$, all of which may vary with
$Re_0$ and boundary/inlet conditions. There seems to be enough
curve-fitting freedom for these forms to account for our data in the
decay regions of our fractal-generated turbulent flows, in particular
figures 5b, 6b, 9, 14, 15, 16 and 17. 
In subsection 4.8 we present a procedure for fitting (37) and (39) to
our data which is robust to much of this curve-fitting freedom. It is
worth noting here, in anticipation of this subsection, that (39) is
consistent with the observation (originally reported in
\citep{HurstVassilicos2007} and \citep{SeoudVassilicos2007}) that
$\lambda$ is approximately independent of $x$ in the decay region, but
provided that ${2\nu a \vert c \vert \over l^{2}(x_{0})U_{\infty}}
(x-x_{0})<<1 $ in much of this region. However, (39) also offers a
possibility to explain the departure from the constancy of $\lambda$
at large enough values of $x/x_{*}$ where $\lambda$ appears to grow
again with $x$ (see figures 14(a) and 17(a)), very much as (39) would
qualitatively predict. Note, in particular, that this departure occurs
at increasing values of $x/x_{*}$ for increasing $Re_0$ (see figures
14(b) and 17(b)) , something which can in principle also be accounted
for by (39). In the following subsection we show that same
observations can be made for $L_u$.


\subsection{The integral length-scale $L_{u}$}

The streamwise evolution of the longitudinal integral length-scale
$L_u$ on the centreline is plotted in figure 18 
for all space-filling fractal square grids as well as for the regular
grid \emph{SRG}. The integral length-scales generated by the fractal
square grids are much larger than the regular grid's even though their
effective mesh sizes are smaller. Comparing data from the
\emph{T-0.46m} tunnel, one can see that $L_u$ appears independent of
the thickness ratio $t_r$. However, there are large differences
between the integral scales generated by fractal grids \emph{SFG17}
and \emph{BFG17} which have the same $t_r$ but fit in different wind
tunnels. This observation suggests that the large-scale structure of
the fractal grid has a major influence on the integral
length-scale. One might in fact expect that the integral length-scale
is somehow linked to an interaction length-scale of the grid. For
regular grids this interaction length-scale is typically the mesh
size, whereas for space-filling fractal square grids a large variety
of interaction length-scales exist, the largest being $L_0$. Figure
\ref{fig:IntegralScale_Norm} supports the view that the scalings of
$L_u$ and its $x$-dependence are mostly determined by $L_0$ and $x_*$,
respectively, though not perfectly.
Figure \ref{fig:Ltr13} suggests that $L_{u}/L_{0}$ is not
significantly dependent on the inlet Reynolds number $Re_{0}$, at
least for the range of $U_{\infty}$ values investigated here. This
figure was obtained for the \emph{SFG13} grid but is representative of
our other three space-filling fractal square grids as well.

\begin{figure}[htbp]
\centering
{\includegraphics[scale=0.5]{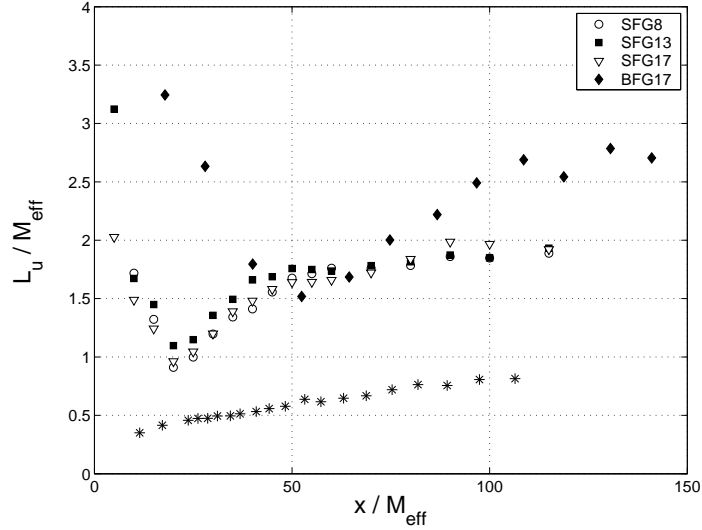}
\label{fig:IntegralScale}}
\caption{Centreline streamwise evolution of the longitudinal integral
length-scale $L_u$ normalized by the mesh size $M_{eff}$ ($U_\infty =
5.2 m/s$). Data obtained for the regular grid \emph{SRG} are also
shown ($\star$).}
\end{figure}

\begin{figure}[htbp]
\centering
\subfigure
{\includegraphics[width=7.5cm]{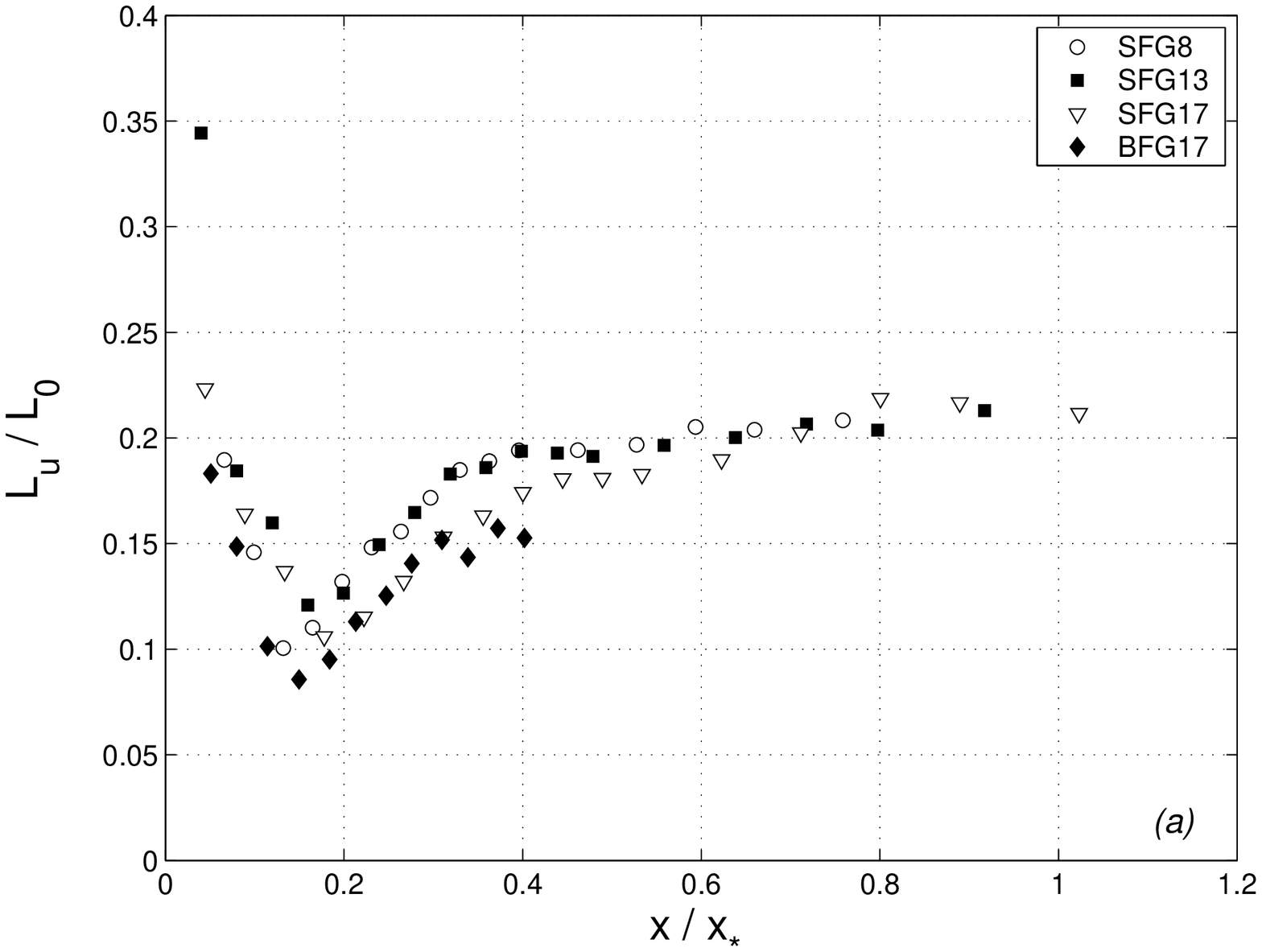}
\label{fig:IntegralScale_Norm}}
\hspace{0.1cm}
\subfigure
{\includegraphics[width=7.5cm]{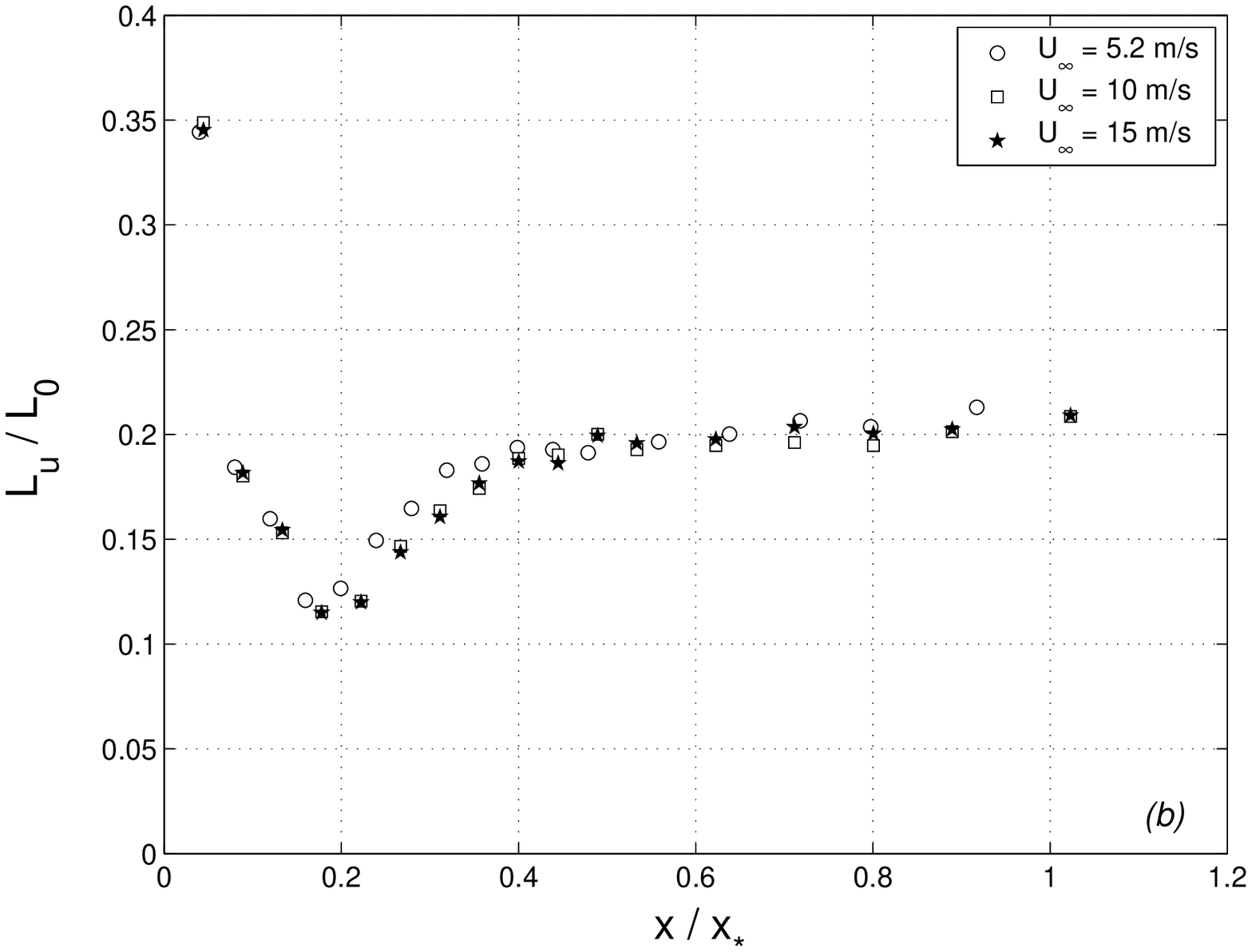}
\label{fig:Ltr13}}
\caption{$L_{u}/L_{0}$ as a function of $x/x_{*}$ along the
centreline. \textit{(a)} Four different fractal grids and $U_\infty =
5.2 m/s$. \textit{(b)} Three different inlet velocities $U_{\infty}$
with the \emph{SFG13} fractal grid. }
\end{figure}

\begin{figure}[htbp]
\centering
\subfigure
{\includegraphics[width=7.5cm]{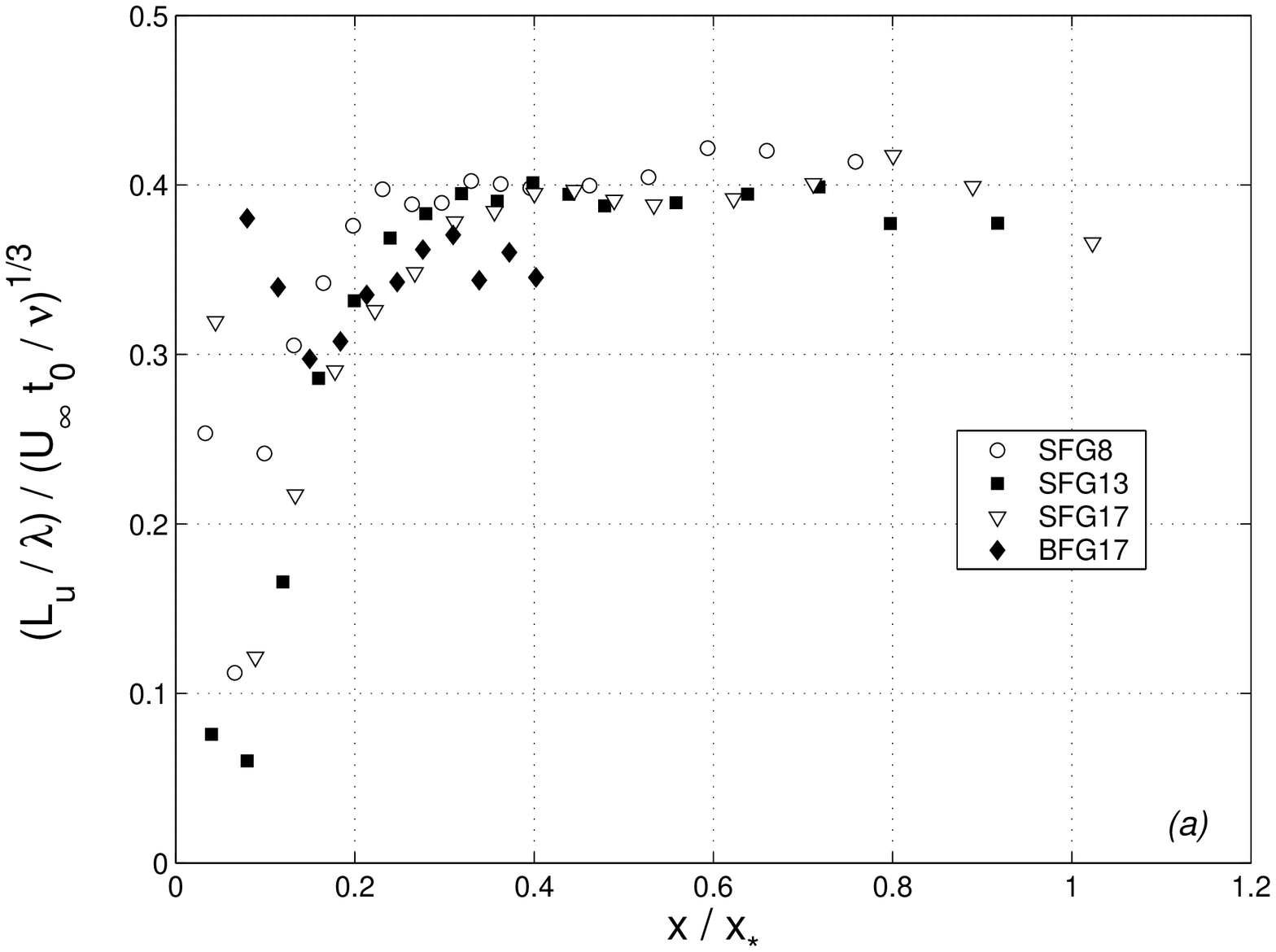}
\label{fig:LulambdaTr}}
\hspace{0.1cm}
\subfigure
{\includegraphics[width=7.5cm]{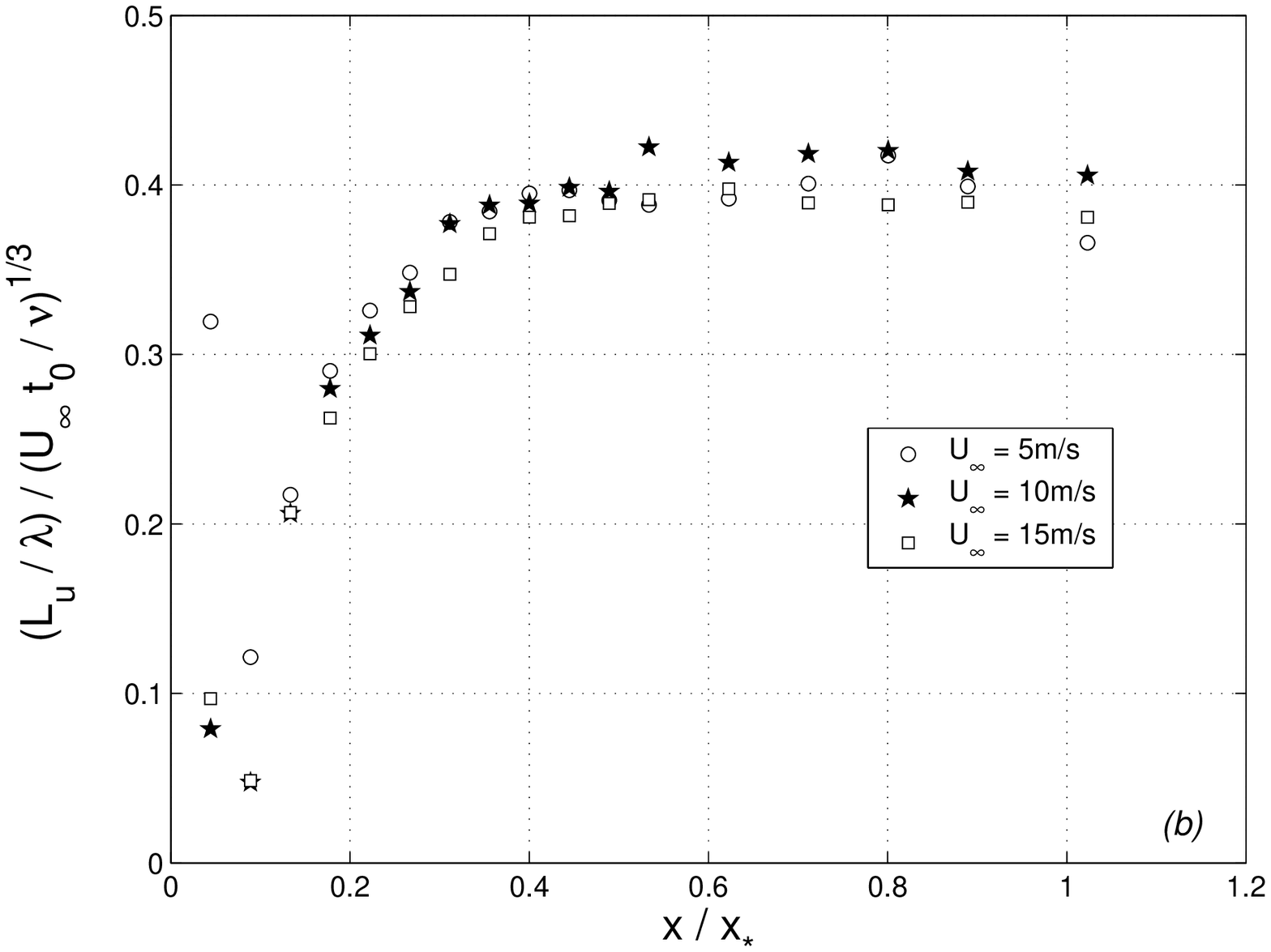}
\label{fig:LulambdaUinf}}
\caption{$\left( L_{u} / \lambda \right) / Re_{0}^{1/3}$ as a function
of $x/x_{*}$ along the centreline \textit{(a)} for the four different
fractal grids and $U_\infty = 5.2 m/s$ and \textit{(b)} for the
fractal grid SFG17 and three different inlet velocities $U_{\infty}$.}
\end{figure}

Irrespective of whether $l$ does or does not depend on streamwise
coordinate $x$, equations (32) and (33) suggest that $L_{u}/\lambda$
is definitely not a function of $x$ but that it can nevertheless be,
in all generality, a function of $Re_{0}$ and of the fractal grid's
geometry. In fact figure 16 is evidence of some dependence on
$U_{\infty}$ at least at the lower $U_{\infty}$ values. Assuming
$L_{u} \sim L_{0}$ as seems to be suggested by figures
\ref{fig:IntegralScale_Norm} and \ref{fig:Ltr13} and using either (34)
or (35) implies either $L_{u}/\lambda \sim Re_{0}^{1/2}$ or
$L_{u}/\lambda \sim Re_{0}^{1/3}$. Of these two implications, it is
the latter which agrees best with our measurements (see figures \ref
{fig:LulambdaTr} and \ref{fig:LulambdaUinf} where we plot
$(L_{u}/\lambda)/Re_{0}^{1/3}$ versus $x/x_{*}$) which is consistent
with the fact that (35) fits our data better than (34).

As reported in \citep{HurstVassilicos2007} and
\citep{SeoudVassilicos2007}, $L_u$ remains almost constant in the decay
region (see figures \ref{fig:IntegralScale_Norm} and
\ref{fig:Ltr13}). The independence on $x$ can be explained in terms of
the $c=0$ single-scale solution of George \& Wang
\citep{GeorgeWang2009}, but it might also be even better explained in
terms of the $c<0$ single-scale solution which yields (32) and (38),
i.e.
\begin{equation}
L_{u}= \alpha (Re_{0}, *) l(x_{0}, Re_{0},*)[1+{4\nu a \vert c \vert
\over l^{2}(x_{0})U_{\infty}}(x-x_{0})]^{1/2}. 
\end{equation}
A careful examination of figures \ref{fig:IntegralScale_Norm} and
\ref{fig:Ltr13} suggests that $L_{u}/ L_0$ may be constant for some of
the way downstream in the decay region till it starts increasing
slightly, very much like the behaviour of the Taylor microscale
$\lambda$ (see figures 14 and 17) and in qualitative agreement with
(41). In fact, the ratio $L_{u}/ \lambda$ is predicted by (39) and
(41) to be independent of $x$ in the decay region, which is in full
agreement with figures 16. 

The fact that $L_{u}/\lambda$ is independent of x (figure 16) in the
decay region where $Re_{\lambda}$ decreases rapidly with increasing
$x$ (figure 15b) is evidence against $\epsilon \sim u'^{3}/L_{u}$, not
only in the context of $c=0$ single-scale solutions of (12), but also
in the context of $c<0$ single-scale solutions of (12). Indeed, (37)
and (41) are consistent with $U_{\infty}{d\over dx} u'^{2} \sim -
u'^{3}/L_{u}$ only if $c=-1$, in which case $Re_{\lambda} =
u'\lambda/\nu$ is independent of $x$ because of (37) and (39) and
therefore in conflict with our experimental observations and those in
\citep{HurstVassilicos2007} and \citep{SeoudVassilicos2007}. In fact,
the downstream decreasing nature of $Re_{\lambda}$ in the decay region
of our fractal-generated turbulent flows imposes $-1< c \le 0$.

\subsection{The energy spectrum $E_{u}(k_{x})$.}

The results of subsections 4.4 to 4.6 imply that the small-scale
turbulence far downstream of low-blockage space-filling fractal square
grids is either fundamentally different from the small-scale
turbulence in documented boundary-free shear flows and decaying wind
tunnel turbulence originating from a regular/active grid, or $\epsilon
\sim u'^{3}/L_{u}$ does not hold in these non-fractal-generated flows
where the length-scale ratio $L_{u}/\lambda$ is proportional to
$Re_{\lambda}$ if one assumes $\epsilon \sim u'^{3}/L_{u}$. 
Our results therefore shed serious doubt on the universality of
$\epsilon \sim u'^{3}/L_{u}$, the cornerstone assumption present
either explicitely or implicitely in most if not all turbulence models
and theories \citep{Batchelor1953}, \citep{TennekesLumley1972},
\citep{Frisch1995}, \citep{Pope2000}, \citep{SagautCambon2008}.  

However, our data do not allow us to educe with full confidence a
formula for the dissipation rate $\epsilon$ in turbulence generated by
space-filling fractal square grids. This issue is related to the fact
that whilst an exponential turbulence decay (equations 10 and 11) fits
our data well, the $Re_{0}$ dependence of $\lambda$ which follows from
it in a self-preserving single-length scale context does
not. Qualitative observations of the x-dependence of $L_u$ and
$\lambda$ may suggest that the turbulence decay is in fact a power-law
of the type (37), rather than exponential, albeit with a power-law
exponent large enough (i.e. $c$ close enough to 0) for the exponential
form to be a good fit. An attempt at addressing this issue is made in
the following and final subsection 4.8. This attempt relies on the
results of our examination of energy spectra and the single-length
scale assumption which we now report.



\begin{figure}[htbp]
\centering
\subfigure
{\includegraphics[width=7.5cm]{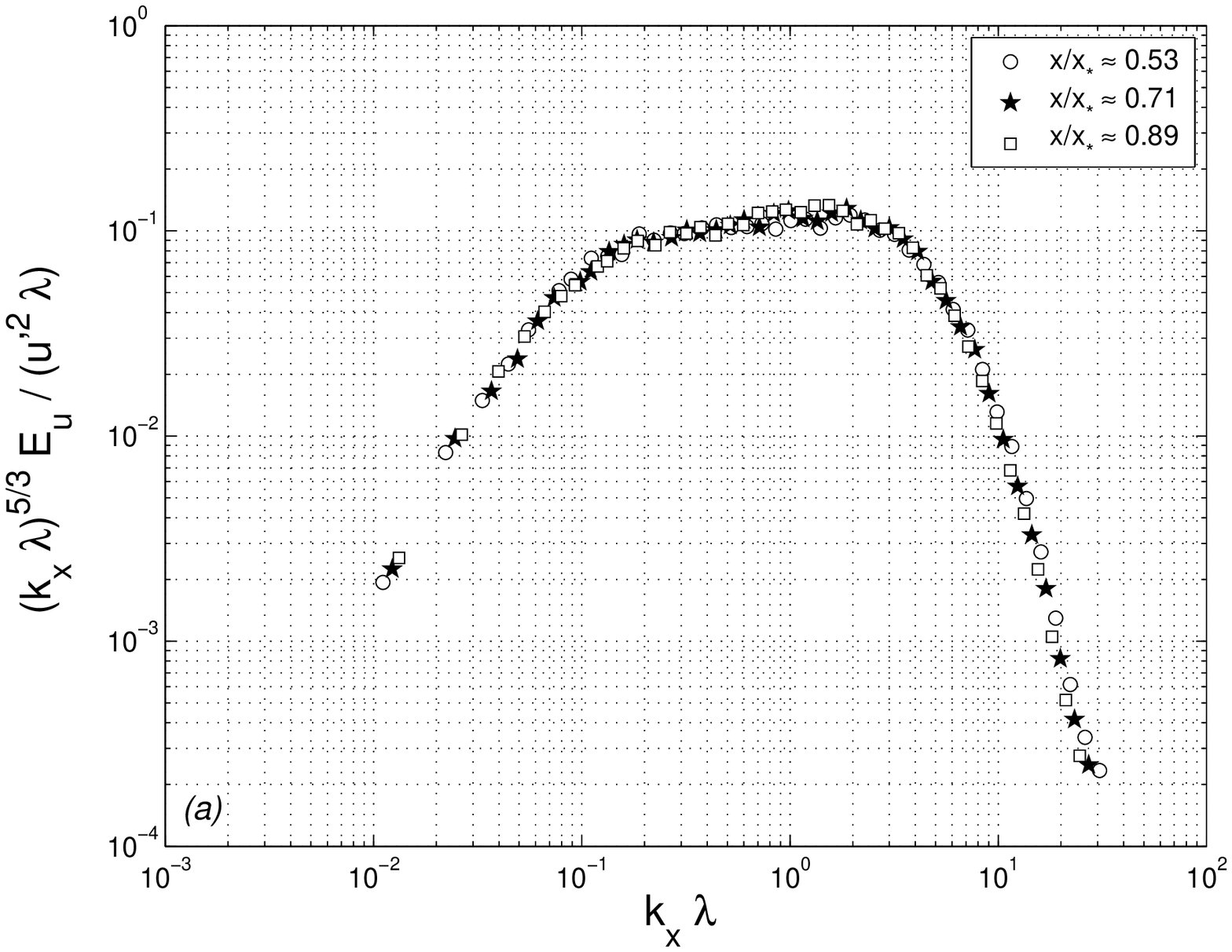}
\label{fig:Eu_x_SFG17_lambda}}
\hspace{0.1cm}
\subfigure
{\includegraphics[width=7.5cm]{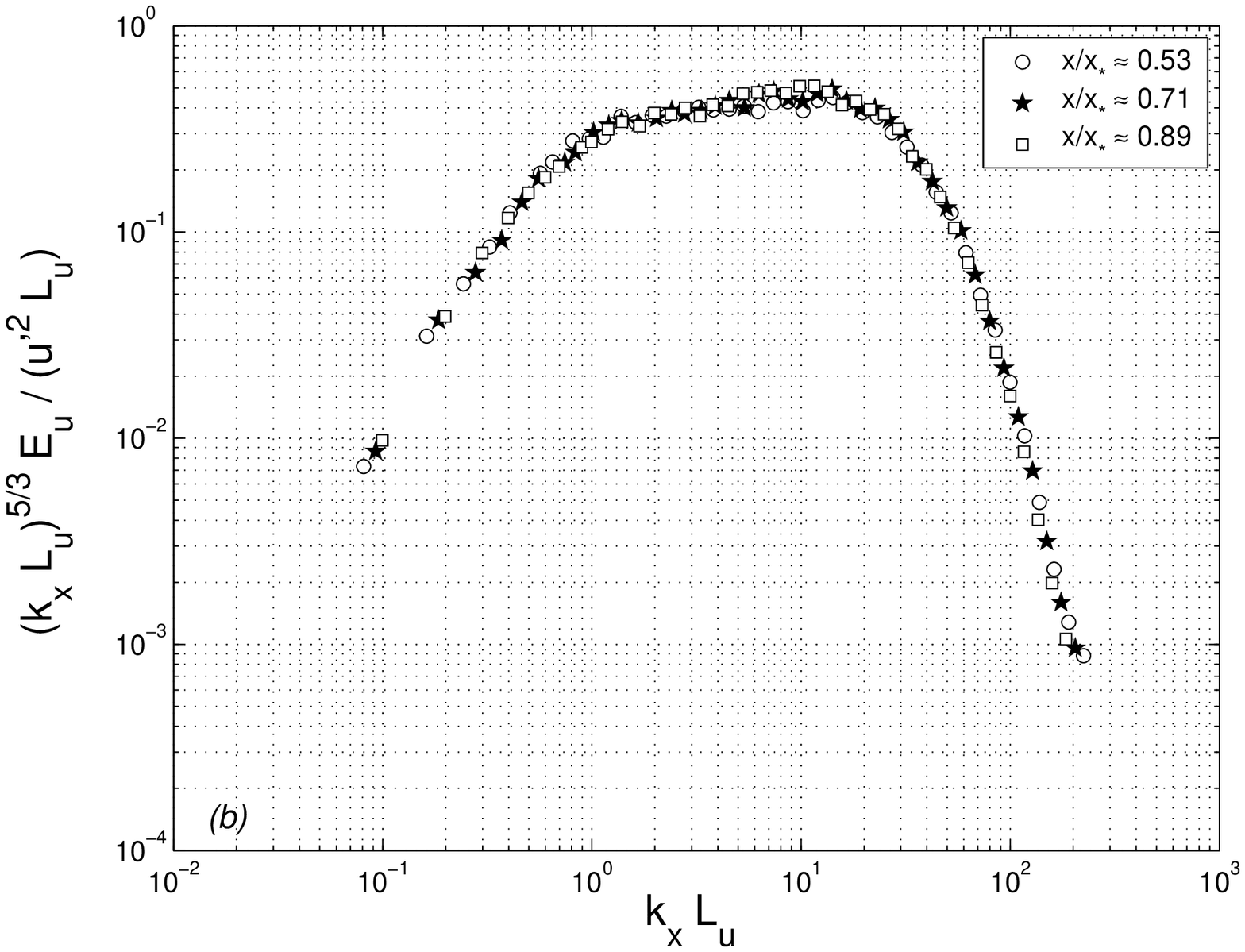}
\label{fig:Eu_x_SFG17_Lu}}
\caption{\emph{SFG17} grid in the \emph{T-0.46} tunnel with
$U_{\infty} = 5.2m/s$ along the centreline. Compensated 1D energy
spectra normalized using \textit{(a)} $u'^{2}$ and $\lambda$ or
\textit{(b)} $u'^{2}$ and $L_u$. Spectra from different centreline
positions $x/x_*$ collapse over all wavenumbers. }
\end{figure}
\begin{figure}[htbp]
\centering
\subfigure
{\includegraphics[width=7.5cm]{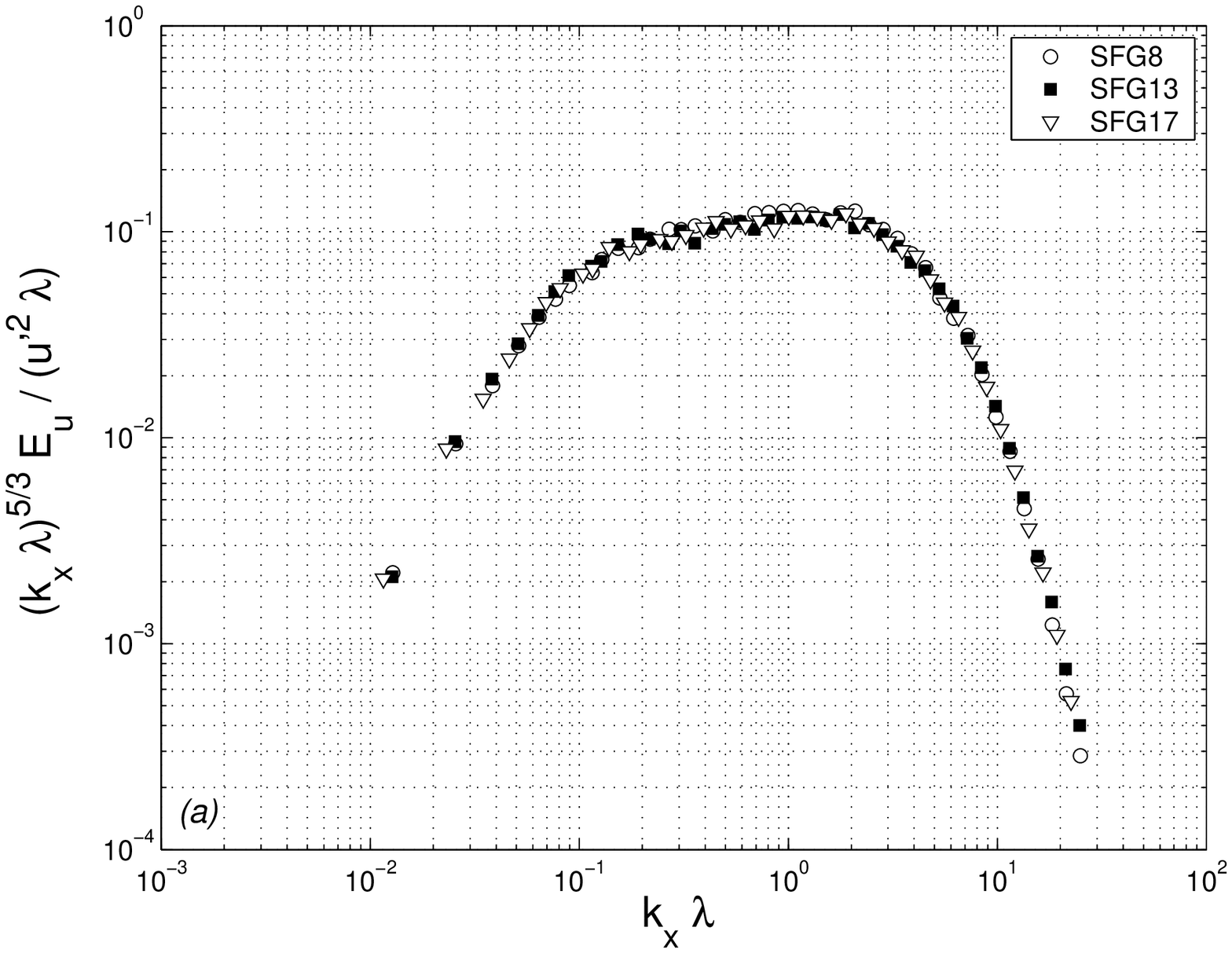}
\label{fig:Eu_tr_lambda}}
\hspace{0.1cm}
\subfigure
{\includegraphics[width=7.5cm]{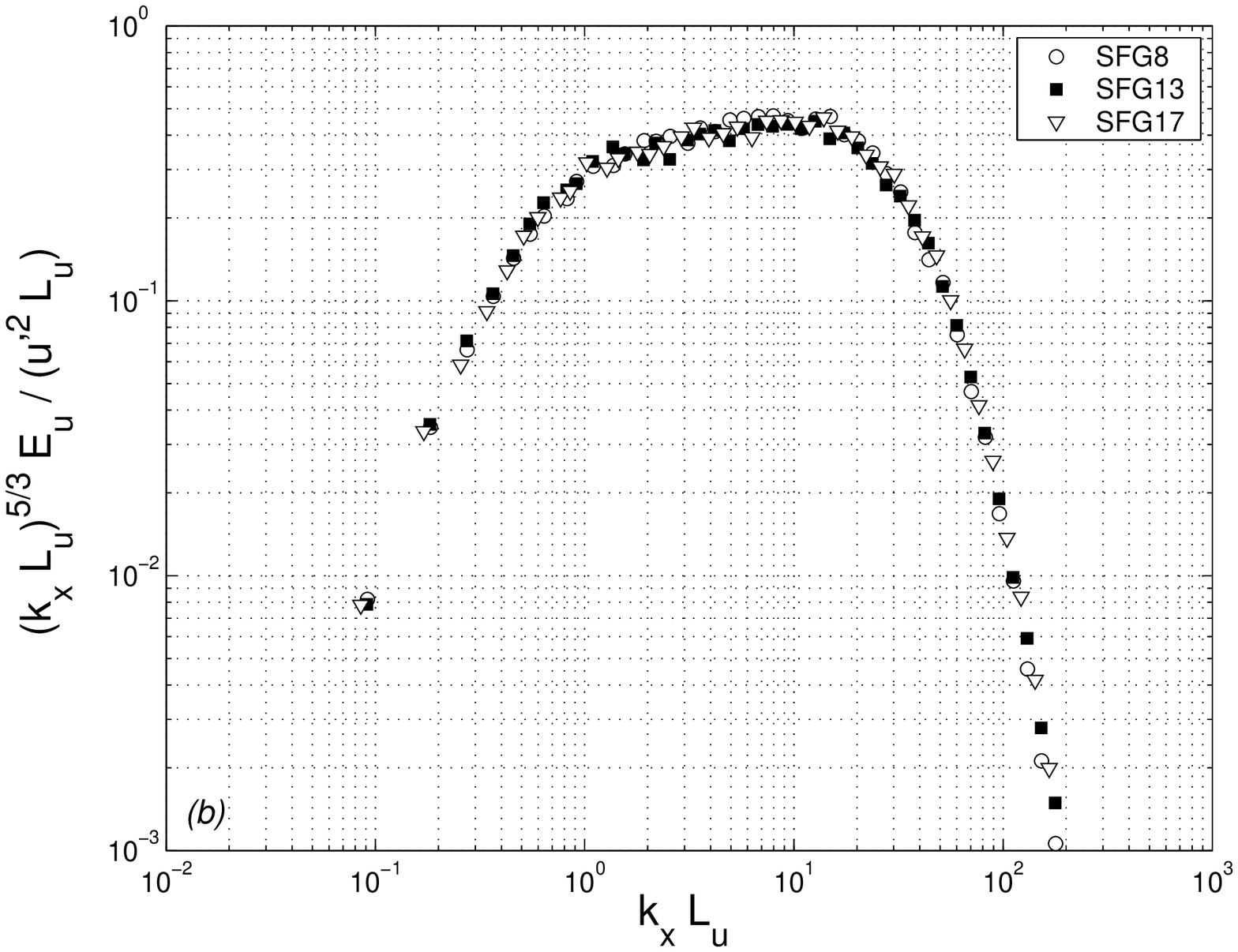}
\label{fig:Eu_tr_Lu}}
\caption{\emph{T-0.46} tunnel with $U_{\infty} = 5.2m/s$ along the
centreline. Compensated 1D energy spectra measured at $x/x_{*} = 0.62$
and normalized using \textit{(a)} $u'^{2}$ and $\lambda$ or
\textit{(b)} $u'^{2}$ and $L_u$. Spectra corresponding to our
different grids appear to collapse over all wavenumbers.}
\end{figure}

Seoud \& Vassilicos \citep{SeoudVassilicos2007} studied the downstream
evolution of the 1D energy spectrum $E_{u} (k_{x},x)$ in the decay
region of space-filling fractal square grids and found that, for a
given velocity $U_{\infty}$, $E_{u} (k_{x},x)$ can be collapsed for
different downstream positions and for all our fractal grids in terms
of (31) where $l$ is replaced by either $\lambda$ or $L_u$. Indeed,
$E_u(k_x) = u'^2 \lambda F_{u} (k_x \lambda )$ and $E_u(k_x) = u'^2
L_u F_{u} (k_x L_u)$ collapse the entire spectral data equally well
at a given inlet velocity $U_{\infty}$, a fact which we confirm in
figures \ref{fig:Eu_x_SFG17_lambda}, \ref{fig:Eu_x_SFG17_Lu},
\ref{fig:Eu_tr_lambda} and \ref{fig:Eu_tr_Lu}. These figures clearly
support George \& Wang's \citep{GeorgeWang2009} single-length scale
assumption (16) and (31).

\begin{figure}[htbp]
\centering
\subfigure
{\includegraphics[width=7.5cm]{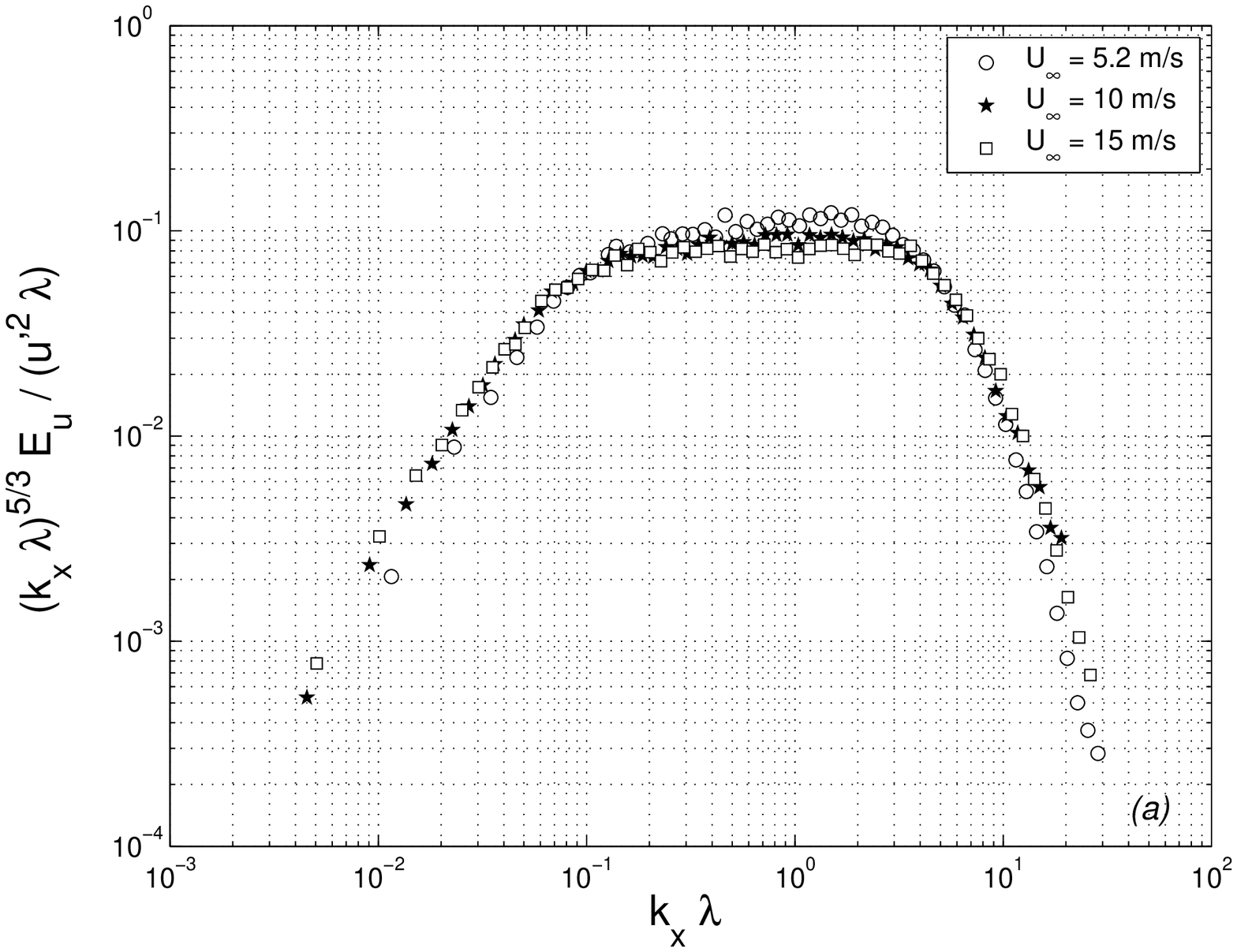}
\label{fig:Spec_SFG17_Uinlet_NormLambda}}
\hspace{0.1cm}
\subfigure
{\includegraphics[width=7.5cm]{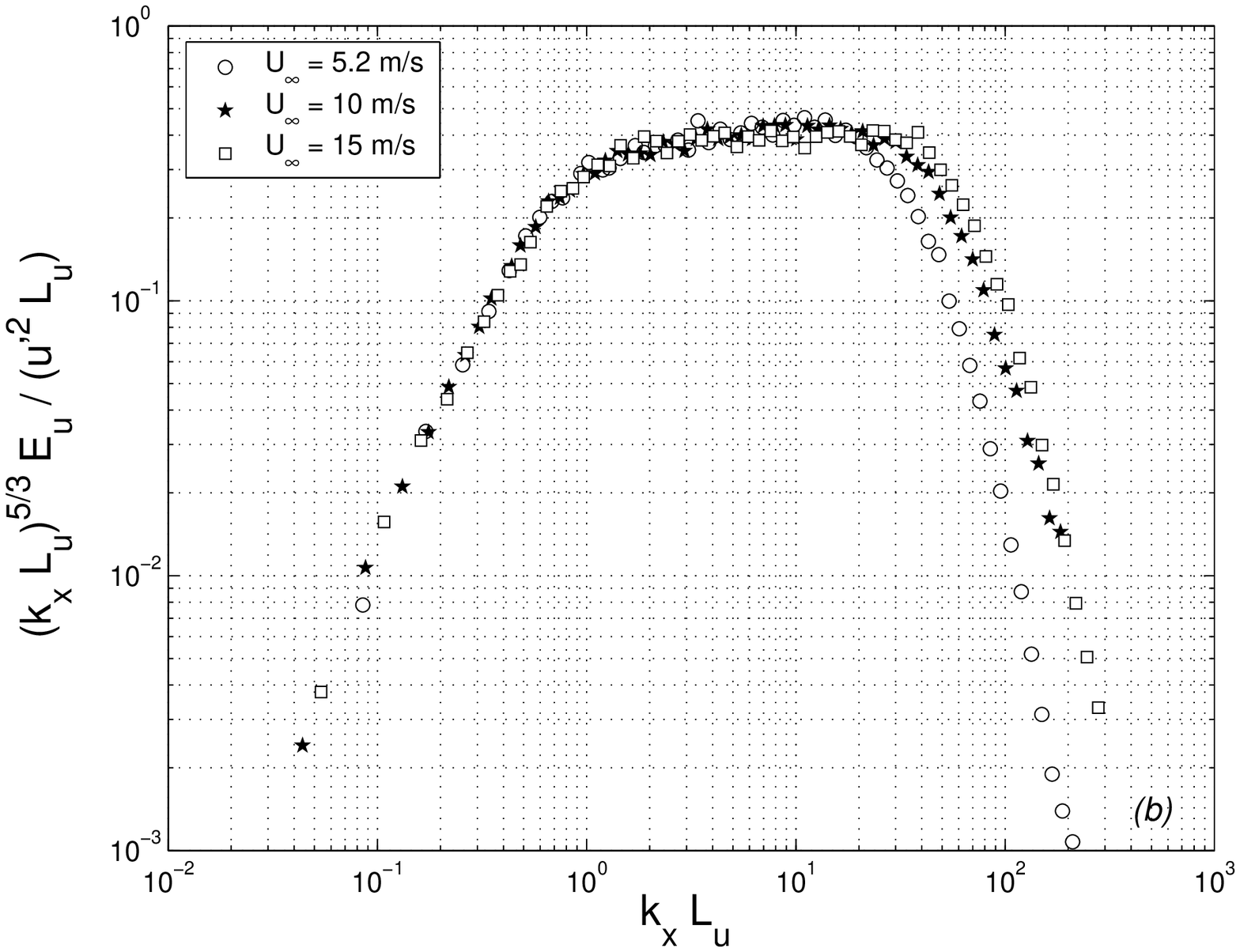}
\label{fig:Spec_SFG17_Uinlet_NormLi}}
\caption{\emph{SFG17} in the \emph{T-0.46} tunnel.  \textit{(a)}
Compensated 1D energy spectra normalized by the Taylor micro-scale
$\lambda$.  \textit{(b)} Compensated 1D energy spectra normalized by
the large-scale properties measured on centreline downstream fractal
grid \emph{SFG17} for different inlet velocities $U_\infty$ ($x /
x_\star \approx 0.62$ on the centreline).}
\end{figure}

However, Seoud \& Vassilicos \citep{SeoudVassilicos2007} did not
attempt to collapse energy spectra for different inlet velocities
$U_{\infty}$. We therefore compare energy spectra obtained at the
same position downstream of the same fractal grid but with different
inlet velocities $U_{\infty}$. In figures
\ref{fig:Spec_SFG17_Uinlet_NormLambda} and
\ref{fig:Spec_SFG17_Uinlet_NormLi} we report such results obtained at
$x/x_\star \approx 0.62$ with the \emph{SFG17} grid. These figures are
representative of all other such results which we obtained with our
other fractal grids and at other positions but which we do not present
here for economy of space.

From these figures, the form $E_u(k_x) = u'^2 \lambda F_{u} (k_x
\lambda )$ may seem to offer a much better collapse for different inlet
velocities $U_{\infty}$ than $E_u(k_x) = u'^2 L_u F_{u} (k_x L_u)$.
The discrepancy of the form $E_u(k_x) = u'^2 L_u F_{u} (k_x L_u)$ is
mostly at the high wavenumbers and is made evident by our compensation
of the spectra by $(k_x L_u)^{5/3}$. It might be tempting to conclude
that $l$ is different from $L_u$ and in fact equal to $\lambda$, but
such a conclusion would be incompatible with $E_u(k_x) = u'^2 \lambda
F_{u} (k_x \lambda )$ because (32) and (33) would then lead to the
inconsistency that $l$ is in fact not different from $L_{u}$. The fact
that $L_{u}/\lambda$ grows with $Re_0$, at least for the range of
$Re_0$ values considered here, suggests that we should be considering
a spectral form $E_u(k_x) = u'^2 l F_{u} (k_x l, Re_{0},*)$ without
neglecting the dependencies on $Re_0$ and perhaps even $*$. 

We have seen in the previous subsection that much of the scaling of
$L_u$ is controlled by $L_0$, i.e. $L_{u}\sim L_{0}$ to a first
scaling approximation, and that $L_u$ does not vary significantly with
$U_{\infty}$. Figure \ref{fig:Spec_SFG17_Uinlet_NormLi} suggests that
the plot of $E_u(k_x) / \left(u'^2 L_u \right) $ versus $k_x L_u$ is
also imperceptibly dependent on $U_{\infty}$ at the lower values of
$k_x L_u$ but not at the higher ones. These three observations can all
be explained if the assumption is made that
\begin{equation}
l(x_{0}, Re_{0}, *) = L_0
\end{equation}
and that 
\begin{equation}
F_{u} (k_x l, Re_{0},*) = f_{u} (k_x l) H_{u} (k_x l Re_{0}^{-n})
\end{equation}
where $n>0$ and $H_u$ is a monotonically decreasing function which is
very close to 1 where $k_x l Re_{0}^{-n} < 1$ and very close to 0
where $k_x l Re_{0}^{-n} >1$. There may be residual dependencies on
the geometry of the fractal grid, i.e. on $*$, but we do not have enough
fractal grids in our disposal to determine them. Once again, this is
an issue for future study. 

Equations (31), (38), (42) and (43) can readily account for the
behaviour observed in figure
\ref{fig:Spec_SFG17_Uinlet_NormLi}. Combined with (32), (42) and (43)
also imply that $L_{u}$ scales with $L_0$ provided that $f_u$ is a
decreasing function of $k_x l$ where $k_x l >1 $. Figures \ref
{fig:LulambdaTr} and \ref{fig:LulambdaUinf} would then suggest that
$\beta$ in (33) scales as $Re_{0}^{-1/3}$. It is the function $H_{u}
(k_x l Re_{0}^{-n})$ in (43) which makes this scaling possible. In
fact, if $f_{u} \sim (k_x l)^{-p}$ where $k_x l >1$, then (33), (38)
and (42) imply $\lambda \sim L_{0} Re_{0}^{-n(3-p)/2}$. Note that the
Kolmogorov-like exponents $p=5/3$ and $n=3/4$ (see \citep{Pope2000},
\citep{TennekesLumley1972}, \citep{Frisch1995}) yield $\lambda \sim
L_{0} Re_{0}^{-1/2}$ identically to (34) which follows from the $c=0$
single-length scale solution of the spectral energy equation (12).

The good collapse in terms of both forms $E_u(k_x) = u'^2 \lambda
F_{u} (k_x \lambda )$ and $E_u(k_x) = u'^2 L_{u} F_{u} (k_x L_{u} )$
in figures \ref{fig:Eu_x_SFG17_lambda} and \ref{fig:Eu_x_SFG17_Lu}
comes from the fact that all data in these figures are obtained for
the same value of $Re_{0}$ and the same fractal grid, and that
$L_{u}/\lambda$ does not vary with $x$. These figures are therefore
also consistent with (42) and (43). The good collapse of the form
$E_u(k_x) = u'^2 L_{u} F_{u} (k_x L_{u} )$ in figure
\ref{fig:Eu_tr_Lu} is mainly a consequence of the fact that the
fractal grids SFG8, SFG13 and SFG17 all have the same value of $L_0$
and can also follow from (42) and (43). However, the apparently good
collapse of the form $E_u(k_x) = u'^2 \lambda F_{u} (k_x \lambda )$ in
figure \ref{fig:Eu_tr_lambda} must be interpreted as being an artifact
of the limited range of values of thicknesses $t_0$ that we have
experimented with (see table 1), more limited than the range of inlet
velocities $U_{\infty}$ which allows the $Re_{0}^{1/3}$ scaling of
$L_{u}/\lambda$ to be picked up by our spectra in figure
\ref{fig:Spec_SFG17_Uinlet_NormLi} but not in figure
\ref{fig:Eu_tr_Lu}.

Returning to figure \ref{fig:Spec_SFG17_Uinlet_NormLambda} we notice
that it does not, in fact, present such a good collapse of the data,
particularly over the range of scales where the collapse in figure
\ref{fig:Spec_SFG17_Uinlet_NormLi} appears good. Within the framework
of (42) and (43), the semblance of a perhaps acceptable collapse in
figure \ref{fig:Spec_SFG17_Uinlet_NormLambda} results from a numerical
circumstance to do with the exponents $n$ and $p$. Chosing $p=5/3$ and
$n=3/4$ for the sake of argument, (31),(38), (42) and (43) would imply
that the quantity plotted in this figure, i.e. $(k_{x}\lambda )^{5/3}
E_{u}(k_{x})/ \left( u'^{2} \lambda \right)$, is in fact equal to
$Re_{0}^{-1/3} H_{u} (k_{x}\lambda Re_{0}^{-1/9} )$ in the range which
would correspond to $k_{x} \lambda \ge 0.1$ in the figure. Over the
range of inlet velocities tried here, $Re_{0}^{-1/9}$ remains about
constant whilst $Re_{0}^{-1/3}$ varies a bit thus producing the effect
seen in figure \ref{fig:Spec_SFG17_Uinlet_NormLambda}: a slight
dependence on $U_{\infty}$ of the plateau and a semblance of a
collapse of the dissipative range of the spectra.

The conclusion of this data analysis is that the self-preserving
spectral form
\begin{equation}
E_u(k_x) = u'^2 l f_{u}(k_x l) H_{u} (k_x l Re_{0}^{-n})
\end{equation}
with 
\begin{equation}
l=L_{0} [1+{4\nu a \vert c \vert \over
L_{0}^{2} U_{\infty}}(x-x_{0})]^{1/2}
\end{equation}
is consistent with the theory of George \& Wang \citep{GeorgeWang2009}
and with our measurements in the decay region in the lee of our
fractal square grids.
We must stress again that future work is required with a wider range
of fractal grids, measurement positions and inlet velocities in order
to reach definitive conclusions confidently valid over a wider range
of parameters.

\subsection{Exponential versus power-law turbulence decay.}

We close section 4 with a discussion of how the exponential turbulence
decay (25) corresponding to the $c=0$ solution and the power-law
turbulence decay (37)-(38)-(39) corresponding to $-1<c<0$ solutions
might fit together into a single framework. We already commented
straight under equation (40) that the power-law form (37) tends to the
exponential form (25) and that the Taylor microscale $\lambda$ becomes
asymptotically independent of $x-x_{0}$ in the limit $c\to 0$. We now
attempt to fit expression (37) to our decay region data and compare it
with our exponential fit (10) (which is consistent with (25) and
(26)). To do this we start by fitting (39) to our Taylor microscale
decay region data using some of the results reached in the previous
subsection for our range of experimental parameters, namely $\beta
\sim Re_{0}^{-1/3}$ and $l(x_{0})=L_{0}$. We therefore reformulate
(39) as follows:
\begin{equation}
\frac{\lambda^2}{L_0^2 Re_0^{-2/3}} = 1 + \gamma \left(\frac{x - x_0}{x_\star}\right)
\end{equation}
where 
\begin{equation}
\gamma = \frac{4 a \left|c\right|}{Re_{0}}.
\end{equation}
We have, in effect, arbitrarily set to 1 a $Re_{0}$-independent
dimensionless parameter (or, if $\beta \sim Re_{0}^{-1/3}$ is somewhat
faulty, perhaps even a weakly $Re_0$-dependent parameter) multiplying
the left hand side of (46). However, most of the potential
$Re_{0}$-dependence remains intact in this relation, in particular as
$\gamma$ is a priori $Re_0$-dependent.

%
%
%
%

\begin{figure}[h]
\centering
\subfigure
{\includegraphics[scale=0.35]{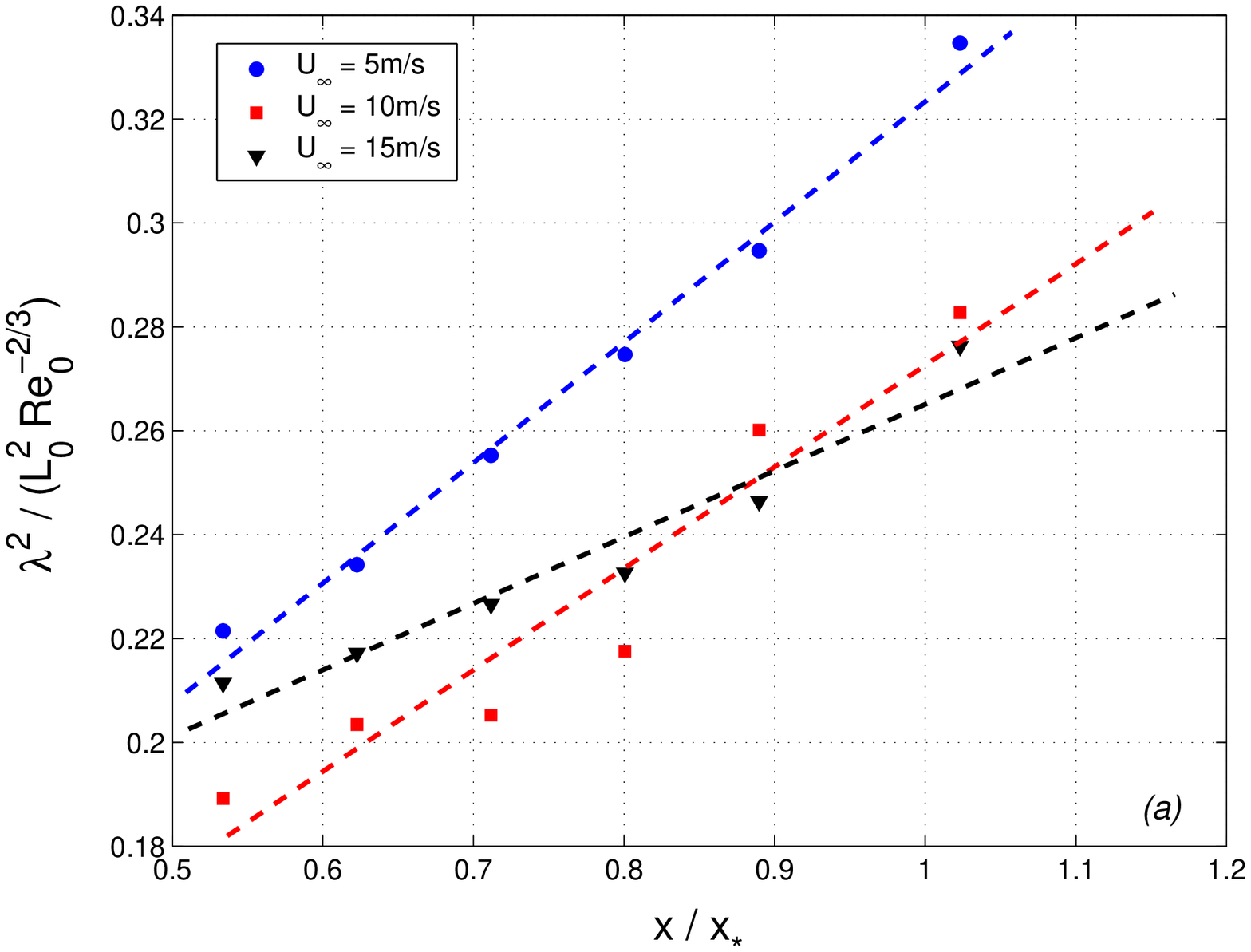}
\label{fig:lambda}}
\subfigure
{\includegraphics[scale=0.35]{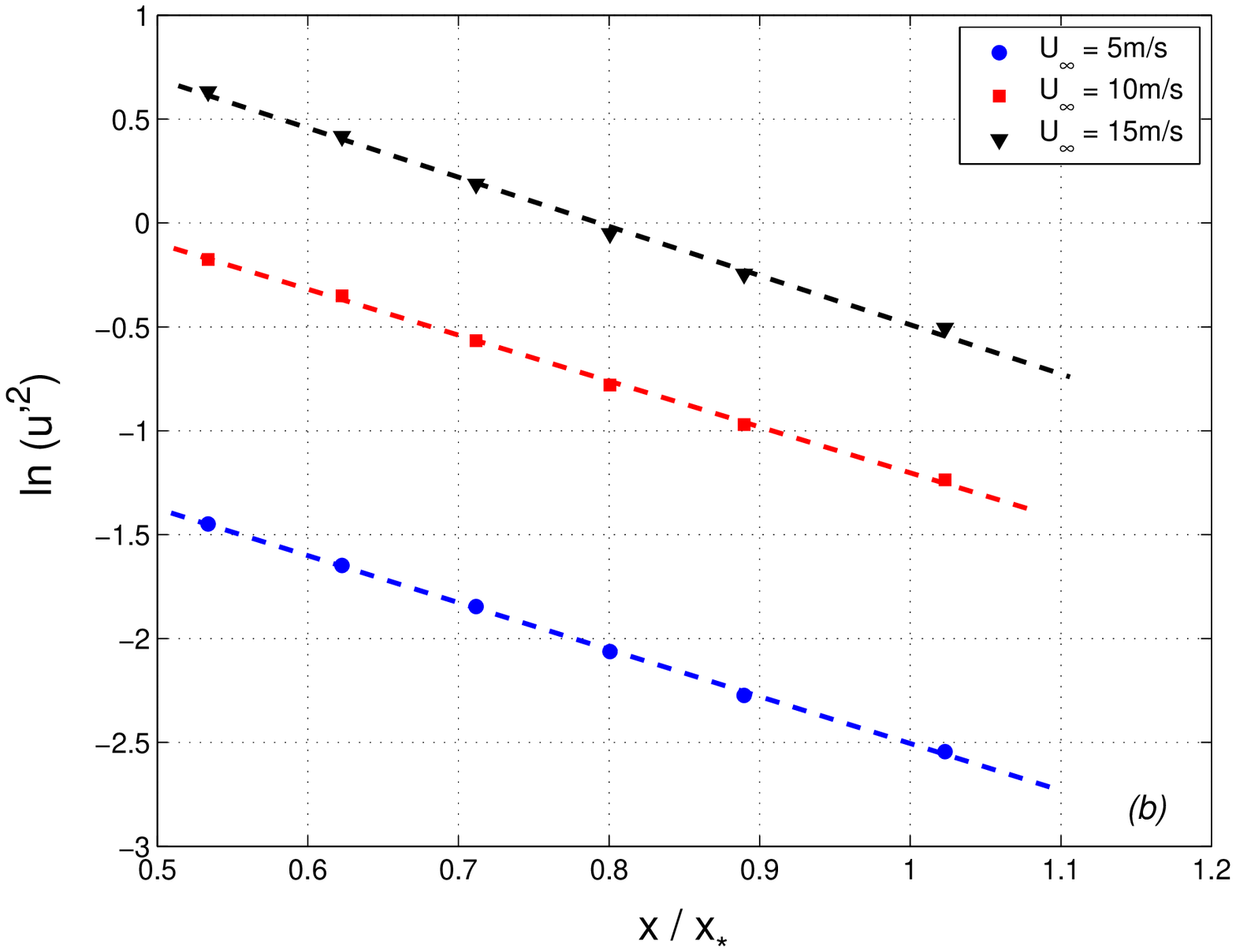}
\label{fig:vrms}}
\caption{\emph{SFG17} grid in the \emph{T-0.46m} tunnel with three
different inlet velocities. \textit{(a)} Fit (46) of centreline
$\lambda$ data.  \textit{(b)} Fit (50) of centreline $u'^2$ data.}
\end{figure}

Relation (46) is plotted in Fig. \ref{fig:lambda} for the SFG17 grid
and for all our inlet velocities $U_\infty$. From these curves it is
easy to estimate $\gamma$ independently of the virtual origin $x_0$,
and we report our results in table 5. 

\begin{table}[htbp]
\begin{center}
\begin{tabular}{c|c|c|c}
\hline
$U_\infty [m/s]$ &	$5.2 m/s$	&	$10 m/s$ & $15 m/s$\\
\hline
\hline
$\gamma$	&	$0.23$	&	$0.2$ & $0.13$\\
\hline
\end{tabular}
\caption{Estimates of the coefficient $\gamma$ for the grid SFG17.}
\label{tab:gamma}
\end{center}
\end{table}

We can now attempt to fit (37) to our turbulence decay data. Equation
(37) can be recast in the form
\begin{equation}
\ln{\left(u'^2\right)} = \ln{\left(\frac{2 E_s(x_0)}{3 l(x_0)}\right)}
+ \frac{1-c}{2c} \ln{\left[1 + \frac{4 \nu a \left|c\right|}{l(x_0)^2
U_\infty}\left(x - x_0\right)\right]}.
\end{equation}
The observed near-constancies of $\lambda$ and $L_u$ in the decay
region suggest from (39) and (41) that $\frac{4 \nu
a\left|c\right|}{l(x_0)^2 U_\infty}\left(x - x_0\right) << 1$.
It is therefore reasonable to consider the first order
approximation of (48), which is 
\begin{equation}
\ln{\left(u'^2\right)} \approx \ln{\left(\frac{2 E_s(x_0)}{3
l(x_0)}\right)} + \left(\frac{1-c}{2c}\right) \left[ \frac{4 \nu a
\left|c\right|}{l(x_0)^2 U_\infty}\left(x - x_0\right)\right]
\end{equation}
and which can be reformulated as
\begin{equation}
\ln{\left(u'^2\right)} \approx \ln{\left(\frac{2 E_s(x_0)}{3 l(x_0)}\right)} + \delta \left(\frac{x - x_0}{x_\star}\right)
\end{equation}
with $\delta = \left(\frac{1-c}{2c}\right) \gamma$. This linear
formula makes it easy to determine $\delta$ from our experimental data
independently of $E_s(x_0)/l(x_{0})$ and $x_0$, as indeed shown in
figure \ref{fig:vrms} where (50) actually appears to fit our data well
for all inlet velocities $U_{\infty}$. Our resulting best estimates of
the dimensionless parameter $\delta$ are reported in table 6. This
parameter appears to be $Re_0$-independent, in agreement with the
$Re_0$-independence of the turbulence intensity reported in figure
\ref{fig:TurbInt_Uinlet}.

\begin{table}[htbp]
\begin{center}
\begin{tabular}{c|c|c|c}
\hline
$U_\infty [m/s]$ &	$5.2 m/s$	&	$10 m/s$ & $15 m/s$\\
\hline
\hline
$\delta$	&	$-2.3$	&	$-2.2$ & $-2.4$\\
\hline
\end{tabular}
\caption{Estimates of the coefficient $\delta$ for the grid SFG17.}
\label{tab:delta}
\end{center}
\end{table}

The dimensionless coefficients $a$ and $c$ can now be obtained from
our estimates of $\gamma$ and $\delta$ using $c= \frac{\gamma}{2
\delta + \gamma}$ and
\begin{equation}
a = {Re_{0}\over 2} \vert \delta +\gamma/2 \vert .
\end{equation}
In table \ref{tab:results} we list the values thus obtained for $c$
and $a$. It is rewarding to see that $c$ turns out to be negative and
in fact larger than $-1$. Of particular interest is the finding that
$c \to 0$ with increasing $Re_0$ and that the values of $c$ are indeed
quite close to $0$ for all our inlet velocities. These results suggest
that the single-length scale power-law turbulence decay (37) tends
towards the exponential turbulence decay (25) with the dimensionless
coefficient $a$ given by (51). Equation (51) is in fact equivalent to
equation (26) which we obtained by fitting our turbulence intensity
data with an exponential decay form. The Taylor microscale $\lambda$
also tends to an $x$-independent form with increasing $Re_0$ because
$\gamma \to 0$, and so does
\begin{equation}
L_{u}=\alpha L_{0}\left[ 1 + \gamma (x-x_{0})/x_{*} \right]^{1/2} 
\end{equation}
(obtained from (41) and (42)). Indeed, we have checked that, in the
decay regions of our fractal-generated turbulent flows, (52) provides
a good fit of our $L_{u}$ data with the same values of $\gamma$ as the
ones listed in table 5 and with a dimensionless constant ($\alpha
\approx 0.34$ in the case of the \emph{SFG17} grid) for all our inlet
velocities $U_{\infty}$.

The dissipation rate $\epsilon$ is given by (40) in the context of the
power-law decaying single-length scale turbulence and it is easy to
check that (40) tends to (29), the dissipation rate form of the $c=0$
exponentially decaying single-length scale turbulence, as $Re_0$
increases. Of course, this assumes that $\gamma$ and $c$ tend to $0$
in that limit as the extrapolation of our fits would suggest.
Equation (36) is incompatible with the view that power-law decaying
single-length scale turbulence tends towards exponentially decaying
single-length scale turbulence in the limit $Re_{0}\to \infty$. 

Similarly, the empirical scaling of equation (35), i.e. $\lambda \sim
L_{0} Re_{0}^{-1/3}$, is also incompatible with such a gradual
asymptotic behaviour. If use is made of (51), or equivalently (26),
equation (39) shows that, as $Re_0$ grows, $\lambda$ tends towards
$\lambda \sim L_{0}Re_{0}^{-1/2}$, the form predicted by the
exponentially decaying single-length scale solution (see equation (34)
and the argument leading to it).

We noted in the previous subsection that an energy spectrum with a
power-law intermediate range, i.e. $f_{u} \sim (k_x l)^{-p}$ where
$k_x l >1$, and a spectral form (31) with (43), (38) and (42) implies
$\lambda \sim L_{0} Re_{0}^{-n(3-p)/2}$. We also noted that the
Kolmogorov-like exponents $n=3/4$ and $p=5/3$ yield $\lambda \sim
L_{0} Re_{0}^{-1/2}$. We are now suggesting that fits of the exponent
$n(3-p)/2$ might tend to $1/2$ as $Re_0$ increases. This seems
consistent with our observation that fits of the intermediate form
$f_{u} \sim (k_x l)^{-p}$ to our spectral data lead to $p=1.50$ for
$U_{\infty}=5 m/s$, $p=1.57$ for $U_{\infty}=10 m/s$ and $p=1.60$ for
$U_{\infty}=15 m/s$ (see figure 25). The exponent $p$ might
indeed be tending towards $5/3$ with increasing $Re_0$, in which case
we might also expect the exponent $n$ to tend towards $3/4$ if
$n(3-p)/2$ tends to $1/2$.


%



\begin{table}[htbp]
\begin{center}
\begin{tabular}{c|c|c|c}
\hline
$U_\infty [m/s]$ &	$5.2 m/s$	&	$10 m/s$ & $15 m/s$\\
\hline
\hline
$c$	&	$-0.053$	&	$-0.048$ & $-0.028$\\
\hline
$a$	&	$20.6$	&	$24.6$ & $31.3$\\
\hline
\end{tabular}
\caption{Estimates of the coefficients $a$ and $c$ for the grid SFG17.}
\label{tab:results}
\end{center}
\end{table}


\begin{figure}[htbp]
\centering
{\includegraphics[scale=0.5]{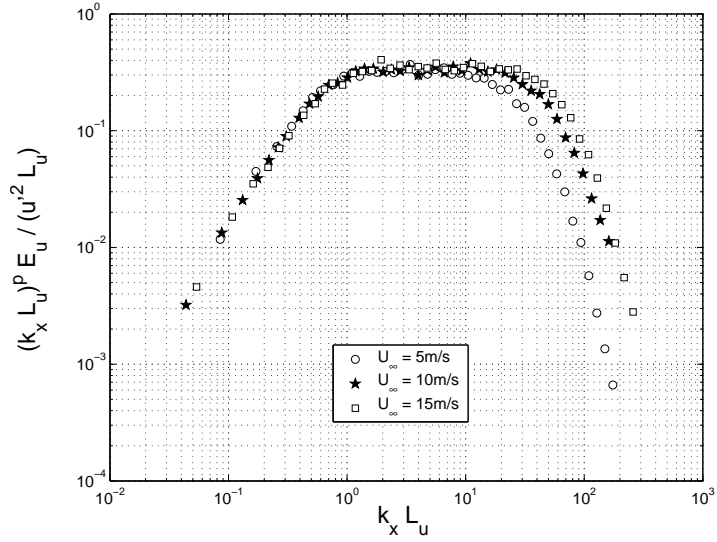}
\label{fig:Lu}}
\caption{ $(k_{x} L_{u})^{p} E_{u}(k_{x})/ \left( u'^{2} L_{u}\right)$
versus $k_{x}L_u$ for the value of the exponent $p$ which gives the
most horizontal plateau. Data obtained with the \emph{SFG17} grid at
$x/x_{*}=0.62$ on the centreline of the \emph{T-0.46m} tunnel with
three different inlet velocities. The exponent $p$ seems to increase
towards $5/3$ with increasing $U_{\infty}$ ($p=1.50$ for $U_{\infty} =
5.2m/s$, $p=1.57$ for $U_{\infty} = 10m/s$, $p=1.60$ for $U_{\infty} =
15m/s$). }
\end{figure}


\section{Conclusions and issues raised}

There are two regions in turbulent flows generated by the low blockage
space-filling fractal square grids experimented with here. The
production region between the grid and a distance about $0.5x_{*}$
from it and the decay region beyond $0.5x_{*}$. In the
production/decay region the centreline turbulence intensity
increases/decreases in the downstream direction. The wake-interaction
length-scale $x_{*}$ is determined by the large scale features of our
fractal grids, $x_{*}= L_{0}^{2}/t_{0}$, but it must be kept in mind
that one cannot change $t_0$ and or $L_0$ without changing the rest of
the fractal structure of these fractal grids. The downstream evolution
of turbulence statistics scales on $x_*$ and can be collapsed with it
for all our grids. However, it must be stressed that we have tested
only four fractal grids from a rather restricted class of
multiscale/fractal grids and we caution against careless
extrapolations of the role of this wake-interaction length-scale to
other fractal grids. For example, Hurst \& Vassilicos
\citep{HurstVassilicos2007} experimented with a low $t_r$ space-filling
fractal square grid which seemed to produce two consecutive peaks of
turbulence intensity instead of one downstream of it. A wider range of
wake-interaction length scales should probably be taken into account
for such a fractal grid, an issue which needs to be addressed in
future work on fractal-generated turbulence.

Whilst the turbulence in the production region is very inhomogeneous
with non-gaussian fluctuating velocities, it becomes quite homogeneous
with approximately gaussian fluctuating velocities in the decay
region. Unlike turbulence decay in boundary-free shear flows and
regular grid-generated wind tunnel turbulence where $L_{u}/\lambda$
and $Re_{\lambda}$ change together so that their ratio remains
constant, in the decay region of our fractal-generated turbulent flows
$L_{u}/\lambda$ remains constant and $Re_{\lambda}$ decreases as the
turbulence decays. This very unusual behaviour implies that
$L_{u}/\lambda \sim Re_{\lambda}$ and the Richardson-Kolmogorov
cascade are not universal to all boundary-free weakly sheared/strained
turbulence. In turn, this implies that $\epsilon \sim u'^{3}/L_{u}$ is
also not universally valid, not even in homogeneous turbulence as our
fractal-generated turbulence is approximately homogeneous in the decay
region. Inlet/boundary conditions seem to have an impact on the
relation between $L_{u}/\lambda$ and Reynolds number. The issue which
is now raised for future studies is to determine what it is in the
nature of inlet conditions and turbulence generation that controls the
relation between the range of excited turbulence scales and the levels
of turbulence kinetic energy. Whilst the general form $L_{u}/\lambda =
Re_{0}^{1/2} fct ({x-x_{0}\over L_{B}})$ may be universal, including
fractal-generated turbulent flows, the actual function of
${x-x_{0}\over L_{B}}$ in this form is not and can even be of a type
which does not allow to collapse the $x$ and $Re_0$ dependencies by
the Richardson-Kolmogorov cascade form $L_{u}/\lambda \sim
Re_{\lambda}$.

This issue certainly impacts on the very turbulence interscale
transfer mechanisms, in particular vortex stretching and vortex
compression which are considered to have qualitatively universal
properties such as the tear drop shape of the Q-R diagram
\citep{Tsinober2009}. Multi-hot wire anemometry \citep{Gulitskietal2007}
applied to turbulence generated by low-blockage space-filling fractal
square grids may have recently revealed very unusual Q-R diagrams
without clear tear-drop shapes
\citep{KholmyanskyTsinober2009}. Fractal-generated turbulence presents
an opportunity to understand these interscale transfer mechanisms
because it offers ways to tamper with them.

The decoupling between $L_{u}/\lambda$ and $Re_{\lambda}$ can be
explained in terms of a self-preserving single-length scale type of
decaying homogeneous turbulence \citep{GeorgeWang2009} but not in terms
of the usual Richardson-Kolmogorov cascade (\citep{Batchelor1953},
\citep{Frisch1995}, \citep{Pope2000}, \citep{SagautCambon2008}) and its
cornerstone property, $\epsilon \sim u'^{3}/L_{u}$.
This self-preserving single-scale type of turbulence allows for
$L_{u}/\lambda$ to increase with inlet Reynolds number $Re_0$, as we
in fact observe. This is a case where the range of excited turbulence
scales depends on a global Reynolds number but not on the local
Reynolds number.  

Our data support the view (both its assumptions and consequences) that
decaying homogeneous turbulence in the decay region of some
low-blockage space-filling fractal square grids is a self-preserving
single-length scale type of decaying homogeneous turbulence
\citep{GeorgeWang2009}. Furthermore, our detailed analysis of our data
suggests that such fractal-generated turbulence might be extrapolated to
have the following specific properties at high enough inlet Reynolds
numbers $Re_{0}$:
\begin{equation} 
E_{u}(k_{x}, x) = u'^{2} (x) L_{0}
(k_{x}L_{0})^{-5/3} H_{u} (k_{x}L_{0} Re_{0}^{-3/4}), 
\end{equation}
\begin{equation} 
u'^{2} (x) \approx u_{0}'^{2} e^{-2x/x_{*}},
\end{equation} 
\begin{equation} 
\epsilon \approx 3 u'^{2} U_{\infty}/x_{*},  
\end{equation} 
\begin{equation} 
L_{u} \sim L_{0} 
\end{equation} 
and 
\begin{equation} 
\lambda \sim L_{0}Re_{0}^{-1/2} 
\end{equation} 
where both $L_u$ and $\lambda$ are independent of $x$. A more detailed
account of our conclusions involves the two types of single-scale
solutions of the spectral energy equation, the $c=0$ and the $-1<c<0$
types introduced in subsections 4.4 and 4.5. In subsection 4.8 we
showed how our data indicate that the turbulence in the decay region
is of the $-1<c<0$ type with a value of $c$ which tends to $0$ as
$Re_0$ increases. This is why we stress the asymptotic extrapolations
(53), (54), (55) and (57) in this conclusion.

Our data require a very clear departure from the usual views
concerning high Reynolds number turbulence \citep{Batchelor1953},
\citep{Frisch1995}, \citep{Lesieur2008}, \citep{Pope2000},
\citep{SagautCambon2008}, \citep{TennekesLumley1972}. There is
definitely a need to investigate these suggested high-$Re_{0}$
properties further. Measurements with a wider range of fractal grids
and a wider range of inlet velocities in perhaps a wider range of wind
tunnels and with a wider range of measurement apparatus: x-wires,
multi-hot wire anemometry \citep{Gulitskietal2007},
\citep{KholmyanskyTsinober2009} and particle image velocimetry. Direct
Numerical Simulations (DNS) of fractal-generated turbulent flows are
only now starting to appear \citep{Nagataetal2008},
\citep{Laizetetal2009} and their role will be crucial. Amongst other
things, these studies will reveal dependencies on inlet/boundary
geometrical conditions $*$ which we have not been able to fully
determine here because of the limited range of fractal grids at our
disposal.

A quick discussion of the features of extrapolations (53)-(57) reveals
the various issues that they raise. The first issue which immediately
arises is the meaning of $Re_{0}\to \infty$. We cannot expect this
limit to lead to (53)-(57) if it is not taken by also increasing the
number $N$ of iterations on the fractal turbulence generator. How do
our results and the extrapolated forms (53)-(57) depend on $N$?

Secondly, in the extrapolated spectral form (53) we have assumed that
the exponent $p$ tends to $5/3$ in the high-$Re_0$ limit and have
therefore, in particular, neglected to consider any traditional
small-scale intermittency corrections (see \citep{Frisch1995}). This
may be consistent with the observation of Stresing et al
\citep{Stresingetal2009} that small-scale intermittency is independent
of $Re_{\lambda}$ in the decay region of our flows. However it is not
clear why $p$ should asymptotically equal $5/3$ in the non-Kolmogorov
context of our self-preserving single-scale decaying homogeneous
turbulence. In particular, the inner length-scale $L_{0}
Re_{0}^{-3/4}$ differs from the Kolmogorov microscale $\left(
\nu^{3}/\epsilon \right)^{1/4}$ which scales as $L_{0} Re_{0}^{-3/4}
(t_{0}/L_{0})^{1/2} (u'/U_{\infty})^{-1/2}$ if account is taken of
(55). If $L_{0} Re_{0}^{-3/4}$ in (53) was to be replaced by this
Kolmogorov microscale, then (57) would fail and the single-length
scale framework of George \& Wang \citep{GeorgeWang2009} would fail
with it. Why is the Kolmogorov microscale absent, or at least
apparently absent, from decaying homogeneous turbulence in the decay
region of some low-blockage space-filling fractal square grids?

Thirdly, (55) suggests that the kinetic energy dissipation rate per
unit mass is proportional to $u'^{2}$ rather than $u'^{3}$ and that
the turnover time scale is the global $x_{*}/U_{\infty}$ rather than
the local $L_{u}/u'$. What interscale transfer mechanisms cause one or
the other dependencies, and what are the implied changes in the vortex
stretching and vortex compression mechanisms hinted at by the recent
preliminary Q-R diagram results of Kholmyansky and Tsinober
\citep{KholmyanskyTsinober2009}? These issues directly address the
universality questions raised in the Introduction and depend on the
mechanisms of turbulence generation in the production region and the
mechanisms which force important features of particular turbulence
generations to be or not to be remembered far downstream from the
initial generator. What is the role of coherent structures, large or
small, in shaping the type of homogeneous turbulence which decays
freely in the decay region?

Fourthly, is it possible that turbulence in various instances in
industry and nature (e.g. in or over forest canopies, coral reefs,
complex mountainous terrains, etc) might appear as a mixture of
single-scale self-preserving turbulence and Richardson-Kolmogorov
turbulence? Could such mixtures of two types of different turbulence
give rise to what may appear as Reynolds number and intermittency
corrections to the usual Richardson-Kolmogorov phenomenology and
scalings?

As a final note, it is worth comparing (55) with the usual estimate
$\epsilon = C_{\epsilon} u'^{3}/L_{u}$, which can also be seen as a
general definition of the dissipation constant $C_{\epsilon}$. One
gets
\begin{equation}
C_{\epsilon} \approx {3\over 5} {(t_{0}/L_{0})\over (u'/U_{\infty})}
\end{equation}
where use has been made of the estimate $L_{u} \approx 0.2 L_{0}$
extracted from figures 19. The dissipation constant $C_{\epsilon}$ is
not only clearly not universal, it can also be given bespoke values by
designing the geometry of the turbulence-generating fractal grid,
i.e. by changing the aspect ratio $t_{0}/L_{0}$. Furthermore, whilst a
constant and universal value of $C_{\epsilon}$ would imply that, given
a value of $L_u$, the level of turbulence dissipation cannot come
without an equivalent pre-determined level of turbulence fluctuations,
(55) and (58) show that it actually is possible to generate an intense
turbulence with reduced dissipation and even design the level of this
dissipation. The implications for potential industrial flow
applications are vast and include energy-efficient mixers (see
\citep{Coffeyetal2009}) and lean premixed combustion gas turbines on
which we will report elsewhere.

%

\vskip 1truecm

\noindent
{\bf Acknowledgements:} We acknowledge Mr Carlo Bruera's and Mr Stefan
Weitemeyer's assistance with the anemometry data collection.

\begin{small}
\bibliography{}

	\end{small}

\end{document}